%% file: draft_jcap_resubmission.tex
\documentclass[a4paper,11pt]{article}
\pdfoutput=1 

\usepackage{jcappub} 

\usepackage[T1]{fontenc} 
\usepackage{color}
\usepackage{textcomp}

\usepackage{hyperref}

\usepackage{soul}
\usepackage{ulem}

\newcommand{\ud}{\mathrm{d}}
\usepackage{import}

\usepackage{comment}

\title{The Dipole of the Astrophysical Gravitational-Wave Background}

\author[a,b]{Lorenzo Valbusa Dall'Armi,}
\author[a,b]{Angelo Ricciardone,}
\author[a,b,c]{and Daniele Bertacca}

\affiliation[a]{Dipartimento di Fisica e Astronomia ``G. Galilei'',
Universit\`a degli Studi di Padova, via Marzolo 8, I-35131 Padova, Italy}
\affiliation[b]{INFN, Sezione di Padova,
via Marzolo 8, I-35131 Padova, Italy}
\affiliation[c]{INAF - Osservatorio Astronomico di Padova, Vicolo dell'Osservatorio 5, I-35122 Padova, Italy.}

\emailAdd{lorenzo.valbusadallarmi@phd.unipd.it}
\emailAdd{angelo.ricciardone@pd.infn.it}
\emailAdd{daniele.bertacca@pd.infn.it}

\abstract{One of the main pillars of the $\Lambda$CDM model  is the Cosmological Principle, which states that our Universe is statistically isotropic and homogeneous on large scales. Here we test this hypothesis using the Astrophysical Gravitational Wave Background (AGWB) expected to be measured by the Einstein Telescope-Cosmic Explorer network; in particular we perform a numerical computation of the AGWB dipole, evaluating the intrinsic contribution due to clustering and the kinematic effect induced by the observer motion. We apply a component separation technique in the GW context to disentangle the kinematic dipole, the intrinsic dipole and the shot noise (SN), based on the observation of the AGWB at different frequencies. We show how this technique can also be implemented in matched-filtering  to minimize the covariance which accounts for both instrumental noise and SN. Since GW detectors are essentially full-sky, we expect that this powerful tool can help in testing the isotropy of our Universe in the next future.}

\begin{document}

\begin{flushleft}
ET-0120A-22
\end{flushleft}

\maketitle

\section{Introduction}
The largest fluctuation in the CMB angular power spectrum is the dipole and it is due to the motion of the observer w.r.t. the CMB rest frame~\cite{Kogut:1993ag,Lineweaver:1996xa,WMAP:2008ydk,Planck:2013kqc,Planck:2020qil}. At the same time, the observed CMB ``anomalies'' at low multipoles (e.g., dipolar asymmetry) could suggest a violation of the statistical isotropy~\cite{Schwarz:2015cma,Planck:2019evm} and they could provide possible hints of new physics; therefore it is quite natural to investigate the dipole of different observables as a consistency check of the $\Lambda$CDM. In the past years, statistical isotropy has been tested by looking at the dipole of Large-Scale Structure (LSS) surveys at various frequencies, for example for 2MASS and 2MRS in the IR~\cite{Gibelyou:2012ri}, for NVSS~\cite{Rubart:2013tx,Tiwari:2015tba,Chen:2015wga} and for TGSS~\cite{Bengaly:2017slg} looking at radio galaxies, and for WISE~\cite{Secrest:2020has} for quasars. Most of these works have found dipoles with direction similar to the CMB one, but with an unexpected large amplitude, which appears to be in conflict with the Cosmological Principle. In LSS surveys, there are several sources of error, such as the Shot Noise (SN), due to the discreteness of the observed sources~\cite{Crawford:2008nh,Rubart:2013tx}, the partial sky coverage of the survey, which generates a bias in the amplitude, a degeneracy between the components of the dipole along different directions, and mode coupling between different multipoles. Many of these issues have been studied systematically in~\cite{Gibelyou:2012ri}. Another trouble in estimating the kinematic dipole in LSS surveys is due to the contamination of the signal due to the intrinsic anisotropies~\cite{Maartens:2017qoa}. In particular, the intrinsic dipole due to clustering can give a non-negligible contribution to the total dipole and it has to be properly removed. In~\cite{Nadolny:2021hti} the intrinsic dipole has been subtracted by combining different observables in a galaxy survey. Recently the physical origin of CMB anomalies has been studied also using the Cosmological Gravitational Wave Background detectable by future GW detectors~\cite{Galloni:2022rgg}.

In this paper we have proposed an alternative way to estimate our peculiar velocity, based on the analysis of the dipole of the astrophysical gravitational wave background (AGWB), generated by the superposition of unresolved signals emitted by astrophysical sources~\cite{Ferrari:1998jf,Ferrari:1998ut,Phinney:2001di,Regimbau:2011rp}. The recent analysis by LIGO/Virgo/KAGRA~\cite{LIGOScientific:2021psn} has shown that the number of mergers of compact objects in the local Universe is quite high, for example for binary black holes (BBH) $R^{\rm BBH}(z=0)=19\, \rm Gpc^{-3}\, yr^{-1}$. The population inferred from this local merger rate exhibits a large enough number of GW sources that can generate a stochastic background that could be detected by upgraded and future GW interferometers~\cite{Maggiore:2019uih,Baker:2019ync,Seoane:2021kkk,Evans:2021gyd}. The upper bound on the total AGWB set by LIGO/Virgo/KAGRA is $\Omega_{\rm AGWB}\leq 3.4\times 10^{-9}$ at 25 Hz~\cite{KAGRA:2021kbb}. Such an AGWB is characterized by a dominant isotropic contribution (the monopole) and by small anisotropies. AGWB anisotropies are generated by different effects: The first contribution is due to the inhomogeneities in the GW sources inherited from the perturbations in the matter distribution of our Universe (intrinsic anisotropies). The computation of these intrinsic anisotropies have been performed for the first time in~\cite{Cusin:2017fwz,Cusin:2017mjm,Cusin:2018rsq,Jenkins:2018kxc,Jenkins:2018uac,Pitrou:2019rjz}, while in~\cite{Bertacca:2019fnt} the authors kept into account for all the relativistic terms in a general covariant setting. A second contribution is given by the fluctuations in the number of the sources that generate the background, which follow a Poisson distribution, generating a SN term~\cite{Jenkins:2019nks,Jenkins:2019uzp}. The last contribution comes from the velocity of the observer w.r.t. the LSS rest frame (kinematic dipole). The AGWB kinematic dipole has been computed using a coordinate independent and gauge-invariant formalism in~\cite{Bertacca:2019fnt}. The dipole induced by the observer motion has been evaluated for the stochastic gravitational wave background of cosmological origin (CGWB) in~\cite{Allen:1996gp,Jenkins:2018nty,Cusin:2022cbb}. In these works, the kinematic dipole has been evaluated as a Doppler boosting of the energy density of the CGWB by using the relative velocity of the observer w.r.t. the CGWB rest frame. In the present work we have quantified the kinematic dipole in a different way, keeping track of the relative velocity between the observer and the sources at different redshifts. This computation leads to the so-called Kaiser-Rocket effect~\cite{Kaiser1987}, where the kinematic dipole depends on the observer's velocity weighted by a Kaiser-Rocket factor, which depends on the Hubble expansion and on the evolution of the sources in time. We have performed a numerical computation of the AGWB intrinsic and kinematic dipole for a population of BBH of masses between $2.5$ and $100$ $M_\odot$, according to the latest LIGO/Virgo constraints~\cite{LIGOScientific:2021psn}. We have taken into account all the terms in the intrinsic dipole, and we have considered redshift-dependent bias and evolution bias up to redshift $z\sim 8$.

Even if the AGWB carries a lot of interesting physical information, both on the astrophysical and on the cosmological side, the low signal-to-noise ratio at present and future interferometers could hinder the power of this observable. The main issue in detecting the anisotropies of stochastic backgrounds is due to the fact that the instrumental noise is larger than the GW spectrum. The standard way to circumvent this problem is to use matched filtering, by convolving the signal in the frequency domain with a filter which is chosen in order to maximize the SNR. This technique has been applied not only to the detection of the monopole~\cite{Maggiore:2007ulw,Flauger:2020qyi}, but also of the polarization~\cite{Orlando:2020oko,Amalberti:2021kzh} and of the anisotropies~\cite{Alonso:2020rar,Mentasti:2020yyd,LISACosmologyWorkingGroup:2022kbp}. While the SNR of the CGWB anisotropies is mainly limited by the instrumental noise, in the AGWB case it has been shown that the SN is at least one order of magnitude larger than the intrinsic anisotropies~\cite{Jenkins:2019nks,Jenkins:2019uzp,Bellomo:2021mer}, therefore it would be hard to measure them. The standard way to reduce the SN is to compute the cumulative SNR for several multipoles or to exploit the  cross-correlation of the AGWB with other cosmological probes, such as LSS~\cite{Canas-Herrera:2019npr,Mukherjee:2019oma,Alonso:2020mva,Yang:2020usq} or the CMB~\cite{Ricciardone:2021kel,Capurri:2021prz}. In~\cite{Alonso:2020mva} it has been shown that neglecting the instrumental noise, the cumulative SNR of the cross-correlation between the AGWB and the galaxy number count is larger than one if we sum the contributions up to $\ell_{\rm max}\gtrsim 10$. In our case, however, we would like to extract information from a single multipole (the dipole) therefore we look for a more efficient way to reduce SN. We have computed indeed the SNR of both auto- and cross-correlation of the AGWB with a galaxy survey, showing that using the standard approach the SNR of the dipole is much smaller than one. So we have exploited the frequency dependence of the three contributions (i.e., intrinsic, kinematic, and SN) to the AGWB anisotropies to perform component separation and to isolate the kinematic dipole with very high accuracy. The underlying idea is that the AGWB is given by the superposition of the GW signal emitted by binary systems, keeping into account for all the evolutionary stages of the binary, the inspiral, the merger, and the ringdown. The inspiral phase gives a $f^{2/3}$ contribution to the monopole, while the other two stages have more complicated parametrizations~\cite{Ajith:2007kx,Ajith:2009bn,Ajith:2012mn}. Since the signals emitted at different stages do not scale in the same way with the frequency, the window function involved in the computation of the AGWB anisotropies depends on the frequency, and so the evolution bias of the GW sources. Thus the kinematic dipole changes in a different way with the frequency w.r.t. the intrinsic and the SN anisotropies, basically due to the Kaiser-Rocket factor. This allows us to isolate the kinematic dipole in the same way galactic foregrounds are removed in CMB experiments~\cite{Planck:2018yye}. We started by performing component separation in the ideal case were the instrumental noise is neglected. We have found that it is possible to isolate the kinematic dipole from the SN and the intrinsic dipole by simply using Internal Linear Combination (ILC)~\cite{Tegmark:1999ke,Tegmark:2003ve}, reducing of more than a factor 10 the error on the kinematic dipole estimate due to SN. Using this technique is possible to generate full maps (i.e., sum of the intrinsic, kinematic, and SN dipole) and to compare the true kinematic dipole map with the cleaned one from the total signal, illustrating that we are able to separate the three contributions. Then, taking into account both the SN and the instrumental noise we have generalized the previous result, deriving a new estimator for the AGWB map. This estimator has the smallest possible covariance and it allows to remove completely the SN. It has been derived for a generic network of GW detectors. However, since one of the best candidate to detect AGWB anisotropies produced by BBHs of solar mass type is the Einstein Telescope (ET)-Cosmic Explorer (CE) network, we have then derived the kinematic dipole estimate for ET+CE network~\cite{Maggiore:2019uih,ETsens,ce}.

The techniques introduced in this paper are not only useful for removing the SN, but they automatically allow to disentangle the kinematic and the intrinsic dipoles. Therefore, for sufficiently large GW monopole amplitudes, we have shown how we can measure, without spurious contaminations due to intrinsic anisotropies, our local velocity.  To our knowledge, we have introduced for the first time in the GW context a component separation technique to disentangle different contributions to the AGWB spectrum. 
In our analysis, we have obtained ${\rm SNR}\approx 10$ for the kinematic dipole by considering SN only, and ${\rm SNR}\approx 2.5$ by considering SN and instrumental noise for ET+CE. In the latter case, the result can be improved for more sensitive interferometers and for different sources considered, for example by looking at the superposition of the AGWB produced by BBH, BNS and BHNS at the same time. 

The structure of the paper is the following: in Section 2 we computed the dipole of the AGWB and the SN contribution; in Section 3 we introduced a new technique to do component separation and we derived the new estimator for the AGWB dipolar map; finally in the Conclusions we summarized our results and we highlighted some possible applications.

\section{Computation of the AGWB Dipole}
\label{Computation of the AGWB dipole}
\subsection{AGWB Anisotropies}
The AGWB is generated by the signal superposition of many unresolved astrophysical sources, which emit GWs with a strain not large enough to be detected with $\rm SNR$ larger than a certain threshold ${\rm SNR}_{\rm thr}$. The value of the threshold ${\rm SNR}_{\rm thr}$ depends on the number of interferometers and on the significance above which we claim a detection~\cite{Finn:1995ah}. In this work we have considered the network ET+CE and for the SNR threshold we have chosen the value $\rm SNR_{\rm thr}=12$~\cite{Maggiore:2019uih}.
Many astrophysical sources can produce an AGWB, such as rotating neutron stars, core collapse supernovas or compact objects coalescences~\cite{Regimbau:2011rp}. In this work we will focus on BBH mergers with masses within the LIGO/Virgo range. This choice is motivated by the fact that BBH mergers are expected to be among the dominant source of AGWB, by looking at the most recent constraints on the BBH merger rate and mass distribution~\cite{LIGOScientific:2021psn}. However, the formalism developed here is completely general and can be adapted to any kind of discrete source of GWs, such as Neutron Star Binaries (BNS)~\cite{Chen:2018rzo,Perigois:2021ovr} or even Primordial Black Holes of both early~\cite{Sasaki:2018dmp} and late type~\cite{Bird:2016dcv}.

The monopole amplitude of the AGWB can be computed by using the energy spectrum emitted by a BBH system in the inspiral, merger, ringdown phases~\cite{Ajith:2007kx,Ajith:2009bn,Ajith:2012mn},
\begin{equation}
\bar{\Omega}_{\rm AGWB}(f_o) = \frac{f_o}{\rho_{\rm c}c^2} \int \frac{\ud z}{(1+z)H(z)}R^{\rm BBH}(z)\int \ud\vec{\theta}\, p(\vec{\theta})\, w(z,\vec{\theta})\, \frac{\ud E}{\ud f_e \ud \Omega_e}(f_e,\vec{\theta})\Biggl|_{f_e=(1+z)f_o}\, ,
\end{equation}
where the window function of the detector $w$ is related to detector efficiency and it represents the fraction of sources described by $R^{\rm BBH}(z)$ that are not individually resolved and thus contribute to the AGWB~\cite{Bertacca:2019fnt, Bellomo:2021mer, LISACosmologyWorkingGroup:2022kbp}. The expression for the energy emitted by a source used here is valid only for short-lived sources, such as BBH mergers in the mass range of ET+CE, while for periodic long-lived sources we should average the GW emission over several periods of the slow evolution of orbital parameters~\cite{Bertacca:2019fnt}.  In our computation we are also averaging w.r.t. the astrophysical parameters $\vec{\theta}$. We have neglected the spin of the BHs since we expect that the spin does not have a great impact on the signal~\cite{Zhu:2011bd}. What we are doing is basically an average w.r.t. the mass distribution of the BBHs only.  As a mass function, we have considered a Power Law + Peak model taking into account the latest constraints~\cite{LIGOScientific:2021psn}.  However, the validity of the technique to extract the kinematic dipole from AGWB measurements that we will present here does not rely on a specific mass distribution. We leave for a future work the discussion of our technique used in a joint-analysis of the resolved sources and the AGWB, estimating the astrophysical parameters and the kinematic dipole together. Following \cite{Bellomo:2021mer}, the merger rate has been computed taking into account the properties of the GW hosts:
we look at the cosmic star-formation rate per halo of mass $M_h$ at redshift $z$ provided by UniverseMachine~\cite{UniverseMachine}, $\langle {\rm SFR}(M_h,z)\rangle_{\rm SF}$, from which we can derive the merger rate of BBH as
\begin{equation}
R^{\rm BBH}(z) = \mathcal{A}_{\rm LIGO}^{\rm BBH}\int \ud t_d \; p(t_d)\int \ud M_h \; \frac{\ud n}{\ud M_h}(z_f,M_h)\;\langle {\rm SFR}(M_h,z_f)\;\rangle_{\rm SF}\, ,
\end{equation}
where $t_d$ is the time delay between the formation of the binary and its merger and $z_f$ is the redshift at which the binary system formed, $z_f(t_d,z) \equiv z(t-t_d)$. For the time delay distribution we have considered an inverse power-law ~\cite{Mapelli:2017hqk}, between $t_d^{\rm min}= 50\, \rm Myr$ and the age of the Universe at the emission of the GWs $t(z)$, 
\begin{equation}
p(t_d) = \ln\left(\frac{t(z)}{t_d^{\rm min}}\right)\; \frac{1}{t_d}\, .
\end{equation}
The halo mass function has been taken from~\cite{Tinker:2008ff}, using also fitting formulas from~\cite{Ludlow:2016ifl,Lahav:1991wc}. The normalization factor $\mathcal{A}_{\rm LIGO}^{\rm BBH}$ has been introduced in order to match the local merger rate with the one estimated by LIGO/Virgo~\cite{LIGOScientific:2021psn}, $R^{\rm BBH}(0) = 19$ $\rm{Gpc}^{-3} \, \rm{yr}^{-1}$.
$\mathcal{A}_{\rm LIGO}^{\rm BBH}$ contains information about the probability that a star becomes a compact object and that a binary system of two compact objects forms.

To compute the AGWB anisotropies we have followed the approach of~\cite{Bertacca:2019fnt,Bellomo:2021mer}, where the Cosmic Ruler formalism has been applied~\cite{Schmidt:2012ne}. In this framework we are able to obtain coordinate independent and gauge invariant results, keeping into account for all possible effects along the past GW-cone. 
In terms of the AGWB density contrast they are given by,
\begin{equation}
\delta_{\rm AGWB}(f_o,\hat{n}) \equiv \frac{\Omega_{\rm AGWB}(f_o,\hat{n})-\bar{\Omega}_{\rm AGWB}(f_o)}{\bar{\Omega}_{\rm AGWB}(f_o)} = \int \ud z \, \tilde{W}(f_o,z)\, \Delta_{\rm AGWB}(f_o,\hat{n},z)\, ,
\label{delta_equation}
\end{equation}
where $\Delta_{\rm AGWB}$ is the source function that encodes contribution from density perturbations, redshift-space distorsions (rsd), GR effects, and, of course, the proper motion of the observer w.r.t. the sources. The window function $\tilde{W}$ weights the contributions of $\Delta_{\rm AGWB}$ to $\delta_{\rm AGWB}$ at the observed frequency $f_o$ and at redshift $z$ ~\cite{Bertacca:2019fnt, Bellomo:2021mer, LISACosmologyWorkingGroup:2022kbp}. It is equal to the energy flux of the GWs emitted by all the sources at redshift $z$ with emitted frequency $f_o(1+z)$, normalized w.r.t. the background monopole amplitude at $f_o$,
\begin{equation}
\tilde{W}(z,f_o) \equiv \frac{f_o}{\rho_c c^2}\frac{1}{\bar{\Omega}_{\rm AGWB}(f_o)}\frac{R^{\rm BBH}(z)}{(1+z)H(z)}\int \ud\vec{\theta}\, p(\vec{\theta}\,)\, w(z,\vec{\theta})\,\frac{\ud E_{\rm GW}}{\ud f_e\ud \Omega_e} (f_e,\vec{\theta})\Biggl |_{f_e=(1+z)f_o}\, .
\label{effective_window_function}
\end{equation}
When we have two stochastic fields, $\delta_X$ and $\delta_Y$, it is useful to work in Fourier space, in order to separate large-scale and small-scale contributions, 
\begin{equation}
\delta_X(\eta,\vec{x})=\int\frac{\ud^3 \vec{k}}{(2\pi)^3}\, e^{i\vec{k}\cdot\vec{x}}\delta_X(\eta,\vec{k})\, .
\end{equation}
In order to factor out the angular dependence w.r.t. to the direction of observation in the sky $\hat{n}$ we expand the Fourier transform of the fields in Legendre polynomials,
\begin{equation}
\Delta_\ell^X(\eta,k)\equiv \int \ud\phi \, \int \ud \mu\, \mathcal{P}_\ell(\mu)\, e^{i\vec{k}\cdot\vec{x}}\,\delta_X(\eta,\vec{k})\, ,  
\end{equation}
where $\mu$ is the angle between $\hat{n}$ and $\hat{k}$, while $\phi$ is the azimuthal angle in the plane perpendicular to $\hat{n}$. The angular power spectrum of the fields $\delta_X$, $\delta_Y$ can be written in terms of the source functions $\Delta_\ell^{\rm X/Y}$ for the various contributions,
\begin{equation}
C_\ell^{XY} = 4\pi \int \frac{\ud k}{k}\, P(k) \, \Delta_\ell^X \Delta_\ell^{Y\, *}\, ,
\label{aps_equation}
\end{equation}
with $P(k)$ primordial scalar power spectrum. Typically the angular power spectrum does not depend on the frequency. On the contrary, in the case of the AGWB, it depends on the frequency and this allows to reduce the SN as we will show.

The detailed computation of the AGWB anisotropies has been done in Appendix \ref{AGWB Anisotropies Computation} and all the contributions to the AGWB $\ell$-source function are listed in Eq. \eqref{source_omega_agwb}.

\subsection{Contributions to the AGWB Anisotropies}
A vector $\vec{d}$ generates a dipolar signature in a map when it contributes to the map with a term like $\hat{n}\cdot\vec{d}$,
since in the harmonic space this becomes
\begin{equation}
    \int \ud \hat{n}\, Y_{\ell m}^*(\hat{n})\,\hat{n}\cdot\vec{d}\propto \delta_{1\ell} \, .
\end{equation}
The AGWB density constrast map is the sum of three different uncorrelated contributions,
\begin{equation}
    \delta_{\rm AGWB}(f,\hat{n}) = \delta_{\rm AGWB}^{\rm int}(f,\hat{n})+\delta_{\rm AGWB}^{\rm SN}(f,\hat{n})+\mathcal{R}(f)\,\vec{v}_o\cdot \hat{n}\, ,
\end{equation}
where the first identifies the anisotropies generated by clustering and GR effects (see Section \ref{Intrinsic Dipole}), while the second one is due to the SN fluctuations of the number of the discrete sources that generate the background (see Section \ref{Shot Noise Computation}), while the third one is the kinematic dipole induced by the local motion of the observer (see Section \ref{Kinematic dipole}). Note that the first two contributions produce anisotropies also at multipoles larger than one, while the kinematic dipole term affects only the $\ell=1$ term for full-sky surveys.

\begin{figure}
    \centering
    \includegraphics[scale=0.5]{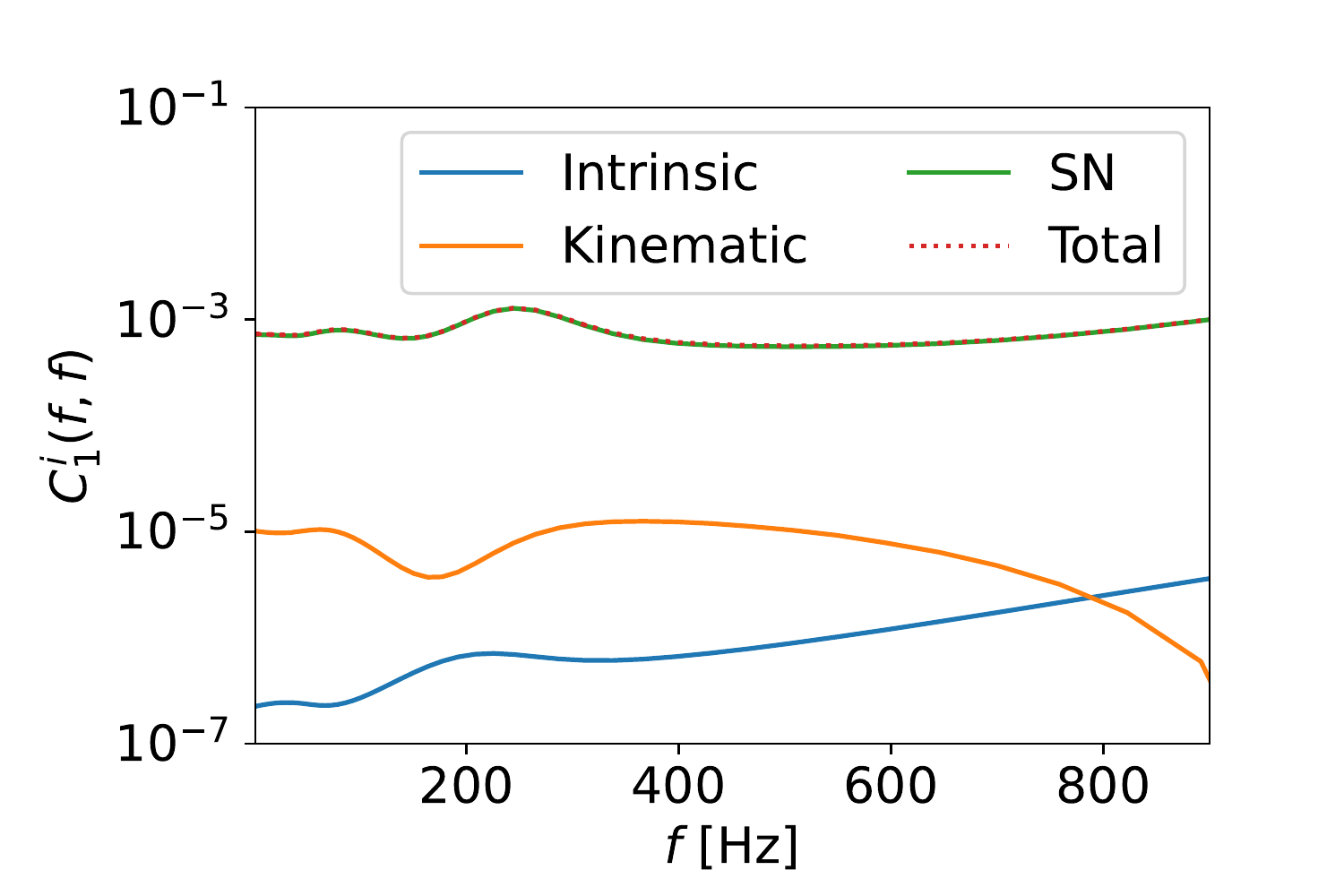}
    \includegraphics[scale=0.5]{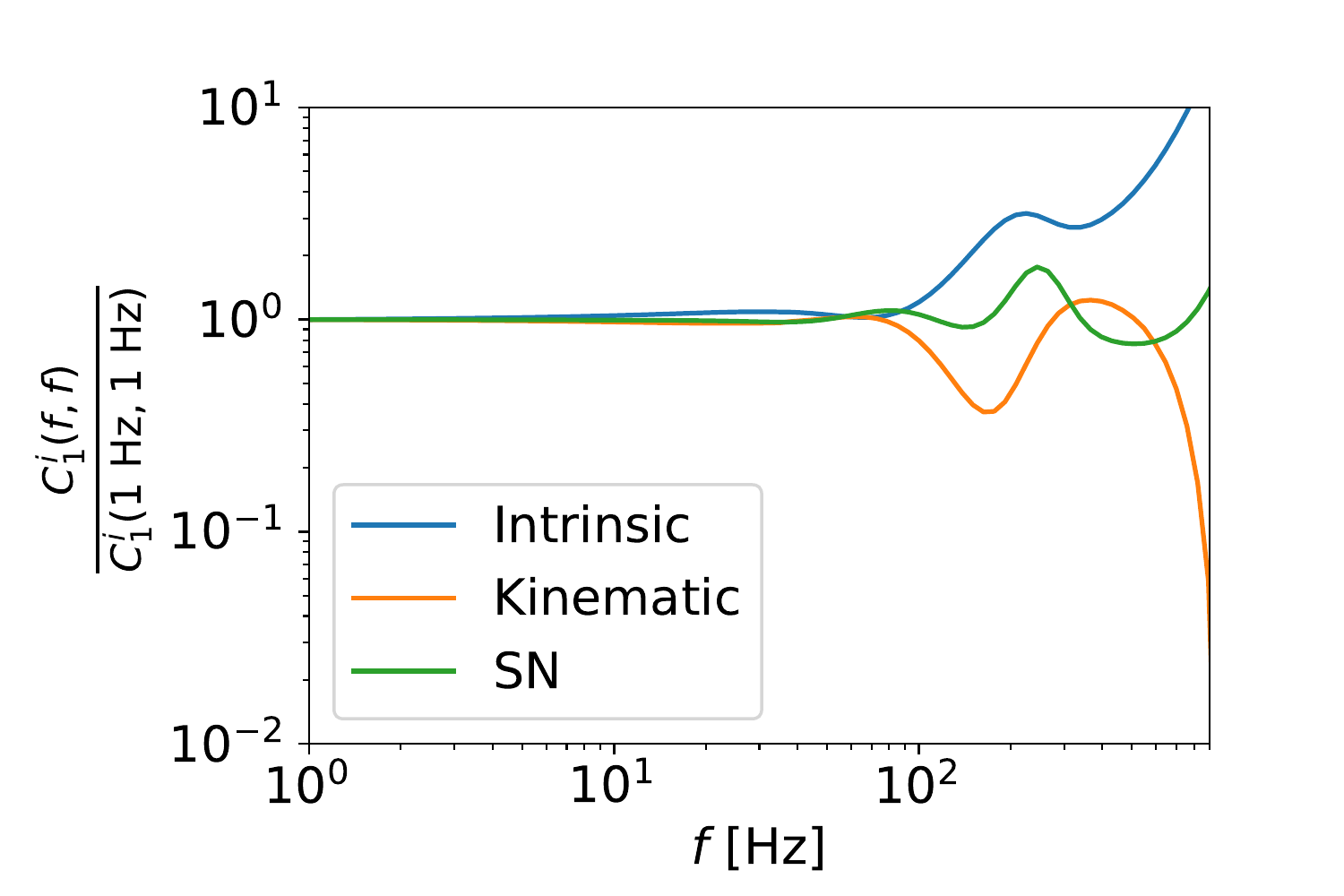}
\caption{{\it Left: plot of the intrinsic, SN, KD, and total contribution to the $\ell=1$ term of the angular power spectrum of the AGWB at different frequencies, for ET+CE. For simplicity, we did not plot the $C^i_1(f,f^\prime)$ spectra, with $f\neq f^\prime$, but we have focused on the auto-correlation only. Right: plot of the intrinsic, SN, KD contributions to the auto-correlation spectra of the AGWB normalized w.r.t. their values at 1 Hz. We have normalized the dipoles at $1\,\rm Hz$ to show more explicitly how they change in frequency. Up to $80\, \rm Hz$ the differences between the contributions are very small, but at higher frequencies the degeneracies are broken and we are able to distinguish between the components of the spectrum.}}
    \label{all_dipoles_figure}
\end{figure}

To connect the configuration space with the angular power spectra space we simply decompose the fields in spherical harmonics,
\begin{equation}
    \delta_{\rm AGWB,\ell m}(f) \equiv \int \ud \hat{n}\, Y_{\ell m}^*(\hat{n}) \, \delta_{\rm AGWB}(f,\hat{n}) =\delta^{\rm int}_{\rm AGWB,\ell m}(f)+\delta^{\rm SN}_{\rm AGWB,\ell m}(f)+\delta^{\rm KD}_{\rm AGWB,\ell m}(f)\, ,
\end{equation}
and we compute the angular power spectrum by using 
\begin{equation}
\begin{split}
    \delta_{\ell \ell^\prime}\delta_{m m^\prime} C_\ell^{\rm AGWB}(f,f^\prime) \equiv & \langle \delta_{\rm AGWB,\ell m}(f)\delta_{\rm AGWB,\ell^\prime m^\prime}(f^\prime)\rangle  \\
    = &\, C_\ell^{\rm AGWB,\rm int}(f,f^\prime)+C_\ell^{\rm AGWB,\rm SN}(f,f^\prime)+C_\ell^{\rm AGWB,\rm KD}(f,f^\prime)\delta_{\ell 1}\, .
\end{split}
\end{equation}
Notice that we have a non-zero AGWB angular power spectrum for $f\neq f^\prime$: this is the property that we use to perform component separation of the kinematic dipole contribution. The plot of the $C^i_1(f,f)$ term at different frequencies and for the three contributions is depicted in the left panel of Figure \ref{all_dipoles_figure}. We can notice that the SN fluctuations are much larger than the KD and the intrinsic ones. Since the SN depends on the sources that generate the background, it cannot be removed by increasing the sensitivity of the interferometers\footnote{Actually, it is the opposite: if the sensitivity of the instrument increases, an higher number of sources is resolved, thus less sources contribute to the AGWB and the SN increases.}, therefore we need to find a strategy to remove it in a statistical way. This is exactly the reason why we will perform a multi-frequency analysis of the AGWB anisotropies, exploiting the fact that we have a non-null cross-correlation between the spectra at different frequencies. In the right plot of Figure \ref{all_dipoles_figure} we have shown the evolution of the $C^i_1(f,f)$ spectrum for the intrinsic, SN, KD contributions, normalized w.r.t. the three contributions evaluated at 1 Hz. In this way it is immediate to see that the evolution in frequency for the three contributions is very different for $f\gtrsim 80\, \rm Hz$. This means that there is no degeneracy at such high frequencies between the three terms, which means that we can use three different templates in frequency to fit the observed signal and separate the three components in the analysis. The reason why up to $f\approx 80\, \rm Hz$ the intrinsic, the SN and the kinematic dipoles does not vary with the frequency (or vary a little and in the same way) is discussed in detail at the end of Section \ref{Kinematic dipole}. Even though this low frequencies would not be useful to separate the kinematic dipole from the other two contributions, they are useful to reduce instrumental noise, as discussed in detail in Section \ref{AGWB Kinematic Dipole Estimate with Shot Noise and Instrumental Noise}.
\subsubsection{Intrinsic Anisotropies}
\label{Intrinsic Dipole}

The intrinsic anisotropies of the AGWB, generated by the evolution of cosmological perturbations, include the perturbation of AGWB sources, due to Cold Dark Matter (CDM) density perturbations, the redshift-space distorsions (RSD), the relative velocities of the sources which emit GWs, and GR effects due to metric perturbations. 

A gauge-invariant computation of the AGWB intrinsic anisotropies has been performed in~\cite{Bertacca:2019fnt}. In Appendix \ref{appendix_agwb_anisotropies_newtonian_gauge} we show that, in the Poisson gauge, the intrinsic anisotropies are
\begin{equation}
\begin{split}
    \delta^{\rm int}_{\rm AGWB} =&\int d\bar{\chi} \, \tilde{\mathcal{W}}\Biggl[b\left(\delta_m-3\mathcal{H}V\right)+(3-b_e)\mathcal{H}V+ \left(3-b_e+\frac{\mathcal{H}^\prime}{\mathcal{H}^2}\right) \Psi +\frac{1}{\mathcal{H}}\Phi^\prime-\frac{1}{\mathcal{H}}\partial_\parallel v_\parallel \\
    &+2\left(b_e-\frac{\mathcal{H}^\prime}{\mathcal{H}^2}-2\right)I-\left(-b_e+\frac{\mathcal{H}^\prime}{\mathcal{H}^2}+2\right)v_\parallel -\frac{1}{\mathcal{H}}\frac{1}{2}h^{\rm TT \, \prime}_{ij}n^in^j\Biggl]\, ,
\end{split}
\end{equation}
where $b$ is the GW bias, while $b_e$ is the evolution bias and it is discussed in detail in Section \ref{Kinematic dipole}. We have introduced the comoving distance $\bar{\chi}$, which is related to the conformal time through $\bar{\chi}\equiv \eta_0-\eta$, with $\eta_0$ the conformal time at the present epoch. The prime here denotes the derivative w.r.t. conformal time. Similar expressions for the electromagnetic (EM) analogues have been derived in~\cite{Challinor:2011bk,Jeong_2012}. Even though from a mathematical point the equations which describe the anisotropies of the AGWB and of the galaxy number count are almost the same, on the physical side there are relevant differences. Different observables depend indeed on a different way from the frequency, and this could be used to perform component separation between various contributions and to estimate more precisely cosmological parameters.

The angular power spectrum of the intrinsic anisotropies is 
\begin{equation}
    C_\ell^{\rm int}(f_o,f_o^\prime)=4\pi\int \frac{dk}{k}P(k)\Delta_\ell^{\rm int}(f_o,\eta,k)\Delta_\ell^{\rm int\, *}(f_o^\prime,\eta,k)\, ,
\end{equation}
where the source function $\Delta_\ell^{\rm int}$ is the sum of the density, RSD, and GR contributions listed in Eq. \eqref{source_omega_agwb}. The angular power spectrum of the intrinsic dipole has been computed with a modified version of CLASSgal~\cite{DiDio:2013bqa} and the result is plotted in Figure \ref{all_dipoles_figure}.

\subsubsection{Kinematic Dipole}
\label{Kinematic dipole}
As for the intrinsic anisotropies, the computation of the AGWB kinematic dipole has been performed in~\cite{Bertacca:2019fnt}, in analogy with what has been done for galaxies in~\cite{Challinor:2011bk,Jeong_2012}. As shown in Appendix \ref{appendix_agwb_anisotropies_newtonian_gauge}, in the Poisson gauge the local velocity of the observer $\vec{v}_{o}$ generates a dipole pattern of the type
\begin{equation}
\delta_{\rm AGWB}^{\rm KD}(f_o,\theta)=\int \ud \bar{\chi}\, \tilde{W}(f_o,z)\left(b_e(f_o,\eta)-\frac{3H^\prime(\eta)}{a(\eta)H^2(\eta)}-3\right)\hat{n}\cdot\vec{v}_o=\mathcal{R}(f_o)\hat{n}\cdot\vec{v}_o\, ,
\label{delta_KD}
\end{equation}
where we have factorized the frequency dependence in the Kaiser-Rocket factor $\mathcal{R}(f_o)$, whose physical meaning will be discussed in detail after Eq. \eqref{kaiser_rocket_factor_equation}. We have introduced the evolution bias in terms of the energy flux of GWs $F$ by using
\begin{equation}
\begin{split}
b_e(f_o,z) \equiv&  \frac{\ud\ln F}{\ud\ln a}(f_o,z) = -\frac{1+z}{F(f_o,z)}\frac{\ud F}{\ud z}(f_o,z)= \\
=& -\frac{1+z}{R^{\rm BBH}(z)\left(\ud E_{\rm GW}/\ud f_e \ud \Omega_e \right)(f_o,z)}\frac{\ud}{\ud z} \left[R^{\rm BBH}(z)\frac{\ud E_{\rm GW}}{\ud f_e \ud \Omega_e}(f_o,z)\right] \, .
\end{split}
\end{equation}
The evolution bias takes into account the anisotropies generated by the creation of new sources. In the case of the AGWB the creation of new sources is weighted w.r.t. the energy emitted by a single source, thus the quantity involved in the evolution bias is not simply the number of sources, but the number of sources at $z$ times the energy emitted by a single source at the frequency $f_o$, which is the total flux emitted at the frequency $f_o$ at redshift $z$,
\begin{equation}
    F(f_o,z)\equiv R^{\rm BBH}(z)\frac{\ud E_{\rm GW}}{\ud f_e \ud \Omega_e}(f_o,z)\, .
\end{equation}
There are two strategies to compute the kinematic dipole~\cite{Hamilton:1997zq}: the first one consists in using the known value of the Local Group (LG) velocity measured from dipole measurements of the CMB, or of quasars or radio galaxies~\cite{Bengaly:2017slg,Secrest:2020has,Siewert:2020krp}; Another possible way exploits the fact that the LG motion is generated by the gravitational pull of the surrounding matter in the Universe. If the density perturbations can be approximated by a linear theory, the peculiar velocity is proportional to the gravitational acceleration,
\begin{equation}
    \vec{v}(t,\vec{r})\equiv \frac{g\mathcal{H}}{4\pi}\int_{\mathcal{V}^R}\ud \vec{r}^{\,\prime}\, \frac{\vec{r}^{\,\prime}-\vec{r}}{|\vec{r}^{\,\prime}-\vec{r}|^3}\delta^R(t,\vec{r}^{\, \prime})\, ,
    \label{eq_vel_lt}
\end{equation}
where $g$ is the rate of growth of perturbations computed in~\cite{Lahav:1991wc}. In the above expression we implicitly require that there is no velocity bias and that the velocity field is mainly determined by Cold Dark Matter (CDM) clustering. To compute the CDM velocity we use linear theory, relating it to the CDM density. In the above expression for the velocity field, we have smoothed rather heavily the density perturbation on small scales~\cite{Yahil1991,Lahav:1991wc}. In this way there is a one-to-one correspondence between redshift and distance~\cite{Fisher:1994xm} and it removes the issues of large velocity dispersions due to the breakdown of linear theory at small scales. The estimates of the two approaches converge in the limit in which the galaxy survey covers a large enough volume. The real observer velocity in a GW experiment is the sum of the LG velocity plus the relative velocity between the Milky Way and the LG, plus the relative velocity of the Sun w.r.t. the Milky Way, plus the relative velocity of the Earth w.r.t. the Sun. These corrections on the observer velocity are non-negligible, since the Sun motion w.r.t. the local group has about half the amplitude and opposite direction w.r.t. the LG velocity, which implies a lower kinematic dipole~\cite{Gibelyou:2012ri,Elkhashab:2021lsk}.
The angular power spectrum of the AGWB kinematic dipole is therefore 
\begin{equation}
C_\ell^{\rm KD}(f_o,f_o^\prime) = 4\pi\int \frac{\ud k}{k}P(k)\Delta_\ell^{\rm KD}(f_o,k)\Delta_\ell^{\rm KD\, *}(f_o^\prime,k)\, ,
\label{C1_kd_equation}
\end{equation}
where the source function of the kinematic dipole has been computed in Eq. \eqref{source_omega_agwb},
\begin{equation}
\Delta_\ell^{\rm KD}(f_o,k) = \frac{\delta^K_{\ell 1} }{2\ell+1} \int_0^{\eta_0} \ud\eta\, \tilde{W}^{[i]}(\eta,f_o)\left[b_e^{[i]}(f_o,\eta)- \frac{H^\prime(\eta)}{a(\eta)H^2(\eta)}-3\right]\frac{1}{k}\theta_{m\, o}(k)\, ,
\label{agwb_kinematic_dipole_definition}
\end{equation}
with $\theta_{m\, o}(k)\equiv \theta_m(\eta_0,k)$  related to the velocity field of CDM $\vec{v}$ through
\begin{equation}
\theta_m(\eta,k)= i\vec{k}\cdot \vec{v}(\eta,k)\, .
\end{equation}
In this work, we are computing the angular power spectrum of the kinetic dipole by using the velocity computed in Eq. \eqref{eq_vel_lt}, which is the velocity of the LG obtained with linear theory. As discussed before, we are not considering the motion of the Earth, of the Sun and of the Milky Way. These relative motions generate a (non-statistical) Doppler shift in the angular power spectrum of the AGWB measured in the LG frame which can be studied with the formalism discussed in~\cite{Jenkins:2018nty,Jenkins:2018uac,Cusin:2022cbb}.
In our work we study the AGWB kinematic dipole by considering only the LG velocity, assuming that the other velocities have already been subtracted.
Due to this, the dependence on the scalar product $\hat{n} \cdot \vec{v}_o$ in the AGWB density contrast 
or, equivalently, the $\delta_{\ell 1}$ factor in the source function, differs from~\cite{Cusin:2022cbb}, where the kinematic dipole of the AGWB is evaluated as a Doppler boosting with the relative velocity of the observer w.r.t. the AGWB rest frame. After the subtraction of the velocities of the Earth, of the Sun and of the Milky Way, the result of the two different approaches should converge in the limit of a SGWB generated on the surface of a sphere, as it is the case of the CMB.
However, our computation, based on the Cosmic Rulers formalism~\cite{Schmidt:2012ne}, takes into account both the emission of GWs at different times and the different velocities of the emitters. The former effect is encoded in the Kaiser-Rocket factor, (see \cite{Bertacca:2019wyg, Bahr-Kalus:2021jvu} where this definition has been used to study the dipole in the LSS)\footnote{In literature, it is also called ``Finger of the observer'' effect, e.g., see~\cite{Elkhashab:2021lsk}.},
\begin{equation}
\mathcal{R}(f_o) = \int \ud\eta\, \tilde{W}(\eta,f_o)\left[b_e(f_o,\eta)- \frac{H^\prime(\eta)}{a(\eta)H^2(\eta)}-3\right]\, ,
\label{kaiser_rocket_factor_equation}
\end{equation}
which represents a Doppler boosting over many infinitesimal shells, one per each $\eta$. This is similar to what has been done for the computation of the kinetic dipole of other astrophysical observables in the literature, e.g. for the EM analogue~\cite{Maartens:2017qoa,Challinor:2011bk,Jeong_2012}. The latter effect is encoded in the redshift-space distorsion (RSD) term, which contributes to the source function of the AGWB intrinsic anisotropies, $\Delta_\ell^{\rm int}$. As already stressed in Section \ref{Intrinsic Dipole}, even though there are analogies in the mathematical expressions between the EM case and the AGWB kinematic dipole computed in~\cite{Bertacca:2019fnt}, there are crucial physical differences, such as the frequency dependence of the observables that will be exploited in the statistical analysis of the dipole.

The frequency dependence of the kinematic dipole (in general this is true for all the anisotropies related to the AGWB) is due to the fact that the stochastic signal considered here is the superposition of the signals emitted by BH binaries of different masses, at different redshifts and at different stages of the evolution of the binary.
This means that when we observe the AGWB anisotropies at a frequency $f_o$, at redshift $z$ a binary system with parameters $M_1$ and $M_2$ contributes to the background with a frequency $f_e=(1+z)f_o$. To different observed frequencies would correspond then different emitted frequencies, which could in principle correspond to GWs emitted at different stages of the evolution of the binary, therefore the energy spectrum integrated w.r.t. all the astrophysical parameters at the redshift $z$ is different for different observed frequencies $f_o$. Since we cannot factorize the frequency dependent part of the energy spectrum with the redshift dependent one, the window function $\tilde{W}$ and the evolution bias $b_e$ depend on the frequency in a non-trivial way. To be more clear, if we would have considered only the inspiral phase of the binary, the integrated energy spectrum would have been
\begin{equation}
\frac{\ud E_{\rm GW}}{\ud f_e \ud \Omega_e}\Bigl |_{f_e=(1+z)f_o} = \int \ud M_1 \ud M_2\, p(M_1,M_2)\, \frac{\ud E_{{\rm GW}}}{\ud f_e \ud \Omega_e}(M_1,M_2,f_o,z)\propto f_o^{-1/3}(1+z)^{-1/3}\, .
\end{equation}
Therefore, since the monopole amplitude goes as $\bar{\Omega}_{\rm AGWB} \propto f_o^{2/3}$,
the window function, which depends on the combination 
   $$ \tilde{W}(z) \propto \frac{f_o \, \left(dE_{\rm GW}/df_e\right)}{ \bar{\Omega}_{\rm AGWB}(f_o)}\,,$$
would have been independent of the frequency. The same argument holds for the evolution bias. This is basically the reason why in Figure \ref{all_dipoles_figure} the intrinsic, the SN and the kinetic dipoles have the same frequency shape for $f\lesssim 80\, \rm Hz$, where all the objects with masses between $2.5 \, M_\odot$ and $100\, M_\odot$ emit GWs during the inspiral stage\footnote{Note however that a non-trivial frequency dependence could appear also at frequencies dominated by GW emission in the inspiral stage, if the proper expression for the energy emitted by continuum sources (when the binary system evolves very slowly) is considered. In this case, instead of using the merger rate times the total energy emitted by a source we should compute the number of sources and the average energy emitted over an orbital period. This computation could generate a non-factorizable frequency dependence in the spectrum, but we will discuss this in more detail in a future work, where sources of different masses and different interferometers will be considered.}. The fact that we cannot disentangle the redshift and the frequeny dependence comes from the fact that we are summing the energy contributions from the inspiral, the merger, and the ringdown,
\begin{equation}
\frac{\ud E_{\rm GW}}{\ud f_e\ud \Omega_e}\Bigl |_{f_e=(1+z)f_o} (z) = \sum_{j=\rm I, M,R} \int dM_1 dM_2\, p(M_1,M_2)\, \frac{\ud E_{{\rm GW}\, j}}{\ud f_e \ud \Omega_e}(M_1,M_2,f_o,z)\, .
\end{equation}
In Figure \ref{all_dipoles_figure} we have plotted the diagonal part of the dipole spectrum, the $C^{\rm KD}_1(f,f)$ term computed in Eq. \eqref{C1_kd_equation}, as a function of the frequency. The features of the spectrum are determined by the Kaiser-Rocket factor only.

In this section, we have computed the angular power spectrum of the AGWB kinetic dipole, $C_\ell(f_o,f_o^\prime)$. We want to stress however that in order to generate the AGWB kinematic dipole map, it is sufficient to generate the AGWB map at a given frequency $f_o$, because the kinematic dipole at any other frequency is univocally determined. This can be seen by the fact that the correlation between two kinematic dipoles at different frequencies is exactly one,
\begin{equation}
r^{\rm KD}(f_1,f_2)\equiv \frac{C_1^{\rm KD}(f_1,f_2)}{\sqrt{C_1^{\rm KD}(f_1,f_1)C_1^{\rm KD}(f_2,f_2)}}= \frac{\mathcal{R}(f_1)\mathcal{R}(f_2)}{\sqrt{\left[\mathcal{R}(f_1)\right]^2\left[\mathcal{R}(f_2)\right]^2}}=1\, ,
\end{equation}
therefore, from a statistical point of view, the two variables are linearly dependent. From a more physical point of view, one can argue that the kinematic dipole is induced by the velocity of the observer $\vec{v}_o$ which depends on the matter distribution, but not on the frequency of the observed GWs. Any information about the frequency of the GWs is indeed encoded in the Kaiser-Rocket factor that can be factorized.

\subsection{Shot Noise}
\label{Shot Noise Computation}

Since the AGWB is generated by the superposition of unresolved astrophysical sources, it is naturally affected by SN, because the sources are discrete events which follow a Poisson distribution~\cite{Peebles,Tegmark:1997yq,Jenkins:2019nks,Jenkins:2019uzp}. The variance associated to the expected number of processes corresponds exactly to the SN. 

Following~\cite{Bellomo:2021mer}, the mean number of GW events per halo is essentially the merger rate of objects per halo, times the probability of having a merger after a time delay $t_d$ w.r.t. the formation of the binary, with astrophysical parameters $\vec{\theta}$ and times the observation time,
\begin{equation}
\bar{N}_{GW|h}(M_h,z,t_d,\vec{\theta}\, ) = p(t_d)p(\vec{\theta}\, )\mathcal{A}_{\rm LIGO}\langle {\rm SFR}(M_h,z_d)\rangle_{\rm SF}\,w(z,\vec{\theta}\, )T_{\rm obs}\frac{\ud V}{\ud z\ud \Omega_e}(z_d)\, ,
\label{N_gw_h_z}
\end{equation} 
where $\mathcal{A}_{\rm LIGO}$ is the LIGO normalization on the local merger rate and $dV/dz$ is the volume element which converts the number density to the number of objects. In the above expression $z$ is the redshift at which GWs are emitted, $t$ is the time at redshift $z$ and $z_d$ is the redshift at\footnote{To compute $z_d$ we invert the relation
\begin{equation}
    t_d = -\int_{z_d}^z d\tilde{z}\frac{1}{(1+\tilde{z})H(\tilde{z})}\, .
    \label{z_d_to_z_equation}
\end{equation}
We consider $t_d$ between $50\, \rm Myr$ and the age of the Universe~\cite{Mapelli:2017hqk}. Note that for very high redshifts the SFR is zero, thus the imprint of very high $z_d$ on the AGWB is zero too.
} $t-t_d$. 

The AGWB anisotropies are described by
\begin{equation}
\begin{split}
\delta_{\rm AGWB}(\hat{n},f_o) =&\frac{1}{\bar{\Omega}_{\rm AGWB}T_{\rm obs}} \frac{f_o}{\rho_c c^2}\\
&\int \frac{dz}{H(z)(1+z)}\int \ud \vec{\theta}\, \frac{\ud E}{\ud \Omega_e \ud f_e}(z,f_o,\vec{\theta}\, ) \\
& \int dM_h \int dt_d \frac{dn}{dM_h}(M_h,z_d)\frac{N_{GW|h}(M_h,z,t_d,\vec{\theta}\,)-\bar{N}_{GW|h}(M_h,z,t_d,\vec{\theta})}{\frac{\ud V}{\ud z\ud \Omega_e}(z_d)}\, . 
\end{split}
\end{equation}
Since the fluctuations due to SN are uncorrelated with fluctuations due to cosmological perturbations, there is no cross-correlation between the SN and the intrinsic anisotropies. The only contribution given by SN is due to fluctuations of $N_{\rm GW|h}$, which follows a Compound Poisson Distribution, whose covariance has been computed in Appendix \ref{Compound Poisson Distribution}. The SN angular power spectrum is independent from the angular scale $\ell$ considered and it is equal to~\cite{Bellomo:2021mer}
\begin{equation} \label{equazione_calcolo_shot_noise}
\begin{split}
C_\ell^{\rm AGWB,SN}=&\int \ud\hat{n} \int \ud \hat{n}^\prime Y_{\ell m}^*(\hat{n}) Y_{\ell m}(\hat{n}^\prime)\left\langle \delta_{\rm AGWB}(\hat{n},f_1)\delta^*_{\rm AGWB}(\hat{n},f_2)\right\rangle_{\rm SN}\\
=&\frac{1}{\rho_c^2 c^4 T^2_{\rm obs}}\frac{f_1 f_2}{\bar{\Omega}_{\rm AGWB}(f_1)\bar{\Omega}_{\rm AGWB}(f_2)}\\
&\int dz\left[\frac{1}{H(z)(1+z)}\right]^2\int \ud\vec{\theta}\,\frac{\ud E}{\ud \Omega_e\ud f_e}(z,f_1,\vec{\theta}\, )\frac{\ud E}{\ud \Omega_e\ud f_e}(z,f_2,\vec{\theta}\, )\\
& \int dM_h \int dt_d \, \frac{dn}{dM_h}(M_h,z_d)\frac{ \bar{N}_{GW|h}(M_h,z,t_d,\vec{\theta}\,)+\bar{N}_{GW|h}^2(M_h,z,t_d,\vec{\theta}\, )}{\left(\frac{\ud V}{\ud z\ud \Omega_e}(z_d)\right)^2}\, .
\end{split}
\end{equation}
In the computation of $C_\ell^{\rm AGWB,SN}$ done here we have assumed that only binaries with the same astrophysical parameters, time delay and formation redshift are correlated. This contribution is expected to be the dominant one, but, for completeness, one has to consider possible SN correlations between binaries with different properties, with a formalism similar to the one used in ~\cite{Smith:2008ut} for overlapping tracers. This has been discussed in Appendix \ref{Compound Poisson Distribution}. The crucial feature of the SN of the AGWB is that it correlates at different frequencies. This correlation comes from the fact that binaries with the same properties and formed at the same time $z_d$ emit at each $z$ GWs with the same frequency, $f(z,t_d,\vec{\theta}\,)$, but to compute the AGWB anisotropies we integrate along the past GW-cone, therefore we can correlate GW signals emitted at $z$ with the ones emitted at $z+\delta z$ by the same kind of sources, which have nonvanishing SN. In this work, we are interested in the late-inspiral, in the merger and in the ringdown, which have a very small duration compared to the timescales over which we are integrating to compute the AGWB, therefore in the interval $\delta z$ the frequency $f(z+\delta z,t_d,\vec{\theta}\,)$ covers over all the frequencies of the merger and the ringdown, thus we are allowed to correlate $f_1$ with $f_2$\footnote{The factor $\ud E/\ud \Omega_e\ud f_e$ weights the contributions to the SN of sources of different masses at different redshifts, hence if $f$ is outside the frequency range over which a binary of parameters $\vec{\theta}$ emits, the energy spectrum is zero and we do not count such binaries.}. Apart from the frequency, the quantities evaluated at $z+\delta z$ are equivalent to the one evaluated at $z$, because $\delta z$ corresponds to a timescale much smaller than the one involved in the AGWB computation, thus we obtain Eq. \eqref{equazione_calcolo_shot_noise}. The assumption that a small $\delta z$ covers all the evolution of the binary system is valid only for burst sources, while for continuous sources, e.g. inspiralling binaries which have not merged during a Hubble time, the evolution of the system in time has to be considered and we have to use a different expression to compute the total energy emitted~\cite{Bertacca:2019fnt}. 

We have depicted the dipole power spectrum of the SN for $f_1=f_2$ for $T_{\rm obs} = 10\, {\rm yrs}$ in Figure \ref{all_dipoles_figure}. As already stressed, the SN is approximately two orders of magnitude larger than the kinematic and the intrinsic dipoles. The result is consistent with~\cite{Jenkins:2019uzp,Bellomo:2021mer}.

There are several strategies to reduce the SN. The first one exploits cross-correlations~\cite{Capurri:2021prz}, which allows in general to obtain higher SNRs w.r.t. the auto-correlation case. However, if the SN is some orders of magnitude larger than the intrinsic anisotropies, as in our case, it is hard to cancel this contribution by using few tracers only. Alternatively, one could use new statistical estimators~\cite{Jenkins:2019nks}, to cancel the offset in the estimate of the angular power spectrum and to reduce as much as possible the SN.

As a new method, we will try to reduce the SN by correlating the AGWB anisotropies at different frequencies, exploiting the different dependence on the frequency of the SN and of the intrinsic w.r.t. the kinematic dipole. 

\section{Component Separation of AGWB Anisotropies} 

\subsection{Detectability of the Kinematic Dipole}

In this Section we want to give an estimate of the detectability of the AGWB kinematic dipole, by using an SNR analysis. We want to show that in principle, if one ignores the frequency dependence of the AGWB anisotropies in an SNR analysis, a detection of the kinematic dipole (and of the other anisotropies too) would be more challenging. If we consider BBH mergers in the mass range of $2.5-100\, M_\odot$, the most promising experiment to detect the anisotropies of the AGWB is the network obtained by the combination of ET and CE, because
with the current bounds on the amplitude of the AGWB monopole aLIGO has a too low sensitivity~\cite{Bellomo:2021mer}. Therefore, from now on, we will focus on the ET+CE case only and we compute the noise angular spectrum using the {\textrm Schnell} code~\cite{Alonso:2020rar}. We consider one ET-D~\cite{ETsens} located in Sardinia and one CE~\cite{ce} with the two interferometers located in Hanford and Livingston\footnote{The state of the art for the CE detectors is one detector in US and one in Australia (see ~\cite{Evans:2021gyd}). However the exact location of the detectors does not affect the analysis and the main results of the paper.}.

To quantify the amount of physical information that we can extract by studying the AGWB dipole we consider as observable the angular power spectrum of the auto- and of the cross-correlation of the AGWB with a galaxy survey, by choosing a specific survey in order to maximize the correlation and so the SNR. One suitable survey to be combined with the AGWB is SKAO2, and in Appendix \ref{appendix_SKA} we report the parametrization that we have used. We will not compute the SNR of the auto-spectrum of the galaxy survey, which is larger than one, since we are interested in discussing only the extra-information we can add by looking at the AGWB. 

The SNR is defined as the ratio between the signal we want to measure and the noise of the detector,
\begin{equation}
    {\rm SNR}^2 \equiv \sum_{\ell=1}^{\ell_{\rm max}}\vec{C}_\ell^{\, \rm T} {\rm cov}_\ell^{-1}\vec{C}_\ell\, ,
\end{equation}
where $\ell_{\rm max}$ identifies the maximum multipole at which we have a non-negligible contribution to the SNR. The vector $\vec{C}_\ell$ represents the observables we are looking at,
\begin{equation}
    \vec{C}_\ell = \begin{pmatrix}
    C_\ell^{\rm AGWB} \\
    C_\ell^{g\times \rm AGWB}
    \end{pmatrix}\, ,
\end{equation}
while ${\rm cov}_\ell$ is the covariance between the pseudo-$C_\ell$'s estimators we are using for the angular power spectrum. We can write the SNR as the sum in quadrature of the SNRs at a given multipole, because for GW experiments $f_{\rm sky}\approx 1$, thus we have no mode-coupling between different multipoles. The covariance of these estimators is given by the sum of the cosmic variance and of instrumental noise plus SN,
\begin{equation}
    {\rm cov}_\ell = \frac{2}{2\ell+1}\begin{pmatrix}
    \left(C_\ell^{i}+N_\ell^{i}\right)^2 & \left(C_\ell^{ i}+N_\ell^{i}\right)\left(C_\ell^{j\times i}+N_\ell^{j\times  i}\right) \\
    \left(C_\ell^i+N_\ell^i\right)\left(C_\ell^{j\times i}+N_\ell^{j\times i}\right) & \frac{\left(C_\ell^{j\times i}+N_\ell^{j\times i}\right)^2+\left(C_\ell^{j}+N_\ell^{j}\right)\left(C_\ell^i+N_\ell^{i}\right)}{2}
    \end{pmatrix}\, ,
\end{equation}
where we have used the compact notation $i=\rm AGWB$, $j=g$ (i.e., galaxy). For the noises in the covariance we have considered
\begin{equation}
\begin{split}
    N_\ell^{\rm AGWB} =& C_\ell^{\rm AGWB,SN}+N_\ell^{\rm AGWB,inst}\, , \\
    N_\ell^{g} =& \frac{1}{\bar{N}_g}\, , \\
    N_\ell^{g\times \rm AGWB} =& N_\ell^{g\times \rm AGWB,SN}\, ,
\end{split}
\end{equation}
where $1/\bar{N}_g$ is the SN term for the galaxy survey and it represents the total number of galaxies observed, while $N_\ell^{g\times \rm AGWB,SN}$ is the SN of the cross-correlation between the galaxy number count and the AGWB~\cite{Canas-Herrera:2019npr,Alonso:2020mva}. In this computation we have however neglected the impact on the SNR of the SN of the cross-correlation, since we want just to show that by looking at the dipole at just one frequency the SNR is much lower than one. Note that in this preliminary computation we have assumed that the integrated response of the instrument can be written in terms of an angular power spectrum. However, in Section \ref{AGWB Kinematic Dipole Estimate with Shot Noise and Instrumental Noise} we will quantify more properly the instrumental noise and we will define a more general estimator for the kinematic dipole, which minimizes both instrumental noise and SN.

\begin{figure}
\includegraphics[scale=0.5]{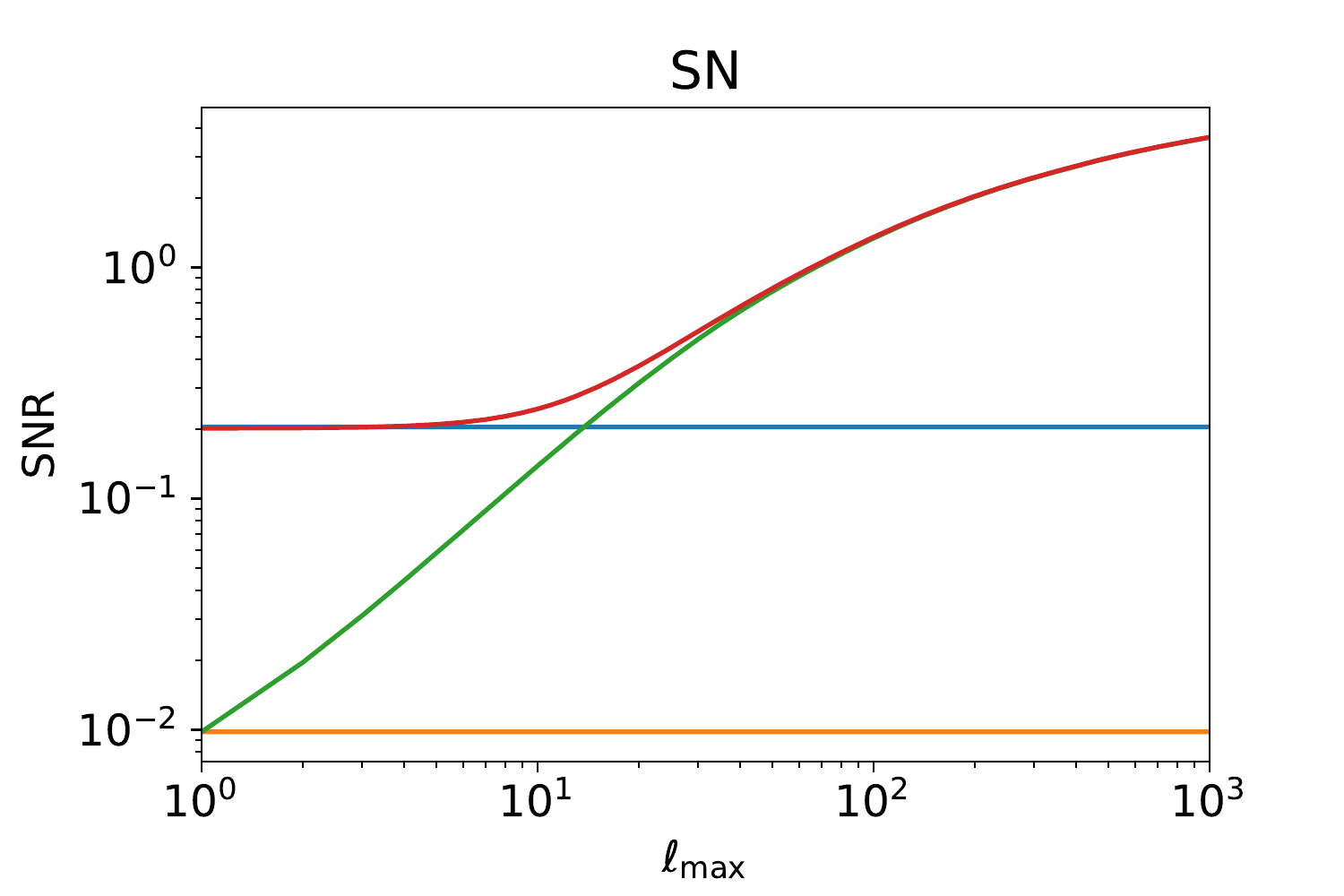}
\includegraphics[scale=0.5]{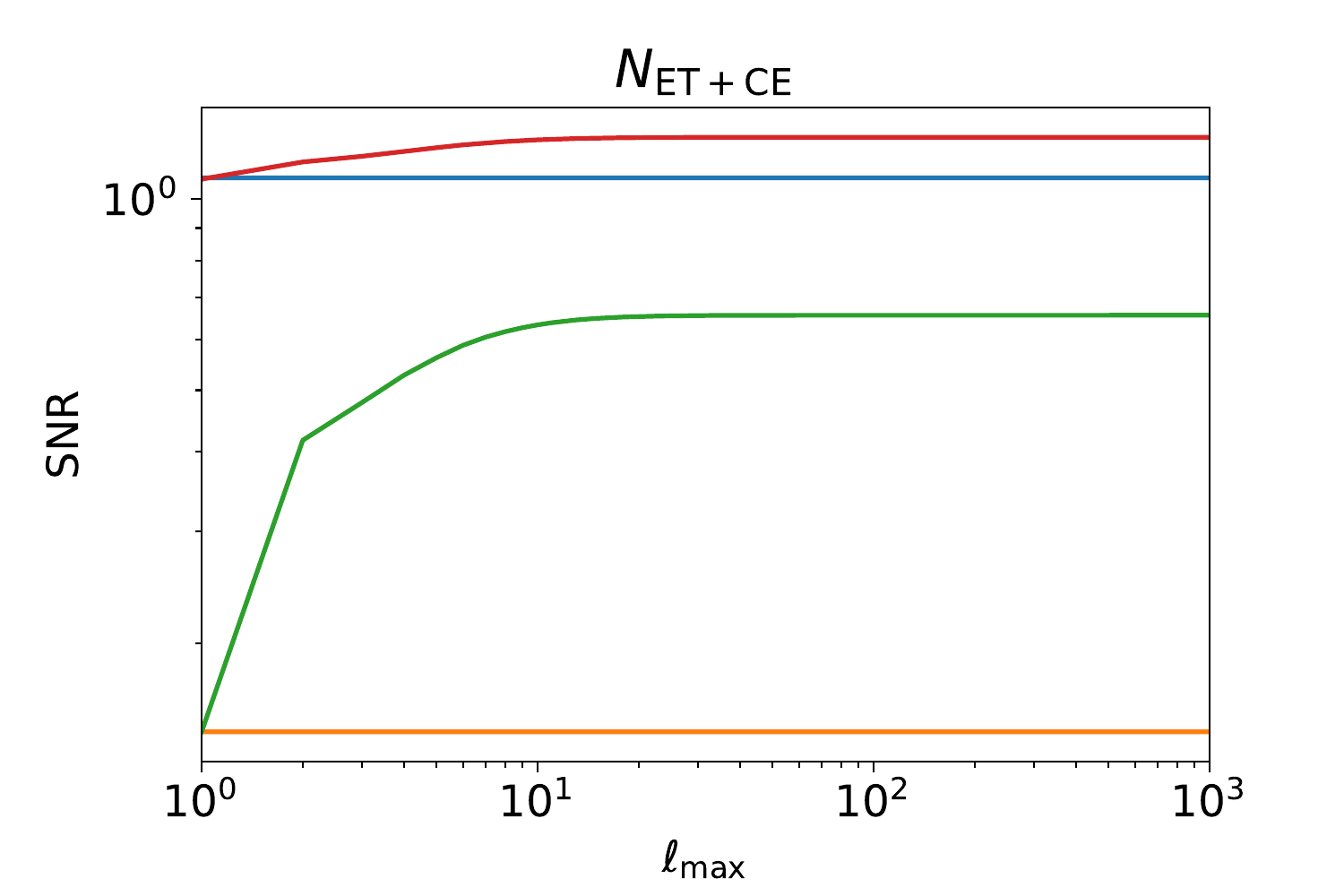}
\includegraphics[scale=0.5]{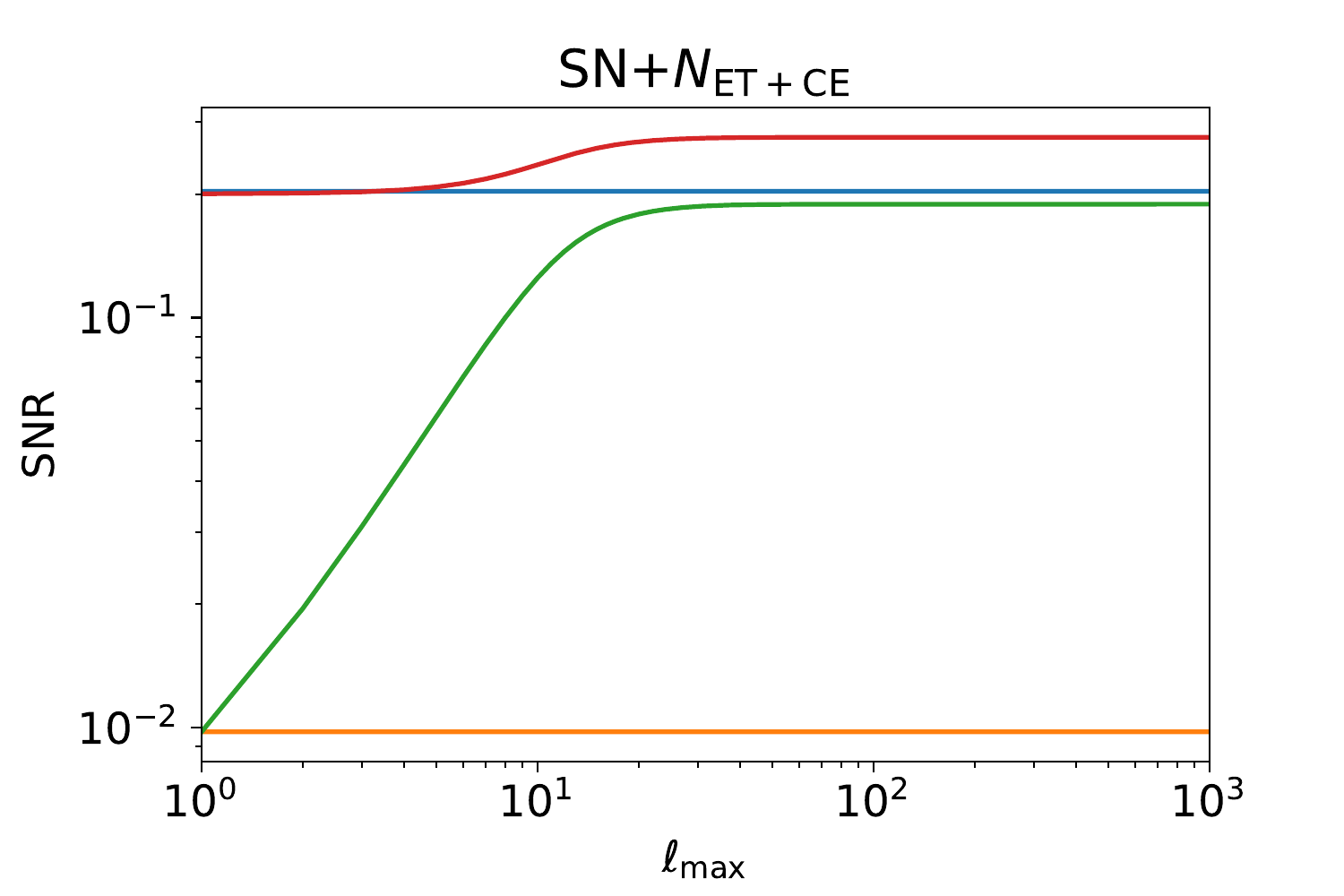}
\includegraphics[scale=0.5]{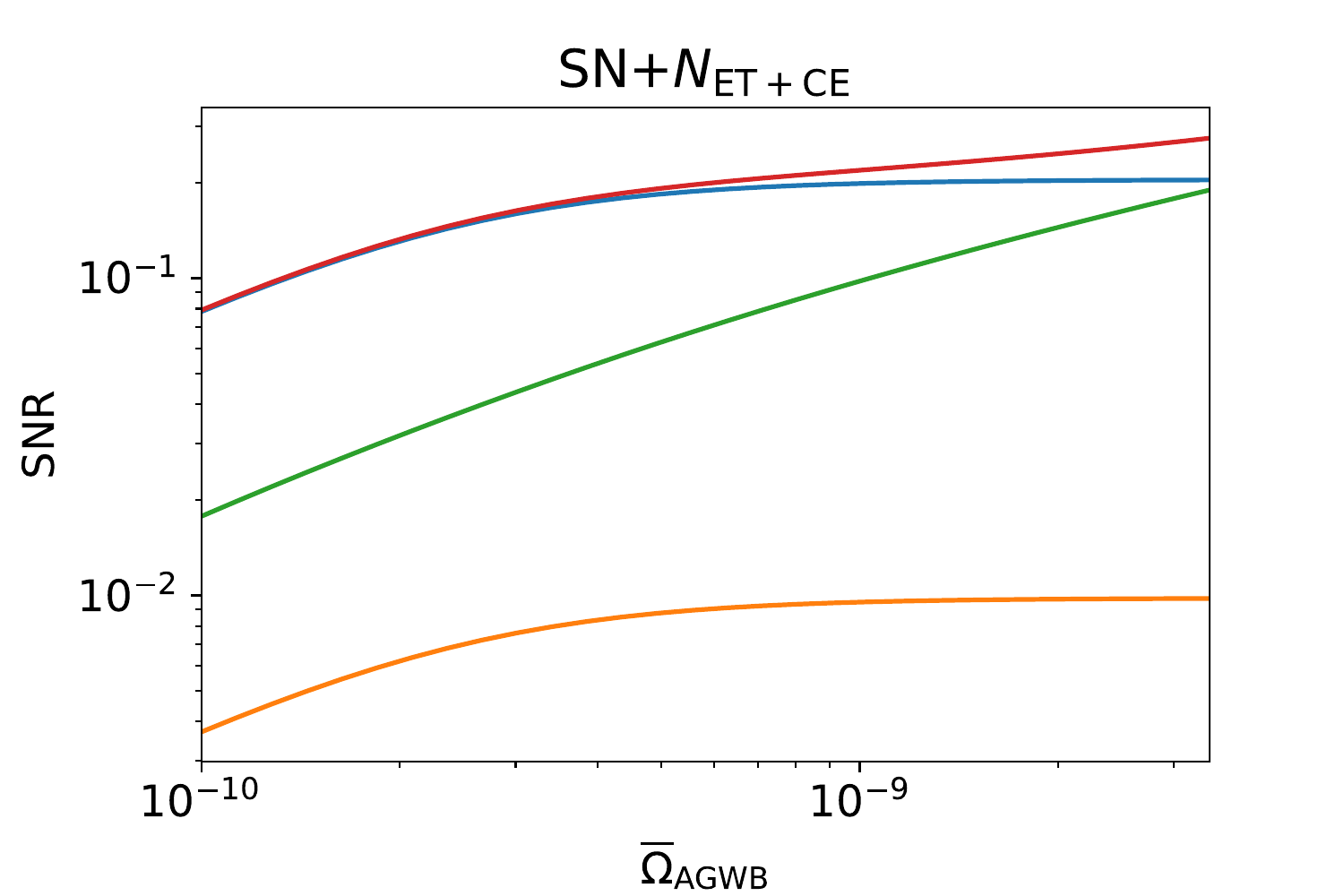}
\caption{{\it Plot of the contributions to the cumulative SNR as a function of the maximum multipole, with SN only (upper left), with instrumental noise only (upper right), with instrumental noise plus SN (lower left). The blue line corresponds to the kinematic dipole, the orange one to the intrisic dipole, the green one to the intrinsic anisotropies and the red one to the total.  When instrumental noise is considered, we have computed the cumulative SNR assuming the maximum monopole amplitude for the AGWB. Lower right: plot of the cumulative SNR for $\ell_{\rm max}=200$ as a function of the monopole amplitude of the AGWB, considering both instrumental noise and SN. We have considered $\ell_{\rm max}=200$, because, when we compute the SNR with instrumental noise, we automatically take into account the angular resolution of the detector, therefore higher multipoles give negligible contribution to the SNR. All the SNRs have been computed for the auto-correlation of the AGWB and for the cross-correlation between the AGWB with the galaxy survey SKAO2. The SNR computed here does not include the auto-correlation of the galaxy survey, because we want to quantify the amount of extra-information added by considering also the AGWB in the analysis.}}
\label{snr_omega_figure}
\end{figure}

In Figure \ref{snr_omega_figure} we have depicted the cumulative SNR of the various contributions to the anisotropies as a function of the maximum multipole considered. We have also plotted the various contribution to the SNR up to  $\ell_{\rm max} = 200$ as a function of the monopole amplitude of the AGWB. Note that when instrumental noise is considered, different choices of $\ell_{\rm max}$ above a certain value do not change the SNR, since the instrumental noise automatically keeps into account for the angular resolution of the detector. We have computed the SNR in three different scenarios: with instrumental noise only, with SN only, and with SN plus instrumental noise.

\subsection{Multi-Frequency Observations}

 The total AGWB map at frequency $f_o$ is the sum of four contributions,
\begin{equation}
\delta^{\rm obs}_{\rm AGWB,\ell m}(f_o,\hat{n}) =\delta_{1 \ell}\, \delta_{\rm AGWB,\ell m}^{\rm KD}(f_o,\hat{n})+\delta_{\rm AGWB,\ell m}^{\rm int}(f_o,\hat{n})+\delta_{\rm AGWB,\ell m}^{\rm SN}(f_o,\hat{n})+n_{\ell m}^{\rm inst}(f_o)\, ,
\end{equation}
where the kinematic dipole term and the intrinsic dipole term contain the astrophysical and cosmological information we would like to extract, while the other two represent a source of uncertainty in our measurements. As discussed in Section \ref{Shot Noise Computation}, the SN is larger than the kinematic/intrinsic anisotropies, therefore it could be a limitation for future GW experiments that plan to look at the physics beyond the monopole of astrophysical backgrounds. At the moment, in the literature the only discussion regarding the SN of the AGWB has been done in~\cite{Jenkins:2019nks,Jenkins:2019uzp,Canas-Herrera:2019npr,Alonso:2020mva,Bellomo:2021mer}, without providing a valid solution to deal with this issue. On the contrary, many techniques have been adopted to reduce instrumental noise at GW interferometers and they exploit the fact that instrumental noise has a different dependence on the frequency w.r.t. the signal, therefore it is possible to choose some proper weights to $\delta_{\rm AGWB}^{\rm obs}$ at different frequencies to minimize the covariance of the AGWB map estimator.

As stressed in Section \ref{Computation of the AGWB dipole}, the AGWB angular power spectrum depends on the observed frequency $f_o$, due to the dependence of the window function $\tilde{W}$ and the different GW bias and GW evolution bias.
The key point here is that the four contributions to the total observed signal depends on the frequency in a different way, therefore it should be possible to break the degeneracy among them by combining the observation for all the available spectra. Note that the same procedure is used for instance to remove galactic foregrounds from CMB maps~\cite{Tegmark:1999ke,Planck:2018yye}. The most commonly used technique to extract the kinetic dipole is to combine different observables~\cite{Nadolny:2021hti}, in order to remove spurious contributions from the intrinsic dipole. For the AGWB it is however natural to use a component separation technique based on multi-frequency observations, because the intrinsic anisotropies and the SN have the same frequency-dependence, because the window function $\tilde{W}$ is very similar to the kernel, which determines the frequency dependence of the SN. This means that, if we introduce an estimator to minimize the covariance of the kinematic dipole, we would be able to remove, at the same time, the intrinsic dipole and the SN, because of their similar frequency shape. 

In this section we will start showing how we can reduce SN and the intrinsic anisotropies contribution in the kinematic dipole estimate by combining the AGWB at few discrete frequencies. For the moment we will neglect instrumental noise, which will be included later on. This part should be useful to intuitively understand the validity of our method and to justify the next step, where we will combine instrumental noise with the total AGWB signal, writing down an expression for an estimator of the kinematic dipole map with minimum covariance.

\subsection{Component Separation with Multi-Frequency Observations}
\label{Component separation with multi-frequency observations}
In this section we want to separate the different contributions to the AGWB by using observations of the anisotropies at different frequencies. 

The SN for two frequencies $f_1$, $f_2$, given the window function $w(z)$, is~\cite{Bellomo:2021mer}
\begin{equation}
\begin{split}
C_\ell^{\rm AGWB,SN}(f_1,f_2)
=&\frac{1}{\left(\rho_c c^2 T_{\rm obs}\right)^2}\int \ud z\left[\frac{1}{(1+z)H(z)}\right]^2\int \ud \vec{\theta}\, K(z,f_1,f_2,\vec{\theta}\,)S(z,\vec{\theta}\,)\, , 
\end{split}
\end{equation}
where the frequency kernel $K$ and the SN fluctuation at redshift $z$ are   
\begin{equation}
\begin{split}
K(z,f_1,f_2,\vec{\theta}\, ) =&\frac{f_1 f_2}{\bar{\Omega}_{\rm AGWB}(f_1)\,\bar{\Omega}_{\rm AGWB}(f_2)}  \frac{\ud E}{\ud f_e\ud \Omega_e}(\vec{\theta},z,f_1)\frac{\ud E}{\ud f_e \ud \Omega_e}(\vec{\theta},z,f_2)\, , \\
S(z,\vec{\theta}\, ) =& \int \ud M_h \int \ud t_d \, \frac{\ud n}{\ud M_h}(M_h,z_d)\frac{ \bar{N}_{GW|h}(M_h,z,t_d,\vec{\theta}\,)+\bar{N}_{GW|h}^2(M_h,z,t_d,\vec{\theta})}{\left(\frac{\ud V}{\ud z\ud \Omega_e}(z_d)\right)^2}\, ,
\end{split}
\end{equation}
and the average number of GW events per halo of mass $M_h$ at redshift $z$ has been defined in Eq. \eqref{N_gw_h_z}. The $S(z,\vec{\theta}\,)$ factor encodes the information about the SN fluctuation of the number of GW sources, while the $K(z,f_1,f_2,\vec{\theta}\,)$ factor weights the contribution to the signal of GW sources with different masses (in general with different astrophysical parameters) in the given frequency bin.
The intrinsic anisotropies of the AGWB depend on the frequency through the window function $\tilde{W}$ defined in Eq. \eqref{effective_window_function}, therefore, as a first approximation, neglecting other possible frequency dependencies due for instance to the GW bias, we can assume that the SN and the intrinsic anisotropies are very similar and cannot be disentangled with this technique. On the other hand, the dominant frequency dependence contribution of the kinematic dipole is given by evolution bias, which depends differently on $f_o$ w.r.t. $\tilde{W}$. 

Intuitively, what we are saying is that if we look at two maps at frequencies $f_1$, $f_2$ the map observed at $f_1$ is constrained by the map observed at $f_2$ by a mean and a covariance given by
\begin{equation}
\begin{split}
\mu_{\ell m}^{\alpha} (f_1|f_2)=&\frac{C_\ell^{\alpha}(f_1,f_2)}{C_\ell^{\alpha}(f_2,f_2)}\delta^{\alpha}_{\ell m}(f_2)\, , \\
C_\ell^\alpha(f_1|f_2)= & C_\ell^\alpha(f_1,f_1)-\frac{\left[C_\ell^\alpha(f_1,f_2)\right]^2}{C_\ell^\alpha(f_2,f_2)}\, ,
\end{split}
\end{equation}
with $\alpha=\{\rm int,SN,KD\}$. The idea is that if we combine the two maps in a proper way, we can cancel the SN bias, and the resulting map will have covariance given by the covariance of the conditioned maps. The point is that if this covariance is sufficiently small, we are able to reduce the impact of SN on our KD estimate. Note that if the kinematic dipole and the SN would have the same frequency dependence, we are not able to separate the two maps, because the linear system would be degenerate, which is approximately what happens for the SN and the intrinsic anisotropies.

The generalization of what we have described for more than two frequencies and with a more formal derivation of the estimator, is the Internal Linear Combination (ILC)~\cite{Tegmark:1999ke,Tegmark:2003ve}, or any other kind of component separation technique.

The ILC does the following: suppose you have some maps at different frequencies, from which you extract the dipole
\begin{equation}
\vec{d}^{\rm obs}_i =\vec{d}^{\rm int}_i+\vec{d}^{\rm KD}_i+\vec{d}^{\rm SN}_i\, ,
\end{equation}
where the vectors refers to the different frequencies, while the index $i$ represents the $x$, $y$, $z$ directions in the sky.  The dipole at a pivot frequency $f_{\rm piv}$ is related univocally to the velocity of the observer through the Kaiser-Rocket factor defined in Eq. \eqref{kaiser_rocket_factor_equation},
\begin{equation}
d_i^{\rm KD}(f) = \mathcal{R}(f)v_{o,i}\, .
\end{equation}
Therefore the total signal is 
\begin{equation}
\vec{d}^{\,  \rm obs}_i =\vec{\mathcal{R}}v_{o,i}+\vec{d}_i^{\, \, \rm int}+\vec{d}_i^{\, \, \rm SN}\, .
\end{equation}
Since the AGWB anisotropies are measured at different frequencies, we can combine the data in a smart way to find an estimator of the observer velocity with a small covariance.  This is done by writing down the most general linear estimate of the observer velocity,
\begin{equation}
\hat{v}_{o,i} \equiv \vec{w}^T \vec{d}^{\rm \, obs}_i,
\label{general_estimator_equation}
\end{equation}
and by choosing the weights $\vec{w}$ of the linear combination that minimize the covariance of the estimator,
\begin{equation}
\frac{\partial}{\partial \vec{w}}\left\langle \left(\hat{v}_{o,i} -v_{o,i} \right)^2\right\rangle = 0\, .
\end{equation}
We require also that our estimator is unbiased, therefore in order to have $\hat{v}_{o,i} \propto v_{o,i}$, we need that $\vec{w}^T\vec{\mathcal{R}} = 1$.
To minimize the differential equation with a constraint we use a Lagrange multiplier. We introduce the Lagrangian function $\mathcal{L}$, 
\begin{equation}
\mathcal{L}(x,\lambda) = \left\langle \left(\hat{v}_{o,i}-v_{o,i} \right)^2\right\rangle-\lambda \left(\vec{w}^T\vec{\mathcal{R}} -1\right)=\vec{w}^T C \vec{w}-\lambda \left(\vec{w}^T\vec{\mathcal{R}} -1\right)\,  ,
\end{equation}
where $C$ is the covariance matrix of the total dipole, where its $(\alpha,\beta)$ entry is defined as,
\begin{equation}
C_{\alpha \beta}\equiv {\rm cov}\left[d_i^{\rm obs}(f_\alpha),d_i^{\rm obs}(f_\beta)\right]=C_1^{\rm int}(f_\alpha,f_\beta)+C_1^{\rm SN}(f_\alpha,f_\beta)\, .
\end{equation}
We impose that the Jacobian of this function is zero, finding
\begin{equation}
\begin{split}
\begin{cases}
\vec{w}^T \vec{\mathcal{R}} = 1 \\
2\vec{w}^T C -\lambda \vec{\mathcal{R}}^T = 0 
\end{cases}
\rightarrow
\begin{cases}
\vec{w}^T = \frac{1}{2}\lambda \vec{\mathcal{R}}^T C^{-1} \\
 \frac{1}{2}\lambda\vec{\mathcal{R}}^T C^{-1} \vec{\mathcal{R}} = 1
\end{cases}
\rightarrow
\begin{cases}
\lambda = \frac{2}{\vec{\mathcal{R}}^T C^{-1} \vec{\mathcal{R}}} \\
\vec{w}^T = \frac{\vec{\mathcal{R}}^T C^{-1}}{\vec{\mathcal{R}}^T C^{-1} \vec{\mathcal{R}}}
\end{cases}
\end{split}
\end{equation}
The estimator of the observer velocity is then computed by substituting in Eq. \eqref{general_estimator_equation} the weights $\vec{w}^T$ computed above,
\begin{equation}
\hat{v}_{o,i} = \frac{\vec{\mathcal{R}}^TC^{-1}\vec{d}_i^{\, \rm obs}}{\vec{\mathcal{R}}^TC^{-1}\vec{\mathcal{R}}}\, .
\end{equation}
The error associated to the estimate is
\begin{equation}
\sigma_{\hat{v}_{o,i}} = \sqrt{\left\langle \left(\hat{v}_{o,i} -v_{o,i} \right)^2\right\rangle} = \sqrt{\vec{w}^T C \vec{w}} = \frac{1}{\sqrt{\vec{\mathcal{R}}^TC^{-1}\vec{\mathcal{R}}}}\, .
\label{final_ILC_error_equation}
\end{equation}
We have computed $\sigma_{\hat{v}_{o,i}}$ for different $\vec{f}$, varying both the total number of frequencies in the ILC analysis and the combination of frequencies, looking for the one with the minimum error. We have compared the analytical estimate of the ILC error, Eq. \eqref{final_ILC_error_equation}, with the Root Mean Square of $M=10^4$ realizations of the system. More specifically, we have generated $M$ realizations of the SN, of the intrinsic and of the kinetic dipole anisotropies. These are uncorrelated, therefore the total map is simply the sum of the three maps generated independently. We then apply the ILC to each of the $M$ maps, finding $\vec{v}_o^{\, \rm est}$, which can be compared with the $\vec{v}_o$ true value, which is given by the realization of the kinetic dipole anisotropies divided by the Kaiser-Rocket factor.

\begin{figure}
\centering
\includegraphics[scale=0.35]{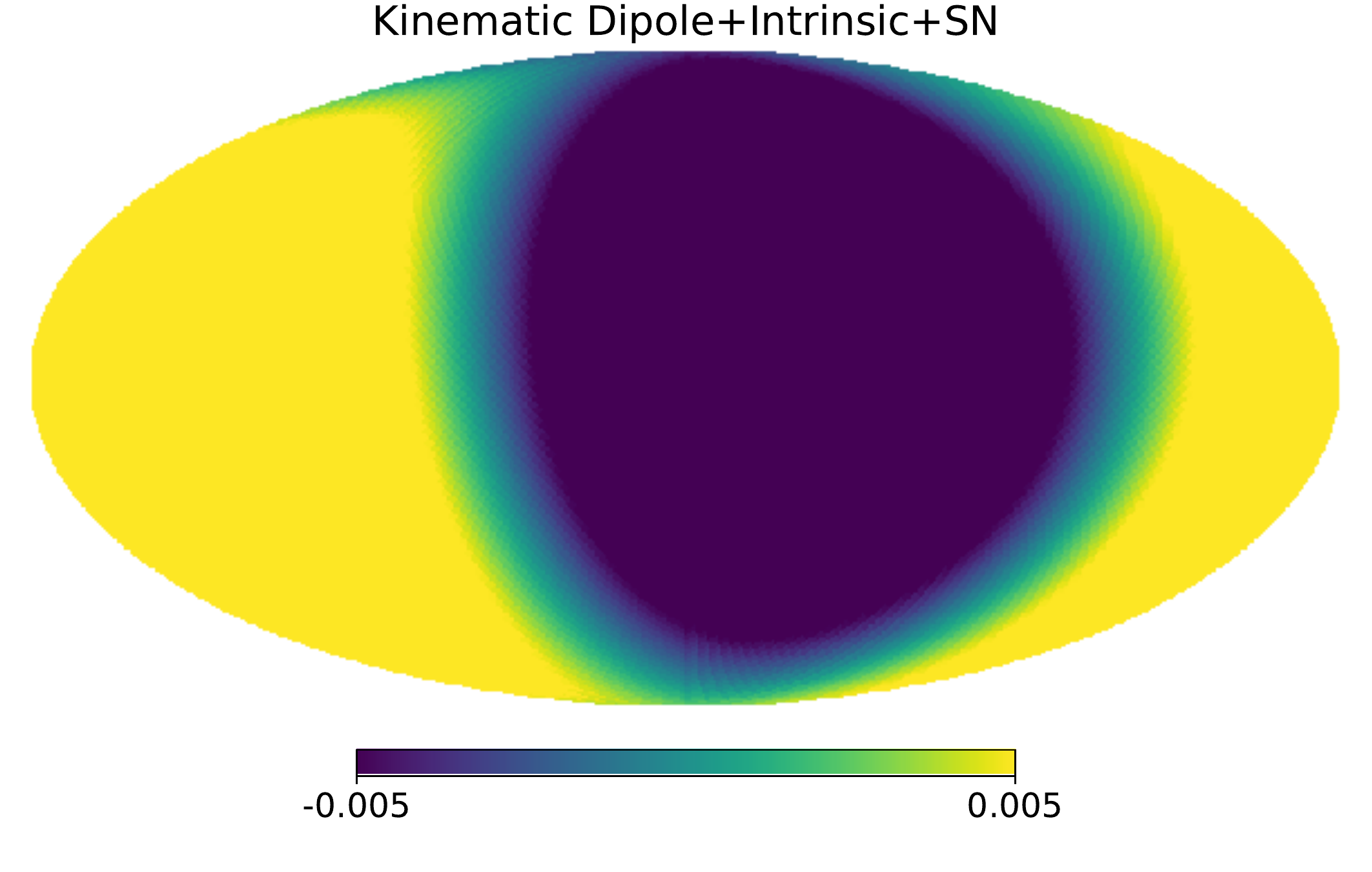}
\includegraphics[scale=0.35]{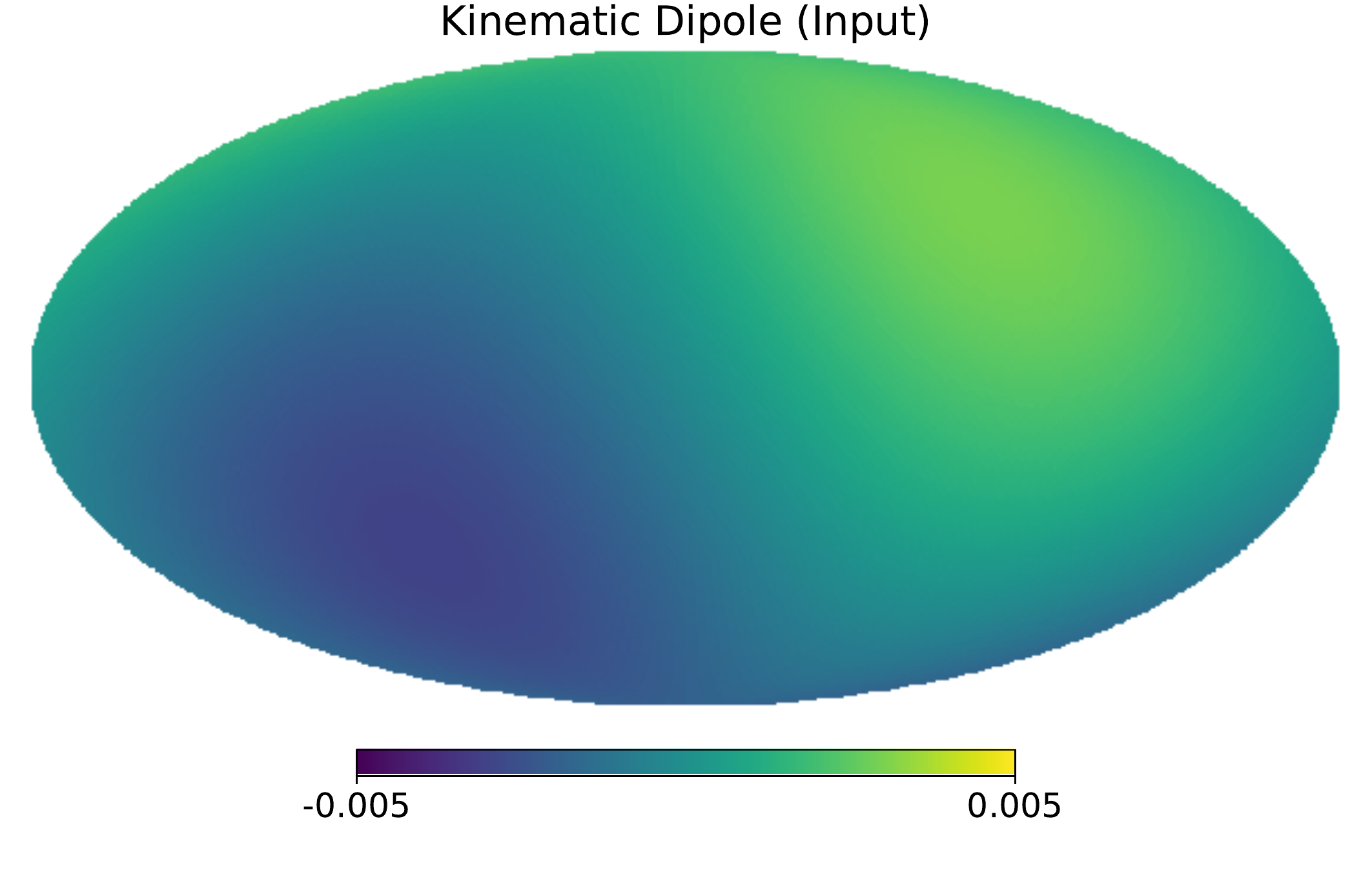}
\includegraphics[scale=0.35]{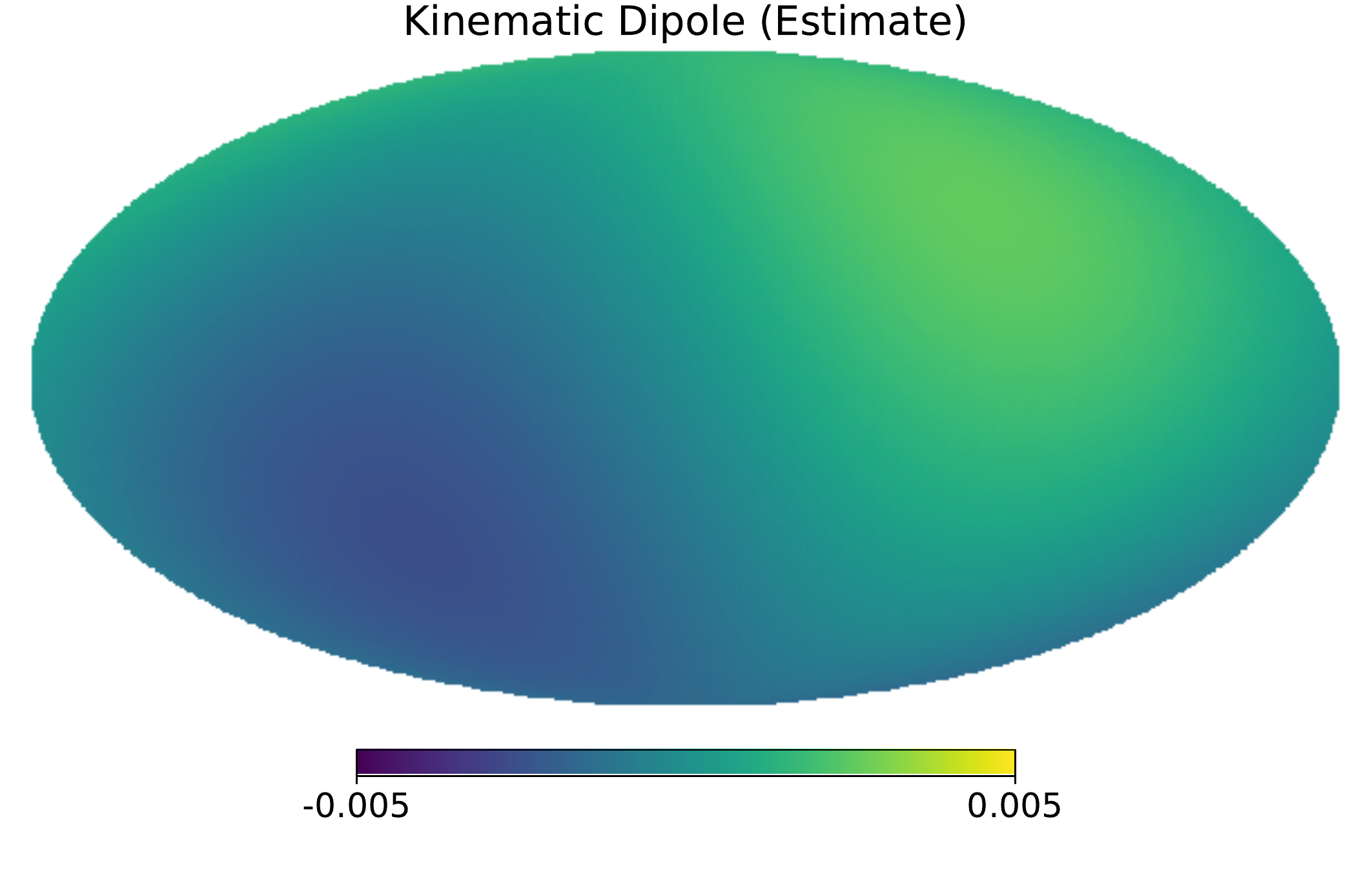}
\caption{{\it Upper Left: AGWB density contrast map at $f_o=30\, \rm Hz$; Upper Right : input velocity map $\hat{n}\cdot \vec{v}_o$; Bottom: reconstructed velocity map $\hat{n}\cdot\vec{v}_o^{\rm est}$.}}
\label{reconstructed_maps}
\end{figure}

In order to show how powerful this technique is, we show explicitly the result of our analysis for one single realization. We have $N=42$ frequencies evenly spaced over the interval $[100,1000]\, \rm Hz$. Compared to the input value we find the following estimate
\begin{equation}
\begin{split}
\vec{v}_{\rm o} = &
\begin{pmatrix}
0.0018 & 0.0032 & 0.0002
\end{pmatrix}\, , \\
\vec{v}_{\rm o}^{\, \rm est} = &
\begin{pmatrix}
0.0020 & 0.0031 & 0.0002
\end{pmatrix}\pm 0.0002\, .
\end{split}
\end{equation}
The velocity here has been computed in natural units, $c=1$, and it represents the velocity of the LG generated from the power spectrum of the density field evaluated at the present epoch. Up to statistical fluctuations due to the fact that we are generating a Gaussian random field, the input velocity $\vec{v}_o$ is consistent with the LG one estimated by Planck, $v_o \approx 600\, \rm km/s$~\cite{Planck:2013kqc}. As stressed in Section \ref{Kinematic dipole}, in this work we are interested in providing a useful tool for the statistical analysis of the AGWB kinematic dipole, therefore we assume that the velocities of the Earth, of the Sun and on the Milky Way have already been subtracted before performing this analysis. Their net effect is a Doppler shift in the angular power spectrum of the AGWB in the LG rest frame, that can be studied in detail as discussed in~\cite{Cusin:2022cbb}.
In Figure \ref{reconstructed_maps} we have given a map explanation of what we are doing: we have plotted the observed map at $f=30\, \rm Hz$, $\delta_{\rm AGWB}^{\rm obs}$, the ‘‘cleaned'' velocity map, $\hat{n}\cdot\vec{v}_o^{\, \rm est}$, and the input velocity map, $\hat{n}\cdot\vec{v}_o$. We can see that without component separation we are not able to distinguish the kinematic dipole imprint on the AGWB dipole, because the SN is much larger, but after our multi-frequency analysis, giving proper weights to the different maps, we are able to disentangle the different contributions, finding that the reconstructed map and the input one are similar at percent level.

To conclude, we have computed the SNR for our new estimator,~\footnote{We have decided to quantify the amount of information on the kinematic dipole we can extract from the AGWB anisotropies in terms of this SNR, summing over all the components in real space of the observer velocity.}
\begin{equation}
    {\rm SNR}^2 = \vec{v}_o^{\, T}\, {\rm cov}^{-1}_{\rm ILC}\, \vec{v}_o \, ,
    \label{SNR_dipole_cov_equation}
\end{equation}
where ${\rm cov}_{\rm ILC}$ in this case is simply a diagonal matrix with entries $\sigma_{\hat{v}_{o,i}}$ defined in Eq. \eqref{final_ILC_error_equation}. The result we have found is ${\rm SNR}\approx 10$, therefore we are able to faithfully reconstruct the local velocity of the observer by considering SN only.

The key assumption we have done here is that we are able to know exactly the Kaiser-Rocket factor $\mathcal{R}(f)$ and the theoretical values of the angular power spectra of the SN and of the intrinsic anisotropies. This is of course a simplification, since there are several uncertainties in the astrophysical models which describe the formation and the evolution of binary systems. However, the point we want to stress is that future detectors like ET will be able to detect more than $10^5$ sources~\cite{Pieroni:2022bbh}, shedding light on the population of compact objects in binary systems. In addition, the component separation introduced here can also be done in a joint-analysis of resolved sources and AGWB. In this way one could marginalize over (some) astrophysical parameters, propagating the error bars on the final estimate of the kinematic dipole.

\subsection{AGWB Kinematic Dipole Estimate with Shot Noise and Instrumental Noise}
\label{AGWB Kinematic Dipole Estimate with Shot Noise and Instrumental Noise}
Now we want to generalize the previous computation to derive the best unbiased estimator for the AGWB kinematic dipole, keeping into account also the instrumental noise. To combine SN and instrumental noise, we use the code Schnell, therefore here we will use the same formalism of~\cite{Alonso:2020rar}. The AGWB is described by 
\begin{equation}
h_{ij}(t,\vec{x}) = \sum_{p} \int df \int d\hat{n}\, h_p(f,\hat{n})e^{2\pi i f (t-\hat{n}\cdot \vec{x})}e_{ij}^p(\hat{n})\, ,
\end{equation}
where $p$ is the GW polarization and $\hat{n}$ the direction of observation. The data measured by a detector $A$ at position $\vec{x}_A$ for an observation time $T$ is 
\begin{equation}
d_{A,T}(t,f) \simeq \int d\hat{n} \sum_p F_A^p(f,\hat{n}) h_p^{\rm tot}(f,\hat{n})+n_{A,T}\, ,
\label{data_definition}
\end{equation}
where we have introduced
\begin{equation}
F_A^p \equiv a_A^{ij}e_{ij}^p e^{-2\pi i f \hat{n}\cdot \vec{x}_A}\, ,
\end{equation}
with $a_A^{ij}$ the detector response function. Note that for the AGWB $h_p^{\rm tot}$ is the sum of three contributions,
\begin{equation}
h^{\rm tot}_p(f,\hat{n}) = h^{\rm KD}_p(f,\hat{n})+h^{\rm int}_p(f,\hat{n})+h^{\rm SN}_p(f,\hat{n})\, .
\end{equation}
The power spectrum of $h_p$ is 
\begin{equation}
\langle h_p(f,\hat{n})\, h_{p^\prime}^*(f^\prime,\hat{n}^\prime)\rangle \equiv \frac{1}{2}\delta(f-f^\prime)\frac{\delta(\hat{n}-\hat{n}^\prime)}{4\pi}\delta_{p p^\prime}I(f,\hat{n})\, ,
\end{equation}
where we have assumed that the only non-null Stokes parameter is the intensity, therefore we are legitimated to introduce $\delta_{p p^\prime}$. The amplitude of the AGWB is related to the intensity through~\cite{Cusin:2018rsq, Contaldi:2020rht}
\begin{equation}
\Omega_{\rm GW}(f,\hat{n}) = \bar{\Omega}_{\rm GW}(f)\left[1+\delta_{\rm GW}(f,\hat{n})\right]= \frac{4\pi^2 f^3}{3H_0^2} I(f,\hat{n})\, .
\label{I_Omega_Equation}
\end{equation}
As stressed in Section \ref{Kinematic dipole}, the frequency dependence of the AGWB kinematic dipole intensity can be factorized,
\begin{equation}
\begin{split}
I^{\rm KD}(\hat{n},f) =& \frac{\Omega_{\rm AGWB}(f)/f^3}{\Omega_{\rm AGWB}(f_{\rm piv})/f_{\rm piv}^3}\frac{C_\ell^{\rm AGWB,KD}(f,f_{\rm piv})}{C_\ell^{\rm AGWB,KD}(f_{\rm piv},f_{\rm piv})}I^{\rm KD}(\hat{n},f_{\rm piv})\\
= & \mathcal{E}^{\rm KD}(f,f_{\rm piv})I^{\rm KD}_{\rm piv}(\hat{n})\, ,
\end{split}
\end{equation}
where $\mathcal{E}^{\rm KD}(f,f_{\rm piv})$ is related to the ratio between the Kaiser-Rocket factors at two frequencies,
\begin{equation}
\mathcal{E}^{\rm KD}(f,f_{\rm piv})\equiv  \frac{\Omega_{\rm AGWB}(f)/f^3}{\Omega_{\rm AGWB}(f_{\rm piv})/f_{\rm piv}^3}\frac{\mathcal{R}(f)}{\mathcal{R}(f_{\rm piv})}\, .
\end{equation}
We have plotted the intensity of the kinematic dipole as a function of the frequency in Figure \ref{all_noises_figure}. In this work we do not consider/propagate the error associated to $\mathcal{E}^{\rm KD}$, but we restrict to the case in which we fix its value. This assumption is not important for our conclusion and a proper way to deal with uncertainties associated to the astrophysical sources is described at the end of Section \ref{Component separation with multi-frequency observations}.  Now we want to build an estimator for the kinematic dipole intensity $I_{\rm piv}^{\rm KD}(\hat{n})$, which is related to the velocity of our frame by
\begin{equation}
I_{\rm piv}^{\rm KD}(\hat{n}) = \mathcal{R}(f_{\rm piv})\,\hat{n}\cdot\vec{v}_o\,.
\end{equation}
A linear estimator in the dipole corresponds to a quadratic estimator in the strain. In our case the optimal estimator is~\cite{Alonso:2020rar,Mentasti:2020yyd,LISACosmologyWorkingGroup:2022kbp}
\begin{equation}
\tilde{I}^{\rm KD}_{\rm piv,\vartheta} = \sum_{A,B,f,f^\prime} d_{f,A}E^{f f^\prime}_{\vartheta,AB}d_{f^\prime,B}-b_\vartheta\, ,
\end{equation}
where the matrix $E$ and the vector $b$ have to be determined by minimizing the covariance and by reducing the bias. In the formalism we are using here the maps are written in terms of discrete pixels $\vartheta$, which correspond to different directions of observation in the sky. More specifically, in our analysis we have used $N_{\rm pixel} = 3072$\footnote{This is equivalent to $N_{\rm side}=16$ in a Healpix map.}, thus each pixel corresponds to a region of the sky of area $\Delta\Omega \equiv 4\pi/N_{\rm pixel}$. Even if we are working in pixel space, our discrete approach is consistent with~\cite{Mentasti:2020yyd,LISACosmologyWorkingGroup:2022kbp}.

The covariance matrix of the data, defined in Eq. \eqref{data_definition}, is 
\begin{equation}
\langle d_{f,A} \, d^*_{f^\prime,B}\rangle = \frac{1}{2}\frac{\delta_{f f^\prime}}{\Delta f}\left[N^{AB}_f+\sum_\vartheta B^{A B,\rm KD}_{f \vartheta}\left(I^{\rm KD}_{\rm piv,\vartheta}+\frac{I_{\vartheta,f}^{\rm int}+I_{\vartheta,f}^{\rm SN}}{\mathcal{E}_f^{\rm KD}}\right)\right]\, ,
\end{equation}
where $I^{\rm KD/int/SN}_{\vartheta}$ are the theoretical kinematic dipole/intrinsic/SN maps respectively, $N_f^{AB}$ is the Power Spectral Density (PSD) of the noise of the interferometers, while the matrix $B$ is
\begin{equation}
B_{f,\vartheta}^{AB,\rm KD} \equiv \Delta \Omega \, \mathcal{E}^{\rm KD}_f \sum_p F_{A,f\vartheta}^p F_{B,f\vartheta}^{p\,*}\, .
\end{equation}
The mean of the estimate that we have found is
\begin{equation}
\begin{split}
\langle \tilde{I}^{\rm KD}_{\rm piv,\vartheta} \rangle =& \sum_{A,B,f,f^\prime} \langle d_{f,A}\, d_{f^\prime,B}\rangle E^{f f^\prime}_{\vartheta,AB}-b_\vartheta =\\
= &  \sum_{A,B,f} \frac{1}{2\Delta f}E_{\vartheta,AB}^{ff}\sum_{\vartheta^\prime}B_{f,\vartheta^\prime}^{AB,\rm KD}\left(I_{\rm piv, \vartheta^\prime}^{\rm KD}+\frac{I_{f, \vartheta^{\prime}}^{\rm int}+I_{f, \vartheta^{\prime}}^{\rm SN}}{\mathcal{E}_f^{\rm KD}}\right)+\sum_{A,B,f} \frac{1}{2\Delta f}E_{\vartheta,AB}^{ff}N_f^{AB,\rm inst}-b_\vartheta\, ,
\end{split}
\end{equation}
from which, requiring the estimator to be unbiased, the bias has to be equal to
\begin{equation}
b_\vartheta = \sum_{A,B,f} \frac{1}{2\Delta f}\,E_{\vartheta,AB}^{ff}\,N_f^{AB,\rm inst}\, .
\end{equation}
The bias we have defined here depends on the instrumental noise only, while the bias given by the SN and by the intrinsic dipole is reduced by the $E^{f f^\prime}_\vartheta$ coefficients. More specifically, we will try to minimize the covariance associated to our estimator and this will give us the full expression for $E^{f f^\prime}_\vartheta$. 

Note that our estimator $\tilde{I}^{\rm KD}_{\rm piv,\vartheta}$ is related to the true kinematic dipole $I^{\rm KD}_{\rm piv,\vartheta}$ by a matrix multiplication in pixel space, therefore, the truly unbiased estimator is
\begin{equation}
\hat{I}_{\rm piv,\vartheta}^{\rm KD} = \sum_{\vartheta^\prime}\left(M^{-1}\right)_{\vartheta\vartheta^\prime}\tilde{I}_{\rm piv,\vartheta^\prime}^{\rm KD}\, ,
\end{equation}
where the matrix $M$ is defined by
\begin{equation}
M_{\vartheta \vartheta^\prime} \equiv \sum_{A,B,f} \frac{1}{2\Delta f}E^{ff}_{\vartheta,AB}B_{f,\vartheta^\prime}^{AB,\rm KD}\, .
\end{equation}
As in the ILC case, we want an unbiased estimator, therefore we require that $\hat{I}_{\rm piv,\vartheta}^{\rm KD}\propto I_{\rm piv,\vartheta}^{\rm KD}$, which means that $M_{\vartheta \vartheta^\prime}$ is diagonal in pixel space, which implies that 
\begin{equation}
M_{\vartheta \vartheta^\prime} = \delta_{\vartheta \vartheta^\prime}\rightarrow \sum_{A,B,f} \frac{1}{2\Delta f}E^{ff}_{\vartheta,AB}B_{f,\vartheta^\prime}^{AB,\rm KD} = \delta_{\vartheta \vartheta^\prime}\, .
\end{equation}
Therefore, the mean value of our estimator can be written as
\begin{equation}
\langle\hat{I}_{\rm piv,\vartheta}^{\rm KD} \rangle = \langle \tilde{I}_{\rm piv,\vartheta}^{\rm KD}\rangle = I_{\rm piv,\vartheta}^{\rm KD}+\frac{1}{2\Delta f}\sum_{f,\vartheta^\prime}  {\rm Tr}\left(E^{ff}_\vartheta B_{f\vartheta^\prime}^{\rm KD}\right)\frac{I_{f, \vartheta^{\prime}}^{\rm int}+I_{f, \vartheta^{\prime}}^{\rm SN}}{\mathcal{E}_f^{\rm KD}}\, .
\end{equation}
The covariance of our estimator is computed w.r.t. the data $d_{A,f}$, but there are two different sources of covariance. The first source of error is given by the fact that our estimator is a linear combination of terms quadratic in the data, therefore the covariance of this object, which has been already computed in the literature~\cite{Alonso:2020rar,Mentasti:2020yyd,LISACosmologyWorkingGroup:2022kbp}, comes from terms quartic in the data and it depends mainly on the instrumental noise. This covariance has to be summed in quadrature with the error given by the fact that the signal $h$ is the sum of different contributions and all the components we are not interested in are foregrounds that have to be removed and that contribute to the total covariance. This kind of error has been computed for the first time in the GW anisotropy context in this work and its interpretation is that we do not know the realization of the maps of the SN and of the intrinsic anisotropies. The calculation of this new term has been done in Appendix \ref{appendix_lagrange_multiplier} and the minimum covariance we have found, which minimizes both the instrumental noise and the contaminations to the signal from the SN and the intrinsic anisotropies is 
\begin{small}
\begin{equation}
{\rm cov}_{\vartheta\vartheta^\prime}=\frac{2\delta_{\vartheta\vartheta^\prime}}{\sum_{f}{\rm Tr}\left(S_{f}^{-1}B_{f,\vartheta}^{\rm KD}S_{f}^{-1}B_{f,\vartheta}^{\rm KD}\right)}+\delta_{\vartheta\vartheta^\prime}\sum_{f,f^\prime,\varphi,\varphi^\prime}\frac{{\rm Tr}\left(S_{f}^{-1}B_{f,\varphi}^{\rm KD}S_{f}^{-1}B_{f,\vartheta^\prime}^{\rm KD}\right){\rm Tr}\left(S_{f^{\prime}}^{-1}B_{f^{\prime}\vartheta^\prime}^{\rm KD}S_{f^{\prime}}^{-1}B_{f^{\prime},\varphi^\prime}^{\rm KD}\right)}{\left[\sum_{f}{\rm Tr}\left(S_{f}^{-1}B_{f,\vartheta}^{\rm KD}S_{f}^{-1}B_{f,\vartheta}^{\rm KD}\right)\right]^2}\frac{C_{ff^\prime,\varphi\varphi^{\prime}}^{\rm tot}}{\mathcal{E}_{f}^{\rm KD}\mathcal{E}_{f^{\prime}}^{\rm KD}}\, ,
\end{equation}
\end{small}
with $C_{ff^\prime,\varphi\varphi^{\prime}}^{\rm tot}$ the covariance matrix of the map of the SN and of the intrinsic anisotropies. The first term is the standard term due to instrumental noise, while the second one is the new term due to SN and the intrinsic anisotropies, that is maximally reduced by the weights in $f$, $f^\prime$ we have chosen. It is worthy to stress that in the traces we have considered all the possible combinations between the interferometers, auto- and cross-correlations. However, the result does not change if the auto-correlations are removed from the analysis, since it is known that the auto-correlation does not play an important role in the detection of stochastic backgrounds.

\begin{figure}
    \centering
    \includegraphics[scale=0.9]{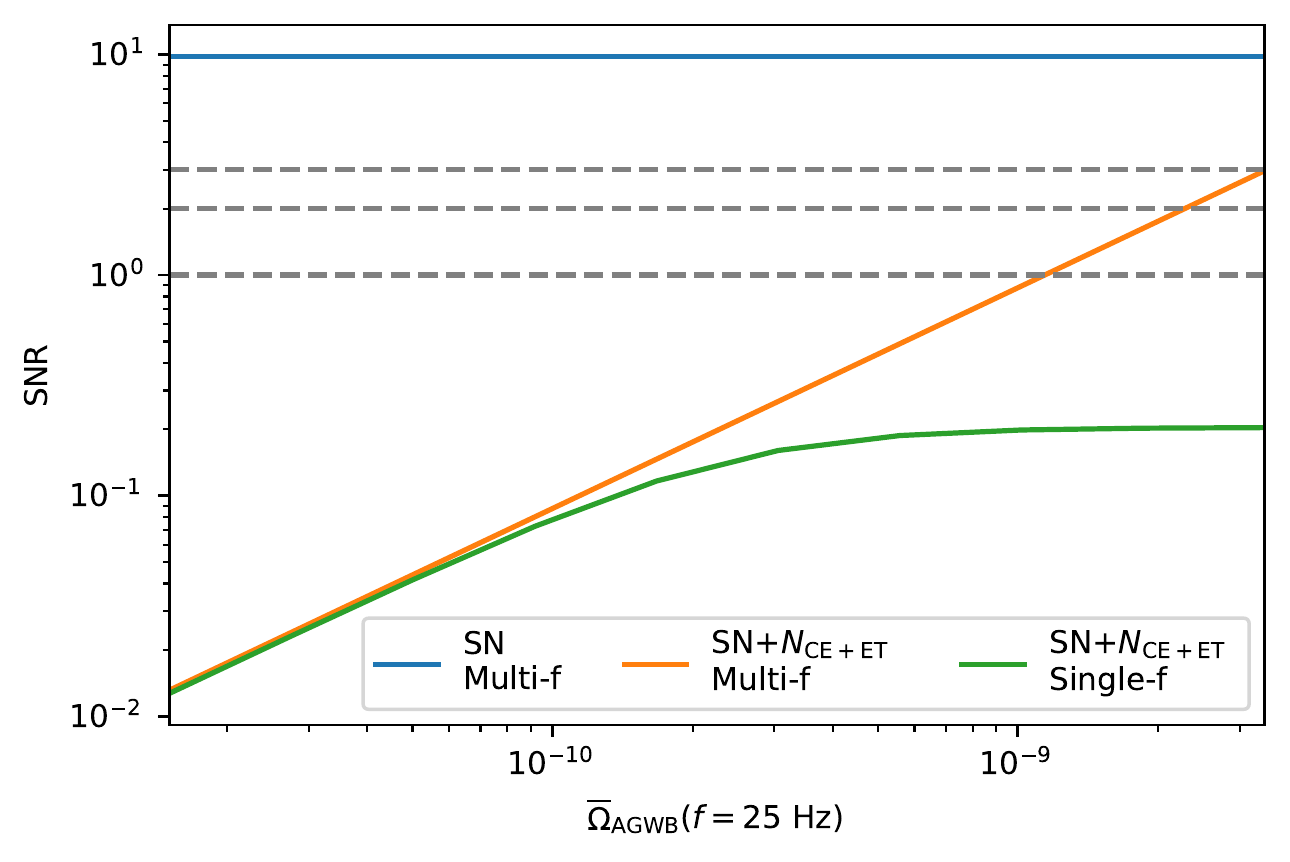}
    \includegraphics[scale=0.9]{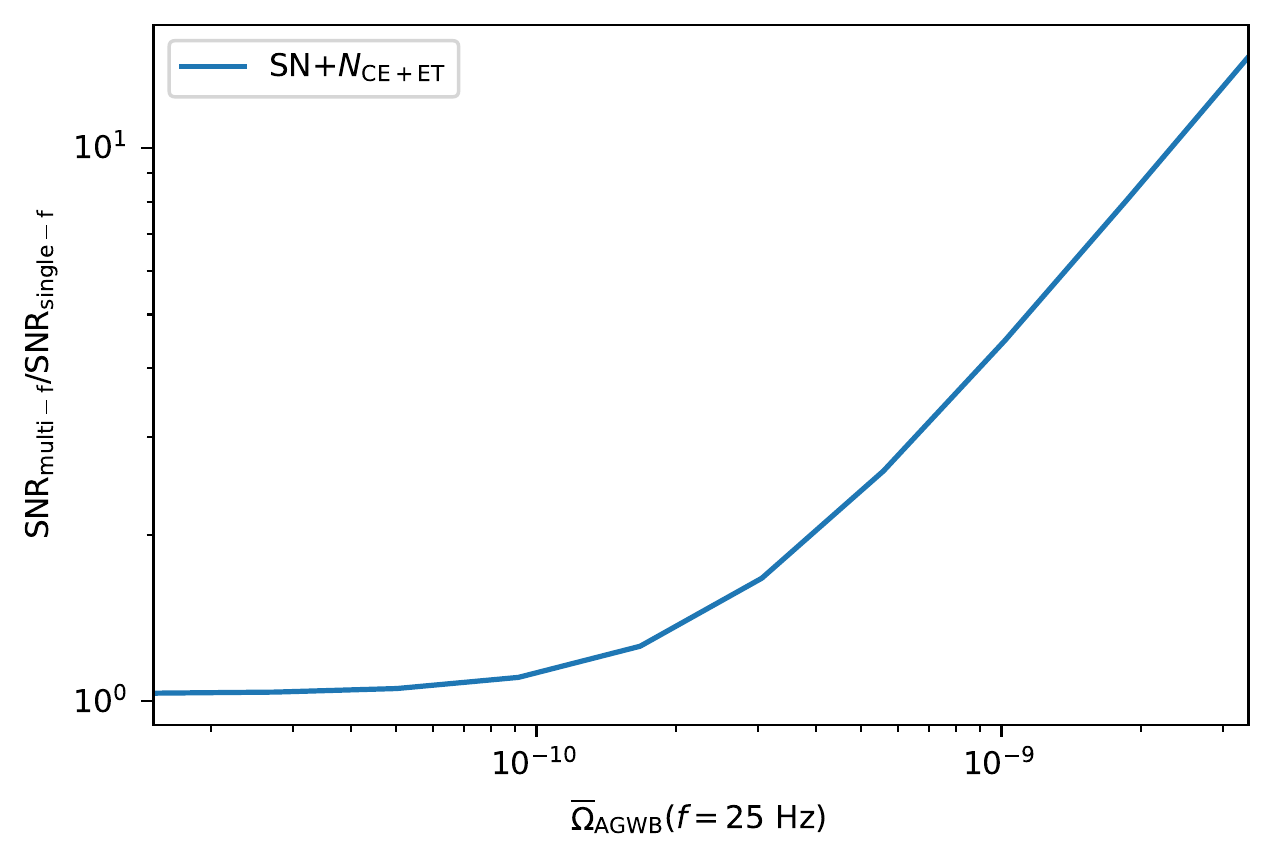}
    \caption{{\it Upper: plot of the SNR of the kinematic dipole as a function of the monopole amplitude of the AGWB by considering SN and SN plus instrumental noise ($N_{ET+CE}$). The horizontal lines show the SNR equal to 1, 2 and 3 respectively. Lower: Plot of the ratio of the SNR obtained by using the multi-frequency and the SNR computed by looking at a single frequency.}}
    \label{snr_improved_figure}
\end{figure}

To connect the covariance of $\hat{I}_{\rm piv}$ to the covariance of $\delta_{\rm AGWB}^{\rm KD}(f_{\rm piv})$ we use Eq. \eqref{I_Omega_Equation},
\begin{equation}
{\rm cov}_{\vartheta\vartheta^{\prime}}^{\rm \delta_{\rm AGWB}}=\left( \frac{1}{\bar{\Omega}_{\rm AGWB}(f_{\rm piv})}\frac{4\pi^2 f^3}{3H_0^2}\right)^2{\rm cov}_{\vartheta\vartheta^\prime}\, .
\end{equation}
Keep in mind that, according to the definitions we have used here, the covariance matrix associated to SN is the covariance matrix of the intensity, which is related to the covariance matrix of the density contrast through
\begin{equation}
C_{f f^\prime}^{\rm SN} = \left(\bar{\Omega}_{\rm AGWB}(f_{\rm piv})\frac{3H_0^2}{4\pi^2 f^3}\right)^2C_{f f^\prime}^{\rm SN,\delta_{\rm AGWB}}\, .
\end{equation}
What we have done until now has been done for a single time-frame. To take into account the duration of the observation ($T_{\rm obs} = 10\, \rm yrs$), we just divide the covariance by $T_{\rm obs}$, neglecting the effect of rigid rotation
\begin{equation}
{\rm cov}^{\rm \delta_{\rm AGWB},tot}_{\vartheta\vartheta^\prime} =\frac{1}{T_{\rm obs}}{\rm cov}_{\vartheta\vartheta^{\prime}}^{\rm \delta_{\rm AGWB}}\, .
\end{equation}
We extract the dipole from a map by using~\cite{Hiratadipole,Gibelyou:2012ri}
\begin{equation}
        v_{o,i} = \sum_\vartheta \Delta\Omega \, \hat{n}^i_\vartheta\, \delta_{\rm AGWB,\vartheta}^{\rm KD}\, ,
\end{equation}
therefore the covariance of the dipole is related to the covariance of the map through
\begin{equation}
    {\rm cov}_{ij} = {\rm cov}(v_{o,i},v_{o,j}) = \sum_\vartheta \Delta \Omega^2 \hat{n}^i_\vartheta \hat{n}^j_\vartheta {\rm cov}_{\vartheta\vartheta}^{\delta_{\rm AGWB,tot}}\, .
\end{equation}
The covariance matrix we have found for the maximum monopole amplitude is 
\begin{equation}
{\rm cov}_{ij} =
\begin{pmatrix}
1.1\times 10^{-6} & -1.2\times 10^{-6} & -2.9\times 10^{-7} \\
-1.2\times10^{-6} & 6.1\times 10^{-7} & 3.1\times 10^{-7}\\
-2.9 \times 10^{-7} & 3.1\times 10^{-7} & 7.3\times 10^{-7}
\end{pmatrix}\, .
\end{equation}
Just to give an order of magnitude of the error on the dipole we marginalize over the $y$ and $z$ directions, obtaining
\begin{equation}
\sigma_{d_x} = \frac{1}{2}\sqrt{\frac{3}{\pi}}{\rm cov}_{10,10^\prime}\approx 0.0005\, .
\end{equation}
In analogy with Eq. (\ref{SNR_dipole_cov_equation}), we have provided an estimate of the SNR of the dipole by using
\begin{equation}
    {\rm SNR}^2 = \vec{v}_o^{\, T}\, {\rm cov}^{-1} \vec{v}_o\, .
\end{equation}
The result is plotted in Figure \ref{snr_improved_figure} as a function of the monopole amplitude of the AGWB at $f=25 \, \rm Hz$. We can see that for values of the monopole of the AGWB within the upper bound of LIGO/Virgo, the estimator is able to reduce the instrumental noise and to give an SNR larger than one. In the figure we have also plotted the SNR by looking at just one frequency, showing that in the limit of large monopoles/low instrumental noise our technique is able to increase the SNR w.r.t. the standard approach.

\section{Conclusions}
One of the most interesting aspects of the AGWB is that the same astrophysical sources contribute to the overall signal in a wide range of frequencies. Since the evolution of a binary system is described by three stages, the inspiral, the merger, and the ringdown, we expect that, when a non-negligible fraction of the sources contribute to the AGWB in the merger and the ringdown stages, the dependence of the AGWB monopole on the frequency is not simply a power law. More specifically, when sources at different redshifts contribute to the overall signal at different stages of the evolution, we are not able to factorize the redshift and the frequency in the contribution to the background. This means that the monopole amplitude, the window function to compute the anisotropies of the AGWB, and the evolution bias are frequency dependent. This allows us to apply a component separation technique between the three contributions to the AGWB dipole: the intrinsic, the kinematic, and the SN, because they have different shapes in the frequency domain.  It is natural therefore to test if the next generation GW observatories are able to extract the velocity of the observer w.r.t. the LSS by looking at the AGWB maps. The analysis of the AGWB kinematic dipole presents some advantages w.r.t. other probes, such as galaxy surveys, because GW interferometers are almost full-sky, therefore the bias induced by partial sky coverage is reduced. Moreover, since interferometers have access to many frequencies, we are able to distinguish between the intrinsic and the kinematic dipole contributions by just using an observable (the AGWB), without introducing cross-correlations between different observables. The only astrophysical information we need to know is the population of the sources that contribute to the AGWB as a function of the redshift and of the mass of the sources. The evolution of the population of binary systems in time can be found for instance by independent experiments which look at the SFR, where the error bars are very small, while the mass distribution of the objects can be extracted by looking at the resolved sources at the interferometers. Even if at the present time we have large uncertainties on the PDF of the masses of the compact objects in binaries, future GW experiments will be able to resolve a lot of events, reducing the error bars on the parameters which describe the mass distributions.

In this work we have quantified the three contributions to the AGWB dipole for a population of BBH with a minimum and a maximum mass of $2.5 \, M_\odot$ and $100 \, M_\odot$ respectively. We have seen that the SN contribution is about one order of magnitude larger than the kinematic one and about two orders of magnitude larger than the intrinsic. Motivated by this, we have performed an analysis on the AGWB dipole in presence of SN only, finding that by using ILC in a frequency range $[100,1000]\, \, \rm Hz$ we are able to extract the kinematic dipole with $\rm SNR \approx 10$. In the more realistic scenario, where also the instrumental noise is considered, the situation is more delicate and the estimator has a more complicated form, since it has to be built starting from the strain of the AGWB and not from the density contrast. By generalizing the formalism of matched-filtering, typically used to minimize the instrumental noise at interferometers, we have built an estimator to reduce the covariance given by instrumental noise and SN. When a network of ET+CE is considered, we are able to extract the kinematic dipole with an SNR $\simeq$ 2.5 for a monopole amplitude close to the upper bound provided by LIGO/Virgo/KAGRA. The only assumptions we have used in our analysis are the following: the covariance due to the SN between objects with different astrophysical properties is negligible, the effect of the marginalization over the binary population parameters is small and all the BBH systems are short-lived sources. Finally we have assumed $T_{\rm obs} = 10\, \rm yrs$ of observation for the network.

The technique introduced in this paper can be extended to other kind of stochastic background with a non-trivial frequency dependence, such as the superposition of the AGWB signals produced by BHNS and BNS, on top of BBH. In this case we expect a larger monopole amplitude, especially at larger frequencies, therefore our analysis would be able to increase the SNR in the case in which both SN and instrumental noise are taken into account. Finally, we have found that the main limitation in determining the observer velocity is given by the instrumental noise, but we expect that with future improvements of interferometers sensitivity, we will be able to measure the kinematic dipole more precisely.

The analysis of the AGWB kinematic dipole with the new method we have introduced has been done by considering the auto-correlation only. We expect that including the cross-correlation with other cosmological probes, such as the galaxy number count, would give an higher precision in the estimate of the observer velocity. Furthermore, here we have considered the contribution to the AGWB given by burst, short-lived sources, neglecting the impact on the signal of early inspirals or of different mass ranges for the BBH population. Adding these signals could help improving the results we have obtained, especially at lower frequencies. Note also that in this preliminary work we have computed the SN under the assumption that only binaries with the same properties contribute. However we have derived a more general expression for the SN, where binaries from different channels contribute. Since in this work we are interested in introducing this new technique, we did not include this extra (subdominant) SN, but in a realistic analysis one has to take into account for it, together with the uncertainties of the astrophysical parameters which describe the population of GW events.

\acknowledgments
We thank N. Bellomo, G. Cusin, A. C. Jenkins, M. Liguori and A. Ravenni for useful discussions and comments on the draft.
D.B. acknowledges partial financial support by ASI Grant No. 2016-24-H.0. A.R. acknowledges funding from MIUR through the ``Dipartimenti di eccellenza'' project Science of the Universe and financial support from the Supporting TAlent in ReSearch@University of Padova (STARS@UNIPD) for the project “Constraining Cosmology and Astrophysics with Gravitational Waves, Cosmic Microwave Background and Large-Scale Structure cross-correlations''.

\appendix

\section{AGWB Anisotropies Computation}
\import{}{Draft_JCAP/Tex_Appendices/appendix_agwb_anisotropies.tex}

\section{Compound Poisson Distribution}
\import{}{Draft_JCAP/Tex_Appendices/appendix_cpd.tex}

\section{SKAO2}
\import{}{Draft_JCAP/Tex_Appendices/appendix_SKA.tex}

\section{Minimum covariance matrix for the kinetic dipole}
\import{}{Draft_JCAP/Tex_Appendices/appendix_lagrange_multiplier.tex}

\end{document}

%% file: Draft_JCAP/Tex_Appendices/appendix_agwb_anisotropies.tex
\label{AGWB Anisotropies Computation}
\label{appendix_agwb_anisotropies_newtonian_gauge}
In this work we compute the AGWB anisotropies in the Poisson gauge, 
\begin{equation}
ds^2 = a^2(\eta)\left[-d\eta^2\left(1+2\Psi\right)+\left(1-2\Phi\right)\delta_{ij}dx^idx^j+h_{ij}^{\rm TT}dx^idx^j\right]\, .
\end{equation}
The observer has a four-velocity 
$u^\mu = [(1-\Psi)/a,v^i/a]$
and we defined the direction of observations as $\hat{n}$.\\ 
The GW density constrast is 
\begin{align} 
\delta_{\rm AGWB} &=\int d\bar{\chi}\,\tilde{\mathcal{W}}\Biggl[b^{[i]}\left(\delta_m-3\mathcal{H}V\right)+(3-b_e^{[i]})\mathcal{H}V+\Psi\left(3-b_e^{[i]}+\frac{\mathcal{H}^\prime}{\mathcal{H}^2}\right)+\nonumber \\
&+2I\left(b_e^{[i]}-\frac{\mathcal{H}^\prime}{\mathcal{H}^2}-2\right)+\left(\delta a_o + \Psi_o-v_{\parallel\, o}\right)\left(b_e^{[i]}-\frac{\mathcal{H}^\prime}{\mathcal{H}^2}-2\right)-v_\parallel\left(-b_e^{[i]}+\frac{\mathcal{H}^\prime}{\mathcal{H}^2}+2\right) \nonumber\\
&+\frac{1}{\mathcal{H}}\Phi^\prime-\frac{1}{\mathcal{H}}\partial_\parallel v_\parallel-\frac{1}{\mathcal{H}}\frac{1}{2}h^{\rm TT \, \prime}_{ij}n^in^j\Biggl]\, ,
\label{map_agwb_equation}
  \end{align}
where we have introduced the following projected quantities along the line-of-sight 
\begin{equation}
\begin{split}
v_\parallel \equiv& \, \hat{n}\cdot \vec{v}\, ,\\
\partial_\parallel \equiv& \, \hat{n}\cdot\vec{\nabla}\, . \\
\end{split}
\end{equation}
$\tilde{W}$ is the window function associated to the AGWB, while the quantity $I$ represents an integrated GR contribution to the AGWB anisotropies,
\begin{equation}
I (\bar \chi)\equiv -\frac{1}{2}\int_0^{\bar{\chi}}\ud \tilde \chi\left(\Psi^\prime+\Phi^\prime-\frac{1}{2}h^\prime_{ij}\right)(\tilde \chi)\, .
\end{equation}
$b$ and $b_e$ are the bias and the evolution bias of the GWs respectively, while $V$ is the velocity potential defined by
\begin{equation}
\vec{v}\equiv \vec{\nabla}V\, .
\end{equation}
The notation $f_o$ indentifies the field $f$ evaluated at the observer, i.e. at coordinates $\bar{\chi}_o=\vec{x}_o=0$. We have denoted with the prime the derivatives w.r.t. the conformal time $\eta$, which is related to the comoving distance $\bar{\chi}$ by
\begin{equation}
\bar{\chi}\equiv \eta_0-\eta\, ,
\end{equation}
where $\eta_0$ is the value of the conformal time at the present.\\
We compute the coefficients of the expansion in Legendre polynomials of the AGWB density contrast,
\begin{equation}
\Delta_\ell^{\rm AGWB} \equiv \int d\phi \int d\mu \, \mathcal{P}_\ell(\mu) \, \delta_{\rm AGWB}(\hat{n})\, .
\end{equation}
In this way the angular power spectrum is simply
\begin{equation}
C_\ell^{XY} = 4\pi \int \frac{dk}{k} P(k)\,  \Delta_\ell^X \, \Delta_\ell^{Y\, *}\, ,
\label{angular_power_spectrum_class}
\end{equation}
where the angular power spectrum is computed w.r.t. the primordial curvature perturbation $\zeta$,
\begin{equation}
\langle \zeta(\vec{k})\,\zeta^*(\vec{k}^\prime)\rangle \equiv (2\pi)^3\delta^{(3)}(\vec{k}-\vec{k}^\prime)\frac{2\pi^2}{k^3}P(k)\, .
\end{equation}
We have computed the different contributions to $\Delta_\ell^{\rm AGWB}$, starting from Eq. \eqref{map_agwb_equation}, separating the stochastic and the deterministic part in each random field in the following way
\begin{equation}
X(\eta,\vec{k})=T_X(\eta,\vec{k})\zeta(\vec{k})\, ,
\end{equation} 
where $T_X$ is the transfer function of the field $X$ which takes into account for its evolution computed by combining the Einstein and the Boltzmann equations.\\
The result we have found is\footnote{We have used the same notation of~\cite{DiDio:2013bqa}.} 
\begin{equation}
\begin{split}
\Delta_\ell^{\rm den} = & \int_0^{\eta_0} d\eta\, \tilde{W}^{[i]}\left(b^{[i]}T_{\delta_m}+3\frac{aH}{k^2}T_{\theta_m}\right)j_\ell(k\bar{\chi})\, , \\
\Delta_\ell^{\rm D1} = & \int_0^{\eta_0} d\eta\, \tilde{W}^{[i]}\frac{1}{k}T_{\theta_m}\left(-b_e^{[i]}+\frac{H^\prime}{aH^2}+3\right)\frac{d}{d[k\bar{\chi}]}j_\ell(k\bar{\chi})\, , \\
\Delta_\ell^{\rm D2} = & \int_0^{\eta_0} d\eta\, \tilde{W}^{[i]}(b_e^{[i]}-3)\frac{aH}{k^2}T_{\theta_m} j_\ell(k\bar{\chi})\, , \\
\Delta_\ell^{\rm rsd} = & \int_0^{\eta_0} d\eta\, \tilde{W}^{[i]}\frac{1}{aH}T_{\theta_m}\frac{d^2}{d[k\bar{\chi}]^2}j_\ell(k\bar{\chi})\, , \\
\Delta_\ell^{\rm G1} = & \int_0^{\eta_0} d\eta\, \tilde{W}^{[i]}T_{\Psi}\left(4-b_e^{[i]}+\frac{H^\prime}{aH^2}\right)j_\ell(k\bar{\chi})\, , \\
\Delta_\ell^{\rm G2} = & 0\, , \\
\Delta_\ell^{\rm G3} = & \int_0^{\eta_0} d\eta\, \tilde{W}^{[i]}\frac{1}{aH}T_{\Phi^\prime}j_\ell(k\bar{\chi})\, ,\\
\Delta_\ell^{\rm G4} = & 0\, , \\
\Delta_\ell^{\rm G5} = & \int_0^{\eta_0} d\eta\, \tilde{W}^{[i]}\left(-b_e^{[i]}+\frac{H^\prime}{a H^2}+3\right)\int_0^{\tilde{\eta}}d\tilde{\eta}j_\ell(k\bar{\chi}) \left(T_{\Phi^\prime}(\tilde{\eta})+T_{\Psi^\prime}(\tilde{\eta})-\frac{1}{2}T_{h,{ij}}^\prime(\tilde{\eta}) n^i n^j\right)\, ,\\
\Delta_\ell^{\rm o\, mon} = &  \int_0^{\eta_0} d\eta \, \tilde{W}^{[i]}\left(T_{\delta a,o} + T_{\Psi,o}\right)\left(b_e^{[i]}- \frac{H^\prime}{a H^2}-3\right)\frac{1}{2\ell+1}\delta_{\ell 0}\, , \\
\Delta_\ell^{\rm KD} = &  \int_0^{\eta_0} d\eta \, \tilde{W}^{[i]}\left(b_e^{[i]}- \frac{H^\prime}{a H^2}-3\right)\frac{1}{k}T_{\theta_m, o}\frac{1}{2\ell+1}\delta_{\ell 1}\, . \\
\label{source_omega_agwb}
\end{split}
\end{equation}

%% file: Draft_JCAP/Tex_Appendices/appendix_cpd.tex
\label{Compound Poisson Distribution}

To compute the SN of the AGWB we follow the approach of~\cite{Jenkins:2019nks,Jenkins:2019uzp}, counting the AGWB sources in the same way of~\cite{Bellomo:2021mer}. 
The average number of stars formed in a time $T_{\rm obs}$ in a halo of mass $M_h$ at redshift $z_d$ is 
\begin{equation}
    \bar{N}_{\star|h}(z_d)\equiv \langle {\rm SFR}_{\rm SF}(M_h,z_d)\rangle T_{\rm obs} \frac{\ud V}{\ud z \ud\Omega_e}(z_d)\, ,
    \label{N_star_average_eq}
\end{equation}
where the dependence on the halo mass is contained implicitly in the subscript $h$. Since the number of stars in a given region of the sky $N_{\star|h}(z_d,\hat{n})$ is a discrete event, it follows a Poisson distribution with mean $\bar{N}_{\star|h}(z_d)$ and covariance
\begin{equation}
    {\rm cov}\left[N_{\star|h}(z_d,\hat{n}),N_{\star|h^\prime}(z_d^\prime,\hat{n}^\prime)\right] = \delta(M_h-M_h^\prime)\delta(z_d-z_d^\prime)\delta(\hat{n}-\hat{n}^\prime)\bar{N}_{\star|h}(z_d)\, ,
\end{equation}
because Poisson fluctuations are associated to the same infinitesimal volume in the sky. We assume then that halos with different masses are not correlated; consequently, the number of stars produced in halos of different masses are uncorrelated. We leave the possibility to relax this assumption for a future work.

We describe the number of binaries formed in a time $T_{\rm obs}$ in a halo of mass $M_h$ at redshift $z_d$ as a function of the number of stars formed,
\begin{equation}
    N_{\rm GW|h}\left(\vec{\theta},t_d,z_d,\hat{n}\right) = \mathcal{A}_{\rm LIGO} p(t_d)p(\vec{\theta})\, w(z,\vec{\theta}\, )\,  \ud \vec{\theta}\, \ud t_d\,  N_{\star|h}(z_d,\hat{n})\, ,
\end{equation}
where $\hat{n}$ represents the direction of observation in the sky, while $t_d$ is the time delay after which the binary merges, which is related to the separation between the compact objects at the formation of the binary, and $\vec{\theta}$ are the astrophysical parameters of the binary (spins, masses, etc.). $\mathcal{A}_{\rm LIGO}$ describes the fraction of stars which become compact objects times the fraction of compact objects which form binary systems and $w$ keeps into account for the fraction of them that contribute to the background. The relation between the redshift at the formation of the binary $z_d$ and the redshift $z$ at which the GWs has been given in Eq. \eqref{z_d_to_z_equation}. It is immediate to see that the average number of GW events we expect from a halo of mass $M_h$ is 
\begin{equation}
    \begin{split}
    \bar{N}_{\rm GW|h}\left(\vec{\theta},t_d,z_d\right) =& \mathcal{A}_{\rm LIGO} p(t_d)p(\vec{\theta})w(z,\vec{\theta}\, )\ud \vec{\theta}\, \ud t_d\bar{N}_{\star|h}(z_d)\, ,
    \label{mean_N_GW_h_eq}\, .
    \end{split}
\end{equation}
The covariance between binaries with the same parameters is the covariance of a Poisson distribution, therefore
\begin{equation}
\begin{split}
    {\rm cov}\left[N_{\rm GW|h}\left(\vec{\theta},t_d,z_d,\hat{n}\right),{N}_{\rm GW|h}\left(\vec{\theta},t_d,z_d^\prime,\hat{n}^\prime\right)\right] = &\delta(\hat{n}-\hat{n}^\prime)\delta(z_d-z_d^\prime)\bar{N}_{\rm GW|h}\left(\vec{\theta},t_d,z_d\right)\, .
    \label{cov_N_GW_h_eq_same_theta}
    \end{split} 
\end{equation}
The two Dirac delta functions reflect the condition that only events in the same volume can be correlated. Notice that we require that the events have to be produced at the same time, but we are not imposing that two GW sources for which we compute the Poisson noise emit at the same time; this will be crucial in further calculations. The computation of the covariance between GW sources which depend on different parameters is more delicate. As shown in~\cite{Smith:2008ut}, the cross-correlation of non-overlapping tracers (independent Poisson samples) is trivially zero. On the other hand, when the two tracers have a common origin, the joint probability can be written in terms of conditional probabilities,
\begin{equation}
    p(N_A,N_B) = \int \ud N_c\, p(N_A|N_c)\, p(N_B|N_c)\, p(N_c)\, .
\end{equation}
In our case, binaries with different time delay and astrophysical parameters share the average SFR per halo, therefore fluctuations in the number of stars produced are expected to affect the number of binaries for every value of the time delay and of the astrophysical parameters, thus we expect an extra correlation w.r.t. the case of equal $t_d$ and $\vec{\theta}$. In analogy with the case of the cross-correlation of the galaxies and the AGWB~\cite{Canas-Herrera:2019npr,Alonso:2020mva}, the cross-correlation of the SN is expected to be the intersection of the events (the number of binaries with different properties),
\begin{equation}
    \begin{split}
    {\rm cov}\left[N_{\rm GW|h},\tilde{N}_{\rm GW|h}\right] = & \delta(M_h-\tilde{M}_h)\delta(z_d-\tilde{z}_d)\delta(\hat{n}-\tilde{\hat{n}})\\
    &p(t_d)p(\tilde{t}_d)p(\vec{\theta}\, )p(\tilde{\vec{\theta}}\,)w(z,\vec{\theta}\, )w(z,\tilde{\vec{\theta}}\, )\,\ud \vec{\theta}\, \ud t_d\, \ud \tilde{\vec{\theta}}\, \ud \tilde{t}_d\mathcal{A}_{\rm LIGO}^2\bar{N}_{\star|h}(z_d)\, .
    \label{cov_N_GW_h_eq}
    \end{split} 
\end{equation}

The total number of binaries formed in halos with mass $M_h$ is obtained by
\begin{equation}
N_{\rm GW|h}^{\rm tot}\left(\vec{\theta},t_d,z_d,\hat{n}\right) \equiv \sum_{i=1}^{N_h(z_d,\hat{n})} N_{\rm GW|h}\left(\vec{\theta},t_d,z_d,\hat{n}\right)\, ,
\end{equation}
where $N_{h}(z_d,\hat{n})$ is the number of halos of mass $M_h$ in a given volume in the sky. The number of halos is a Poisson random variable with mean and covariance given by
\begin{equation}
\begin{split}
\bar{N}_{h} =& \frac{\ud n}{\ud M_h}(M_h,z_d) \ud M_h\, , \\
{\rm cov}\left(N_h,N_h^\prime \right)=& \delta(M_h-M_h^\prime)\delta(z_d-z_d^\prime)\frac{\ud n}{\ud M_h}(M_h,z_d) \ud M_h\, .
\label{mean_cov_N_h_eq}
\end{split}
\end{equation} 

We can connect the quantities written above to the AGWB by using
\begin{equation}
\Omega_{\rm AGWB}(f,\hat{n}) = \frac{f}{\rho_c c^2} \int \frac{\ud z}{(1+z)H(z)} \sum_{\vec{\theta},M_h,t_d}\frac{\ud E}{\ud f_e \ud\Omega_e}(\vec{\theta},f_e)\frac{N_{\rm GW|h}^{\rm tot}\left(\vec{\theta},t_d,z_d,\hat{n}\right)}{T_{\rm obs}\frac{\ud V}{dz \ud \Omega_e}(z_d)}\, ,
\end{equation}
where $f_e = f(1+z)$ and $z$ is the redshift which corresponds to the GW emission, evaluated at the time $t$, while $z_d$ is the time at the formation of the binary, evaluated at $t-t_d$. To compute the average value of the AGWB signal along a specific direction in the sky, $\bar{\Omega}_{\rm AGWB}(f)$ we need to compute the mean value of $N_{\rm GW|h}^{\rm tot}\left(\vec{\theta},t_d,z_d,\hat{n}\right)$. The total number of GW events per halo of mass $M_h$ at $z_d$ follows a Compound Poisson Distribution (CPD) and its mean is computed by using the law of total expectation\footnote{The law of total expectation says that if we have $x$, $y$ random variables, the expectation value of $x$ is $$E[x] = E\left[E(x|y)\right]\, ,$$ where the first expectation value is computed w.r.t. the probability $p(x|y)$, while the second expectation value is computed w.r.t. $p(y)$.}
\begin{equation}
\begin{split}
\bar{N}_{\rm GW|h}^{\rm tot}\left(\vec{\theta},t_d,z_d\right) =& E\left[E\left(N^{\rm tot}_{\rm GW|h}\left(\vec{\theta},t_d,z_d,\hat{n}\right)\bigl |N_h(z_d)\right)\right] = \bar{N}_{\rm GW|h}\left(\vec{\theta},t_d,z_d\right)E[N_h(z_d)] =\\
=& \bar{N}_{\rm GW|h}\left(\vec{\theta},t_d,z_d\right) \bar{N}_h(z_d)\, .
\end{split}
\end{equation}
The law of total expectation simplifies the computation, because when we fix the number of halos to $N_h$ and we compute the expectation value of $N_{\rm GW|h}^{\rm tot}$ we have just to sum $N_h$ independent expectation values, all of them equal to $\bar{N}_{\rm GW|h}$, thus the expectation value is $\bar{N}_{\rm GW|h}N_h$. The AGWB monopole is then
\begin{equation}
\bar{\Omega}_{\rm AGWB}(f) = \frac{f}{\rho_c c^2} \int \frac{\ud z}{(1+z)H(z)} \sum_{\vec{\theta},M_h,t_d}\frac{\ud E}{\ud f_e \ud \Omega_e}(\vec{\theta},f_e) \frac{\bar{N}_{\rm GW|h}^{\rm tot}\left(\vec{\theta},t_d,z_d\right)}{T_{\rm obs}\frac{\ud V}{\ud z \ud\Omega_e}(z_d)}\, .
\end{equation}
The AGWB density contrast is defined as 
\begin{equation}
\delta_{\rm AGWB}(f,\hat{n}) \equiv \frac{\Omega_{\rm AGWB}(f,\hat{n})-\bar{\Omega}_{\rm AGWB}(f)}{\bar{\Omega}_{\rm AGWB}(f)} \, ,
\end{equation}
therefore its covariance receives a contribution due to the SN fluctuation of $N_{\rm GW|h}^{\rm tot}$, 
\begin{equation}
\begin{split}
\delta_{\rm AGWB}(f,\hat{n}) = \frac{f}{\rho_c c^2\bar{\Omega}_{\rm AGWB}(f)} \int &  \frac{\ud z}{(1+z)H(z)} \sum_{\vec{\theta},M_h,t_d}\frac{\ud E}{\ud f_e \ud \Omega_e}(\vec{\theta},f_e) \frac{\delta N_{\rm GW|h}^{\rm tot}\left(\vec{\theta},t_d,z_d,\hat{n}\right)}{T_{\rm obs}\frac{\ud V}{\ud z \ud \Omega}(z_d)}\, ,
\end{split}
\end{equation}
where we have defined the fluctuation of the total number of GW events per halo, 
\begin{equation}
\delta N_{\rm GW|h}^{\rm tot}\left(\vec{\theta},t_d,z_d,\hat{n}\right) \equiv N_{\rm GW|h}^{\rm tot}\left(\vec{\theta},t_d,z_d,\hat{n}\right)-\bar{N}_{\rm GW|h}^{\rm tot}\left(\vec{\theta},t_d,z_d\right)\, .
\end{equation}
By definition, this quantity has zero mean and its covariance is computed by using the law of total covariance\footnote{The law of total covariance says that $${\rm cov}(x,y) = E\left[{\rm cov}\left(x|z,y|z\right)\right]+{\rm cov}\left(E[x|z],E[y|z]\right)\, ,$$ where the first expectation value is computed w.r.t. the conditional probabilities $p(x|z)$, $p(y|z)$, while the second expectation value is computed w.r.t. $p(z)$.},
\begin{equation}
\begin{split}
{\rm cov}\left[\delta N_{\rm GW|h}^{\rm tot},\delta \tilde{N}_{\rm GW|h}^{\rm tot}\right] = & E\left\{{\rm cov}\left[\delta N_{\rm GW|h}^{\rm tot},\delta\tilde{N}_{\rm GW|h}^{\rm tot}\Bigl |N_h\right]\right\}+\\
&+{\rm cov}\left\{E\left[\delta N_{\rm GW|h}^{\rm tot}|N_h\right],E\left[\delta\tilde{N}_{\rm GW|h}^{\rm tot}|\tilde{N}_h\right]\right\}\, .
\end{split}
\end{equation}
where the quantities with the tilde are evaluated for different parameters w.r.t. to the other ones. We compute separately the two contributions,
\begin{equation}
\begin{split}
E\left\{{\rm cov}\left[\delta N_{\rm GW|h}^{\rm tot},\delta\tilde{N}_{\rm GW|h}^{\rm tot}\Bigl |N_h\right]\right\} = & E\left\{ {\rm cov}\left[N^{\rm tot}_{\rm GW|h}-\bar{N}_{\rm GW|h}\bar{N}_h,\tilde{N}_{\rm GW|h}-\tilde{\bar{N}}_{\rm GW|h}\tilde{\bar{N}}_h\Bigl |N_h\right]\right\} =\\
= & {\rm cov}\left[N_{\rm GW|h},\tilde{N}_{\rm GW|h}\right] E[N_h]=\\
=&{\rm cov}\left[N_{\rm GW|h},\tilde{N}_{\rm GW|h}\right]\bar{N}_h\, ,  \\
{\rm cov}\left\{E\left[\delta N_{\rm GW|h}^{\rm tot}|N_h\right],E\left[\delta\tilde{N}_{\rm GW|h}^{\rm tot}|\tilde{N}_h\right]\right\} = & \bar{N}_{\rm GW|h}\tilde{\bar{N}}_{\rm GW|h}{\rm cov}\left(N_h,\tilde{N}_h\right) = \\
= & \delta(M_h-\tilde{M}_h)\delta(z_d-\tilde{z}_d)\bar{N}_{\rm GW|h}\tilde{\bar{N}}_{\rm GW|h} \bar{N}_h\, .
\end{split}
\end{equation}
The SN term for $t_d = \tilde{t}_d$, $\vec{\theta} = \tilde{\vec{\theta}}$ is 
\begin{equation}
\begin{split}
{\rm CSN} = & {\rm cov}\left[\delta_{\rm AGWB}(f_1,\hat{n}),\delta_{\rm AGWB}(f_2,\hat{n}^\prime)\right] = \\
= & \frac{f_1 f_2}{(\rho_c c^2)^2\bar{\Omega}_{\rm AGWB}(f_1)\bar{\Omega}_{\rm AGWB}(f_2)}\delta(\hat{n}-\hat{n}^\prime) \\
&\int   \frac{\ud z}{(1+z)H(z)} \int \ud \vec{\theta}\frac{\ud E}{\ud f_e \ud \Omega_e}(\vec{\theta},f_1,z)\int   \frac{\ud z^\prime}{(1+z^\prime)H(z^\prime)} \, \frac{\ud E}{\ud f_e \ud \Omega_e}(\vec{\theta},f_2,z^\prime) \\
& \int \ud M_h \int \ud t_d \frac{\ud n}{\ud M_h}(M_h,z_d)\frac{1}{\left(T_{\rm obs}\frac{\ud V}{\ud z \ud \Omega_e}(z_d)\right)^2}\\
&\left[\bar{N}_{\rm GW|h}\left(\vec{\theta},t_d,z_d\right)+\bar{N}^2_{\rm GW|h}\left(\vec{\theta},t_d,z_d\right) \right]\delta(z_d-z_d^\prime)\, .
\label{eq_CSN_theta_equal_nondef}
\end{split}
\end{equation}
The above equation shows the cross-correlation of the SN at different frequencies $f_1$, $f_2$, for binaries with the same astrophysical parameters $\vec{\theta}$ and time delay $t_d$. The binaries of the two $\delta_{\rm AGWB}$ have the same $z_d$ and the same $t_d$, thus they emit GWs of the same frequency at the same redshift. In this work we are considering the GW sources in the late-inspiral, in the merger and in the ringdown stages, therefore we assume that all the GW are emitted in an infinitesimal time around their merger (at $z$). However, the GW evolution in time is determined by the time delay and the astrophysical parameters, therefore the redshift at which a source emits GWs at a certain frequency is known, 
\begin{equation}
    f = f(z,\delta z,t_d,\vec{\theta}\,)\, ,
\end{equation}
where $\delta z$ is the shift in redshift\footnote{To give an example, if the frequency of GWs emitted is $f_{\rm merge}$, then $\delta z = 0$. As already stressed, $\delta z$ has to be much smaller than one for short-lived sources, thence it is used here just to show how we correlate GW events of different frequencies emitted at different times, but it plays no role in the numerical computation of the AGWB.} between the merger $z$ and the redshift at which the GWs of frequency $f$ are emitted. When we try to compute the correlation of the SN at frequencies $f_1$, $f_2$ we require that $z_d$ is the same, which is equivalent, for the same $t_d$, to impose that $z$ is the same,
\begin{equation}
\begin{split}
    \delta(z_d-z_d^\prime)=&\delta\left[z_d\left(z,\delta z(f_1),t_d,\vec{\theta}\,\right)-z_d^\prime\left(z^\prime,\delta z^\prime(f_2),t_d,\vec{\theta}\,\right)\right]=\\
    = & \frac{\delta\left\{z^\prime- {z_d^\prime}^{-1}\left[z_d\left(z,\delta z(f_1),t_d,\vec{\theta}\,\right),\delta z^\prime (f_2),t_d,\vec{\theta}\right]\right\} }{\left | \frac{dz_d^\prime}{dz^\prime}\left(z^\prime,\delta z^\prime(f_2),t_d,\vec{\theta}\,\right)\right|_{z^\prime = {z_d^\prime}^{-1}\left[z_d\left(z,\delta z(f_1),t_d,\vec{\theta}\,\right),\delta z^\prime(f_2),t_d,\vec{\theta}\right]}} =\Biggl|_{\delta z,\delta z^\prime \rightarrow 0} \delta(z^\prime-z)\, .
\end{split}
\end{equation}
The latter equality holds because in this work we are computing the AGWB generated by burst sources, for whom the $\delta z$ gives correction to the redshift and to $dz_d^\prime/dz^\prime$ of the order $10^{-9}$, thus the integration w.r.t. $z^\prime$ in the computation of $\rm CSN$ gives $z^\prime = z$. A rigorous expression for the SN of the AGWB would require the substitution in Eq. \eqref{eq_CSN_theta_equal_nondef}
\begin{equation}
\begin{split}
    z\rightarrow & z+\delta z(f_1,z,t_d,\vec{\theta}\, )\, , \\
    z^\prime \rightarrow & z^\prime+\delta z^\prime(f_2,z^\prime,t_d,\vec{\theta}\,)\, ,
\end{split}
\end{equation}
but since the stages of the evolution of the binaries we are considering last much less than the timescales over which the quantities on which the AGWB depends vary, we can take the limit $\delta z\rightarrow 0$, $\delta z^\prime \rightarrow 0$. By plugging these results in Eq. \eqref{eq_CSN_theta_equal_nondef} we find Eq. \eqref{equazione_calcolo_shot_noise}. For $t_d \neq \tilde{t}_d$, $\vec{\theta} \neq \tilde{\vec{\theta}}$, by using \eqref{N_star_average_eq},~\eqref{mean_N_GW_h_eq},~\eqref{cov_N_GW_h_eq},~\eqref{mean_cov_N_h_eq}, the SN covariance matrix is equal to 
\begin{equation}
\begin{split}
{\rm CSN} = & {\rm cov}\left[\delta_{\rm AGWB}(f_1,\hat{n}),\delta_{\rm AGWB}(f_2,\hat{n}^\prime)\right] = \\
= & \frac{f_1 f_2}{(\rho_c c^2)^2\bar{\Omega}_{\rm AGWB}(f_1)\bar{\Omega}_{\rm AGWB}(f_2)} \\
&\int   \frac{\ud z}{(1+z)H(z)} \int \ud \vec{\theta} \frac{\ud E}{\ud f_e \ud \Omega_e}(\vec{\theta},f_1,z)\int   \frac{\ud z^\prime}{(1+z^\prime)H(z^\prime)} \int \ud \vec{\theta}^\prime\, \frac{\ud E}{\ud f_e \ud \Omega_e}(\vec{\theta},f_2,z^\prime) \\
& \int \ud M_h \int \ud t_d \frac{\ud n}{\ud M_h}(M_h,z_d)p(t_d)\int \ud t_d^\prime\, p(t_d^\prime)\frac{p(\vec{\theta})p(\vec{\theta}^\prime)}{\left(T_{\rm obs}\frac{\ud V}{\ud z \ud \Omega_e}(z_d)\right)^2}\\
&\left[\left(w(z,\vec{\theta}\, ){\rm SFR}_{\rm SF}(M_h,z_d)\rangle T_{\rm obs} \frac{\ud V}{\ud z \ud \Omega_e}(z_d)\right)+\left(w(z,\vec{\theta}\, ){\rm SFR}_{\rm SF}(M_h,z_d)\rangle T_{\rm obs} \frac{\ud V}{\ud z \ud \Omega_e}(z_d)\right)^2 \right]\\
\\
&\delta\left[z_d\left(z,\delta z(f_1),t_d,\vec{\theta}\,\right)-z_d^\prime\left(z^\prime,\delta z^\prime(f_2),t_d^\prime,\vec{\theta}^\prime\,\right)\right]\delta(\hat{n}-\hat{n}^\prime)\, .
\end{split}
\end{equation}
In this case the Dirac delta cannot be written in general as $\delta(z-z^\prime)$, because we have $t_d\neq t_d^\prime$, thus we keep the expression for the Dirac delta in terms of $z_d$, leaving for a future work the detailed computation. For simplicity we rewrite the above equation in terms of the following quantities,
\begin{equation}
\begin{split}
S = &  \int \ud M_h \int \ud t_d \frac{\ud n}{dM_h}(M_h,z_d)\int \ud t_d^\prime\, \frac{p(t_d)p(t_d^\prime )}{\left(T_{\rm obs}\frac{\ud V}{\ud z \ud \Omega_e}(z_d)\right)^2}\\
&\hspace{5.5em}\delta\left[z_d\left(z,\delta z(f_1),t_d,\vec{\theta}\,\right)-z_d^\prime\left(z^\prime,\delta z^\prime(f_2),t_d^\prime,\vec{\theta}^\prime\,\right)\right]\\
&\hspace{5.5em}\biggl[\left(w(z,\vec{\theta}\, ){\rm SFR}_{\rm SF}(M_h,z_d)\rangle T_{\rm obs} \frac{\ud V}{\ud z \ud \Omega_e}(z_d)\right)+\\
&\hspace{7em}+\left(w(z,\vec{\theta}\, ){\rm SFR}_{\rm SF}(M_h,z_d)\rangle T_{\rm obs} \frac{\ud V}{\ud z \ud \Omega_e}(z_d)\right)^2 \biggl] \, , \\
K= & \frac{f_1 f_2}{\bar{\Omega}_{\rm AGWB}(f_1)\bar{\Omega}_{\rm AGWB}(f_2)} \\
& \int d\vec{\theta}\,  p(\vec{\theta})\frac{\ud E}{\ud f_e \ud\Omega_e}(\vec{\theta},f_1,z) \int d\vec{\theta}^\prime\, p(\vec{\theta}^\prime)\frac{\ud E}{\ud f_e \ud \Omega_e}(\vec{\theta},f_2,z^\prime) \, ,
\end{split}
\end{equation}
where $S(z,z^\prime)$ encodes the information about the Poisson fluctuation of the sources, while the SN kernel $K(z,z^\prime,f_1,f_2)$ factorizes the frequency dependence due to the energy spectrum of the sources. The covariance due to SN can be written as
\begin{equation}
\begin{split}
{\rm CSN} = & \frac{1}{(\rho_c c^2)^2} \int   \frac{\ud z}{(1+z)H(z)} \int   \frac{\ud z^\prime}{(1+z^\prime)H(z^\prime)} K(z,z^\prime,f_1,f_2)S(z,z^\prime)\delta(\hat{n}-\hat{n}^\prime) \, .
\end{split}
\end{equation}
The angular power spectrum of the SN of the AGWB is constant in $\ell$, because\footnote{The angular dependence is factorized in $$\int d\hat{n}\,d\hat{n}^\prime\, Y_{\ell m}^*(\hat{n}) Y_{\ell m}(\hat{n}^\prime) \delta(\hat{n}-\hat{n}^\prime) = \int d\hat{n}\, |Y_{\ell m}(\hat{n})|^2 = 1\, . $$}
\begin{equation}
\begin{split}
C_\ell^{\rm AGWB,SN}(f_1,f_2)=& \int \ud\hat{n} \ud\hat{n}^\prime Y_{\ell m}^*(\hat{n})Y_{\ell m}(\hat{n}^\prime) {\rm CSN}(f_1,f_2,\hat{n},\hat{n}^\prime)= \\
= &  \frac{1}{(\rho_c c^2)^2} \int   \frac{\ud z}{(1+z)H(z)} \int   \frac{\ud z^\prime}{(1+z^\prime)H(z^\prime)} K(z,z^\prime,f_1,f_2)S(z,z^\prime)\, .
\end{split}
\end{equation}

%% file: Draft_JCAP/Tex_Appendices/appendix_SKA.tex
\label{appendix_SKA}
The parametrization of the futuristic SKAO ‘‘phase two'' is described in~\cite{Maartens:2021dqy}, 
\begin{equation}
\begin{split}
\frac{\ud N}{\ud \Omega_e \ud z} =& 10^{c_1(S_c)}z^{c_2(S_c)}{\rm exp}\left[-c_3(S_c)z\right]\, \rm deg^{-2}\, , \\
c_1 = & 6.55\, , \, \, c_2 = 1.93\, , \, \, c_3 = 6.12\, ,  \\
Q(z) =& 0.28z^4-1.18z^3+1.76z^2+1.367z\, , \\
b^g_e(z) =& 0.08z^5-5.47z^4+16.4z^3-19.6z^2+7.35z+0.22e^{89.2z^4-169.2z^3-102.5z^2+15.5z+0.24}\, .
\end{split}
\end{equation}
There are basically two reasons why we have chosen SKAO2. The first one is that this survey has an high sky coverage, $f_{\rm sky}^{\rm SKAO}\approx 72 \%$. In addition, the SKAO2 window function peaks in a similar redshift range of the window function of the AGWB $\tilde{W}$. This means that the cross-correlation is very high and this increases the SNR. 

%% file: Draft_JCAP/Tex_Appendices/appendix_lagrange_multiplier.tex
\label{appendix_lagrange_multiplier}
The first source of error in our estimator is the covariance associated to fluctuations of the strain amplitude $h$~\cite{Alonso:2020rar,Mentasti:2020yyd,LISACosmologyWorkingGroup:2022kbp}, computed assuming that the noise and the signal are Gaussian, therefore the four-point function of the strain can be written as the sum of the product of two-point functions,
\begin{equation}
\begin{split}
{\rm cov}\left(\hat{I}_{\rm piv,\vartheta}^{\rm KD},\hat{I}_{\rm piv,\vartheta^\prime}^{\rm KD}\right)_h=& \Biggl \langle \left(\frac{1}{2\Delta f} \sum_{f f^\prime,A,B}d_{f,A} E^{f f^\prime}_{\vartheta,AB}d_{f^\prime,B}-E^{f f^\prime}_{\vartheta,AB}\langle d_{f,A}d_{f^\prime,B}\rangle\right)
\\ & \times
\left(\frac{1}{2\Delta f} \sum_{f^{\prime\prime} f^{\prime\prime\prime},C,D}d_{f^{\prime\prime},C} E^{f^{\prime\prime} f^{\prime\prime\prime}}_{\vartheta^\prime,CD}d_{f^{\prime\prime\prime},D}-E^{f^{\prime\prime} f^{\prime\prime\prime}}_{\vartheta^\prime,CD}\langle d_{f^{\prime\prime},C}d_{f^{\prime\prime\prime},D}\rangle\right) 
 \Biggl \rangle=\\
 = & \frac{1}{(2\Delta f)^2}\sum_{f ,f^\prime, f^{\prime\prime},f^{\prime\prime\prime}}E^{f f^\prime}_{\vartheta,AB}E^{f^{\prime\prime}f^{\prime\prime\prime}}_{\vartheta^\prime,CD}\biggl(\langle d_{f,A}d_{f^\prime,B}d_{f^{\prime\prime},C}d_{f^{\prime\prime\prime},D}\rangle \\&\hspace{15em}-\langle d_{f,A}d_{f^\prime,B}\rangle\langle d_{f^{\prime\prime},C}d_{f^{\prime\prime\prime},D}\rangle\biggl)\, .
\end{split}
\end{equation}
Now we note that when we take the four-point function by coupling $A$ with $B$ and $C$ with $D$, we obtain a term that cancels the second one in the sum. If we define the total covariance matrix of the strain as
\begin{equation}
S_f^{AB} = N^{AB}_f+\sum_{\vartheta^\prime} B^{A B,\rm KD}_{f \vartheta^\prime}\left(I_{\rm piv, \vartheta^\prime}^{\rm KD}+\frac{I_{f, \vartheta^{\prime}}^{\rm int}+I_{f, \vartheta^{\prime}}^{\rm SN}}{\mathcal{E}_f^{\rm KD}}\right)\, ,
\end{equation}
we have that the covariance generated by fluctuations in $h$ is 

\begin{eqnarray}
{\rm cov}\left(\hat{I}_{\rm piv,\vartheta}^{\rm KD},\hat{I}_{\rm piv,\vartheta^\prime}^{\rm KD}\right)_h &=& \frac{1}{(2\Delta f)^2}\sum_{f ,f^\prime, f^{\prime\prime},f^{\prime\prime\prime}}E^{f f^\prime}_{\vartheta,AB}E^{f^{\prime\prime}f^{\prime\prime\prime}}_{\vartheta^\prime,CD}\left(S_f^{AC}S_{f^\prime}^{BD}\delta_{f f^{\prime\prime}}\delta_{f^\prime f^{\prime\prime\prime}}+S_f^{AD}S_{f^\prime}^{BC}\delta_{f f^{\prime\prime\prime}}\delta_{f^\prime f^{\prime\prime}}\right)\nonumber\\
& = &  \frac{1}{2\Delta f^2}\sum_{f,f^\prime}{\rm Tr}\left(S_f E_{\vartheta}^{f f^\prime}S_{f^\prime}E^{f^\prime f}_{\vartheta^\prime}\right)\, .
\label{eq:standard_cov_fluctuations_in_h}
\end{eqnarray}

\begin{figure}
\centering
\includegraphics[scale=0.75]{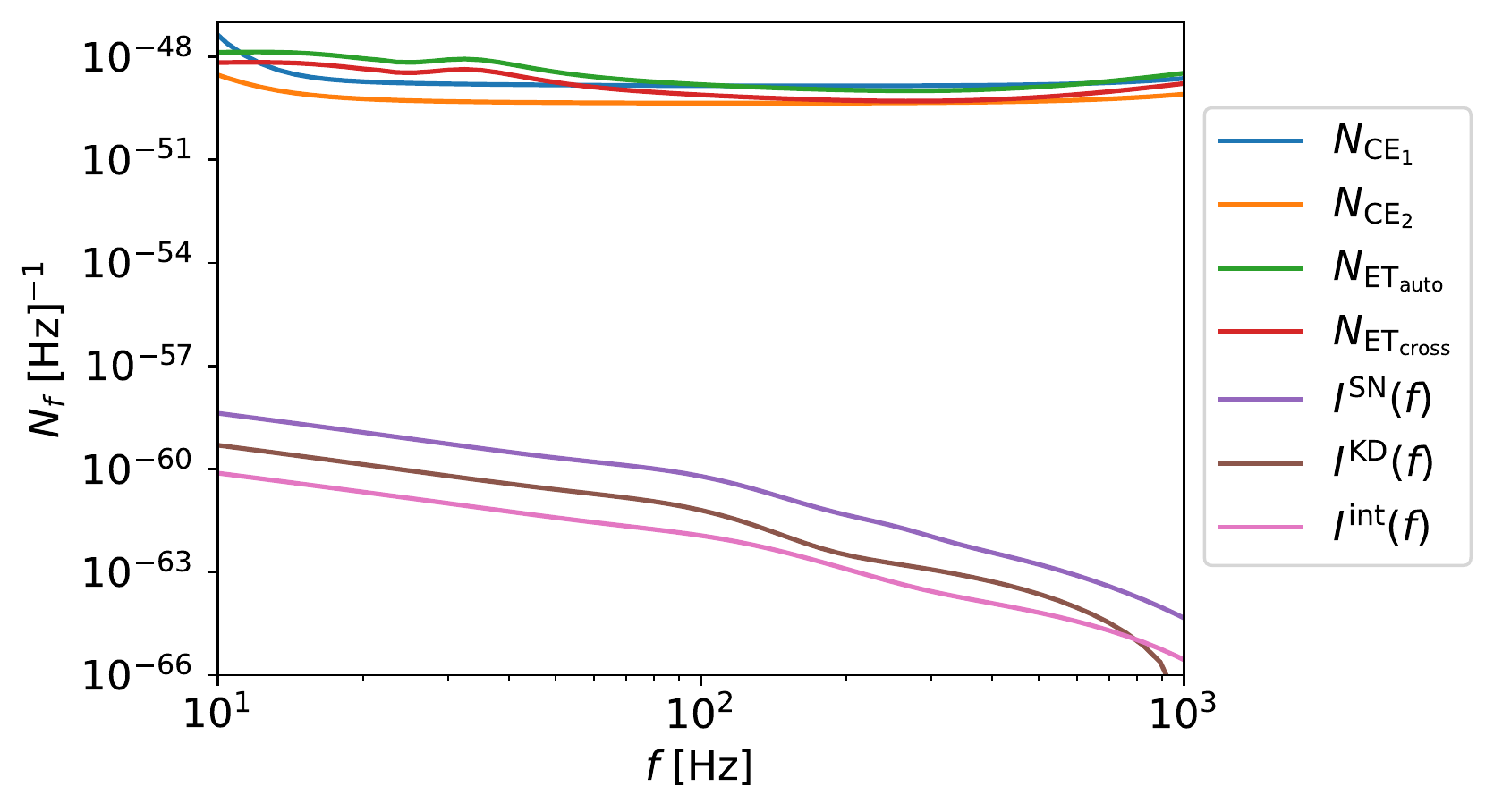}
\caption{{\it PSDs of the noise of the detectors ET, CE1, CE2, and of the possible cross-correlations between the ET channels. We have plotted also the intensity of the SN at different frequencies. Note that SN (and so the signal) are much smaller than instrumental noise, but with this (standard) technique, we are able to clean the signal.}}
\label{all_noises_figure}
\end{figure}

The contributions to the matrix $S_f$ are plotted in Figure \ref{all_noises_figure}. There is however another source of error in our estimator: here we are not trying to estimate just the total map $I_{\vartheta}^{\rm tot}$, but we are trying to perform component separation between different contributions in the map. This point is crucial, because in this step we want to quantify the amount of uncertainty in our measurement provided by the SN. In our analysis we have an estimate for the kinematic dipole $\hat{I}^{\rm KD}_{\rm piv,\vartheta}$ whose mean value $\langle \hat{I}^{\rm KD}_{\rm piv,\vartheta}\rangle$ differs from the true map $I_{\rm piv,\vartheta}^{\rm KD}$ because of SN and intrinsic anisotropies. These fluctuations are quantified by the cosmic variance and they can be computed in the following way, 
\begin{eqnarray}
&&{\rm cov}\left(\left\langle\hat{I}_{\rm piv,\vartheta}^{\rm KD}\right\rangle,\left\langle\hat{I}_{\rm piv,\vartheta^\prime}^{\rm KD}\right\rangle\right)_{cv}=  \left\langle \left(\langle \hat{I}^{\rm KD}_{\rm piv,\vartheta}\rangle_h - I^{\rm KD}_{\rm piv,\vartheta}\right) \left(\langle \hat{I}^{\rm KD}_{\rm piv,\vartheta^\prime}\rangle_h - I^{\rm KD}_{\rm piv,\vartheta^\prime}\right)\right\rangle_{cv} \nonumber\\
&& \quad =  \left\langle \sum_{f,\varphi}\frac{1}{2\Delta f}{\rm Tr}\left(E^{ff}_{\vartheta}B^{\rm KD}_{f \varphi}\right)\frac{I_{f, \varphi}^{\rm int}+I_{f, \varphi}^{\rm SN}}{\mathcal{E}_f^{\rm KD}}\sum_{f^\prime,\varphi^\prime}\frac{1}{2\Delta f}{\rm Tr}\left(E^{f^\prime f^\prime}_{\vartheta^\prime}B^{\rm KD}_{f^\prime \varphi^\prime}\right)\frac{I_{f^\prime, \varphi^\prime}^{\rm int}+I_{f^\prime, \varphi^\prime}^{\rm SN}}{\mathcal{E}_{f^\prime}^{\rm KD}}\right\rangle \nonumber\\
&&\quad =  \frac{1}{(2\Delta f)^2}\sum_{f,f^\prime,\varphi\varphi^\prime}{\rm Tr}\left(E^{ff}_\vartheta B^{\rm KD}_{f \varphi}\right){\rm Tr}\left(E^{f^\prime f^\prime}_{\vartheta^\prime} B^{\rm KD}_{f^\prime \varphi^\prime}\right)\frac{C_{f f^\prime,\varphi\varphi^\prime}^{\rm int}+C_{f f^\prime,\varphi\varphi^\prime}^{\rm SN}}{\mathcal{E}_f^{\rm KD}\mathcal{E}_{f^\prime}^{\rm KD}}\, ,\nonumber\\
\label{eq:cov_fluctuations_in_cv}
\end{eqnarray}
where the covariance matrices in the last row are related to the angular power spectra through
\begin{equation}
C_{f f^\prime,\vartheta \vartheta^\prime}^{j} = \sum_\ell (2\ell+1)C_\ell^j(f,f^\prime)\mathcal{P}_\ell^{\vartheta \vartheta^\prime}\, ,
\end{equation} 
with $\mathcal{P}_\ell$ the Legendre polynomials. Note that here we have assumed that the SN and the intrinsic anisotropies are uncorrelated. So, the total covariance matrix is the sum of the (standard) covariance matrix used in the literature, given by Eq. \eqref{eq:standard_cov_fluctuations_in_h}, plus the new source of error we have introduced here, computed in Eq. \eqref{eq:cov_fluctuations_in_cv},
\begin{equation}
\begin{split}
{\rm cov}_{\vartheta\vartheta^\prime} = \sum_{f,f^\prime}\Biggl[&\frac{1}{2\Delta f^2}{\rm Tr}\left(E^{f f^\prime}_{\vartheta}S_f E_{\vartheta^\prime}^{f^\prime f}S_{f^\prime}\right)+\\
&+\frac{1}{(2\Delta f)^2}\sum_{\varphi,\varphi^\prime}{\rm Tr}\left(E^{f f}_{\vartheta}B^{\rm KD}_{f,\varphi}\right){\rm Tr}\left(E^{f^\prime f^\prime}_{\vartheta^\prime}B^{\rm KD}_{f^\prime,\varphi^\prime}\right)\left(\frac{C_{f f^\prime,\varphi\varphi^\prime}^{\rm int}+C_{f f^\prime,\varphi\varphi^\prime}^{\rm SN}}{\mathcal{E}_f^{\rm KD}\mathcal{E}^{\rm KD}_{f^\prime}} \right)\Biggl]\, ,
\end{split}
\end{equation}
where we have summed in quadrature the two errors because they are uncorrelated. To simplify the notation we define
\begin{equation}
    C_{ff^\prime,\varphi\varphi^\prime}^{\rm tot}\equiv C_{f f^\prime,\varphi\varphi^\prime}^{\rm int}+C_{f f^\prime,\varphi\varphi^\prime}^{\rm SN}\, . 
\end{equation}
The SN is proportional to $\delta_{\varphi\varphi^\prime}$ (the angular power spectrum is constant in $\ell$), but the intrinsic anisotropies depend on the cosine between the two directions of observation, therefore the covariance matrix becomes
\begin{equation}
\begin{split}
{\rm cov}_{\vartheta\vartheta^\prime} = \frac{1}{2\Delta f^2}\sum_{f,f^\prime}\Biggl[&{\rm Tr}\left(E^{f f^\prime}_{\vartheta}S_f E_{\vartheta^\prime}^{f^\prime f}S_{f^\prime}\right)+\frac{1}{2}\sum_{\varphi,\varphi^\prime}{\rm Tr}\left(E^{f f}_{\vartheta}B^{\rm KD}_{f,\varphi}\right){\rm Tr}\left(E^{f^\prime f^\prime}_{\vartheta^\prime}B^{\rm KD}_{f^\prime,\varphi^\prime}\right)\frac{C_{ff^\prime,\varphi\varphi^\prime}^{\rm tot}}{\mathcal{E}_f^{\rm KD}\mathcal{E}^{\rm KD}_{f^\prime}}\Biggl]\, .
\label{eq_cov_to_minimze}
\end{split}
\end{equation}
To minimize the covariance given by Eq. \eqref{eq_cov_to_minimze}, we use a Lagrange multiplier,
\begin{equation}
\mathcal{L} = {\rm cov}_{\vartheta\vartheta^\prime}-\lambda_\vartheta\left[\frac{1}{2\Delta f}\sum_{f} {\rm Tr}\left(E^{ff}_{\vartheta}B^{\rm KD}_{f\vartheta^\prime}\right)-\delta_{\vartheta \vartheta^\prime}\right]\,,
\end{equation}
and by imposing that the derivative of $\mathcal{L}$ w.r.t. $E^{f f^\prime}_{\vartheta^\prime,AB}$ is zero we have 
\begin{equation}
\begin{split}
\sum_{C,D}S_f^{AC}E^{f^\prime f}_{\vartheta,CD}S_{f^\prime}^{DB}+&\frac{\delta_{f f^\prime}}{2}\sum_{f^{\prime\prime},\varphi^\prime}B^{\rm KD,AB}_{f,\varphi}{\rm Tr}\left(E^{f^{\prime\prime} f^{\prime\prime}}_{\vartheta}B^{\rm KD}_{f^{\prime\prime},\varphi^\prime}\right)\frac{C_{ff^\prime,\varphi\varphi^\prime}^{\rm tot}}{\mathcal{E}_f^{\rm KD}\mathcal{E}^{\rm KD}_{f^{\prime\prime}}}-\Delta f\lambda_\vartheta \delta_{f f^\prime} B^{AB,\rm KD}_{f,\vartheta} = 0 \, .
\end{split}
\end{equation}
Without writing explicitly the detector indices, we have
\begin{equation}
S_f E_{\vartheta}^{f f^\prime} S_{f^\prime} + \frac{\delta_{f f^\prime}}{2}\sum_{f^{\prime\prime},\varphi^\prime}B_{f,\varphi}^{\rm KD} {\rm Tr}\left(E^{f^{\prime\prime}f^{\prime\prime}}_\vartheta B^{\rm KD}_{f^{\prime\prime},\varphi^\prime}\right)\frac{C_{ff^\prime,\varphi\varphi^\prime}^{\rm tot}}{\mathcal{E}_f^{\rm KD}\mathcal{E}^{\rm KD}_{f^{\prime\prime}}}=\Delta f\lambda_\vartheta\delta_{f f^\prime} B^{\rm KD}_{f,\vartheta} \, ,
\end{equation}
so, when $f\neq f^\prime$, we find
\begin{equation}
S_f E_{\vartheta}^{f f^\prime} = 0\, ,
\end{equation}
which means that $E_\vartheta^{f f^\prime}=0$ or that $E_\vartheta^{f f^\prime}$ belongs to the kernel of $S_f$. 

For the moment we are interested in $f=f^\prime$, therefore we find
\begin{equation}
E_{\vartheta}^{f f}+ \frac{1}{2}\sum_{f^{\prime\prime},\varphi,\varphi^\prime}S_f^{-1}B_{f,\varphi}^{\rm KD}S_f^{-1} {\rm Tr}\left(E^{f^{\prime\prime}f^{\prime\prime}}_\vartheta B^{\rm KD}_{f^{\prime\prime},\varphi^\prime}\right)\frac{C_{ff^\prime,\varphi\varphi^\prime}^{\rm tot}}{\mathcal{E}_f^{\rm KD}\mathcal{E}^{\rm KD}_{f^{\prime\prime}}}=\Delta f\lambda_\vartheta S_f^{-1} B^{\rm KD}_{f,\vartheta}S_f^{-1} \, .
\end{equation}
Motivated by Figure \ref{all_noises_figure}, we solve this equation in perturbation theory, expanding at first order in $C^{\rm SN}_{f f^{\prime\prime}}$.
The zero order solution is 
\begin{equation}
E^{ff\, (0)}_\vartheta = \Delta f \lambda_\vartheta S_f^{-1}B_{f,\vartheta}^{\rm KD}S_f^{-1}\, .
\end{equation}
Now we substitute this solution in the trace, finding that the first order solution is
\begin{equation}
E^{ff\, (1)}_\vartheta =-\frac{1}{2} \Delta f \lambda_\vartheta \sum_{f^{\prime\prime},\varphi,\varphi^\prime}S_f^{-1}B_{f,\varphi}^{\rm KD}S_f^{-1}{\rm Tr}\left(S_{f^{\prime\prime}}^{-1}B_{f^{\prime\prime}\vartheta}^{\rm KD}S_{f^{\prime\prime}}^{-1}B_{f^{\prime\prime},\varphi^\prime}^{\rm KD}\right)\frac{C_{ff^\prime,\varphi\varphi^\prime}^{\rm tot}}{\mathcal{E}_f^{\rm KD}\mathcal{E}_{f^{\prime\prime}}^{\rm KD}}\, .
\end{equation}
The full solution is simply given by the sum of the two contributions,
\begin{equation}
E^{ff}_\vartheta = \lambda_\vartheta\,\Delta f \left[ S_f^{-1}B_{f,\vartheta}^{\rm KD}S_f^{-1}-\frac{1}{2}\sum_{f^{\prime\prime},\varphi,\varphi^\prime}S_f^{-1}B_{f,\varphi}^{\rm KD}S_f^{-1}{\rm Tr}\left(S_{f^{\prime\prime}}^{-1}B_{f^{\prime\prime}\vartheta}^{\rm KD}S_{f^{\prime\prime}}^{-1}B_{f^{\prime\prime},\varphi^\prime}^{\rm KD}\right)\frac{C_{ff^\prime,\varphi\varphi^\prime}^{\rm tot}}{\mathcal{E}_f^{\rm KD}\mathcal{E}_{f^{\prime\prime}}^{\rm KD}}\right]\, .
\end{equation}
To find the parameter $\lambda_\vartheta$ we use the condition given by the Lagrange multiplier with $\vartheta = \vartheta^\prime$,
\begin{equation}
\begin{split}
\lambda_\vartheta = & \frac{2}{\sum_f{\rm Tr}\left(S_f^{-1}B_{f,\vartheta}^{\rm KD}S_f^{-1}B_{f,\vartheta}^{\rm KD}\right)}+\sum_{f^{\prime\prime},\varphi,\varphi^\prime}\frac{\sum_f{\rm Tr}\left(S_f^{-1}B_{f,\varphi}^{\rm KD}S_f^{-1}B_{f,\vartheta}^{\rm KD}\right){\rm Tr}\left(S_{f^{\prime\prime}}^{-1}B_{f^{\prime\prime}\vartheta}^{\rm KD}S_{f^{\prime\prime}}^{-1}B_{f^{\prime\prime},\varphi^\prime}^{\rm KD}\right)}{\left[\sum_f{\rm Tr}\left(S_f^{-1}B_{f,\vartheta}^{\rm KD}S_f^{-1}B_{f,\vartheta}^{\rm KD}\right)\right]^2}\frac{C_{ff^\prime,\varphi\varphi^\prime}^{\rm tot}}{\mathcal{E}_f^{\rm KD}\mathcal{E}_{f^{\prime\prime}}^{\rm KD}}\, ,
\end{split}
\end{equation}
and the final expression for the weights to give to the signals measured at interferometers are
\begin{small}
\begin{equation}
\begin{split}
\frac{E^{ff}_\vartheta}{\Delta f} 
= & \frac{2S_f^{-1}B_{f,\vartheta}^{\rm KD}S_f^{-1}}{\sum_{f^\prime}{\rm Tr}\left(S_{f^\prime}^{-1}B_{f^\prime,\vartheta}^{\rm KD}S_{f^\prime}^{-1}B_{f^\prime,\vartheta}^{\rm KD}\right)}\\
&+S_f^{-1}B_{f,\vartheta}^{\rm KD}S_f^{-1}\sum_{f^\prime,f^{\prime\prime},\varphi,\varphi^\prime}\frac{{\rm Tr}\left(S_{f^\prime}^{-1}B_{f^\prime,\varphi}^{\rm KD}S_{f^\prime}^{-1}B_{f^\prime,\vartheta}^{\rm KD}\right){\rm Tr}\left(S_{f^{\prime\prime}}^{-1}B_{f^{\prime\prime}\vartheta}^{\rm KD}S_{f^{\prime\prime}}^{-1}B_{f^{\prime\prime},\varphi^\prime}^{\rm KD}\right)}{\left[\sum_{f^\prime}{\rm Tr}\left(S_{f^\prime}^{-1}B_{f^\prime,\vartheta}^{\rm KD}S_{f^\prime}^{-1}B_{f^\prime,\vartheta}^{\rm KD}\right)\right]^2}\frac{C_{ff^\prime,\varphi\varphi^\prime}^{\rm tot}}{\mathcal{E}_{f^\prime}^{\rm KD}\mathcal{E}_{f^{\prime\prime}}^{\rm KD}}\\
&+(-1)\sum_{f^{\prime\prime},\varphi,\varphi^\prime}S_f^{-1}B_{f,\varphi}^{\rm KD}S_f^{-1}\frac{{\rm Tr}\left(S_{f^{\prime\prime}}^{-1}B_{f^{\prime\prime}\vartheta}^{\rm KD}S_{f^{\prime\prime}}^{-1}B_{f^{\prime\prime},\varphi^\prime}^{\rm KD}\right)}{\sum_{f^\prime}{\rm Tr}\left(S_{f^\prime}^{-1}B_{f^\prime,\vartheta}^{\rm KD}S_{f^\prime}^{-1}B_{f^\prime,\vartheta}^{\rm KD}\right)}\frac{C_{ff^\prime,\varphi\varphi^\prime}^{\rm tot}}{\mathcal{E}_f^{\rm KD}\mathcal{E}_{f^{\prime\prime}}^{\rm KD}}\, .
\end{split}
\end{equation}
\end{small}
The covariance matrix up to first order in $C^{\rm SN}_{f f^{\prime\prime}}$ is therefore
\begin{small}
\begin{equation}
\begin{split}
{\rm cov}_{\vartheta\vartheta^\prime}
=& \frac{2\delta_{\vartheta\vartheta^\prime}}{\sum_{f^\prime}{\rm Tr}\left(S_{f^\prime}^{-1}B_{f^\prime,\vartheta}^{\rm KD}S_{f^\prime}^{-1}B_{f^\prime,\vartheta}^{\rm KD}\right)}+\\
&+\delta_{\vartheta\vartheta^\prime}\sum_{f^\prime,f^{\prime\prime},\varphi,\varphi^\prime}\frac{{\rm Tr}\left(S_{f^\prime}^{-1}B_{f^\prime,\varphi}^{\rm KD}S_{f^\prime}^{-1}B_{f^\prime,\vartheta^\prime}^{\rm KD}\right){\rm Tr}\left(S_{f^{\prime\prime}}^{-1}B_{f^{\prime\prime}\vartheta^\prime}^{\rm KD}S_{f^{\prime\prime}}^{-1}B_{f^{\prime\prime},\varphi^\prime}^{\rm KD}\right)}{\left[\sum_{f^\prime}{\rm Tr}\left(S_{f^\prime}^{-1}B_{f^\prime,\vartheta^\prime}^{\rm KD}S_{f^\prime}^{-1}B_{f^\prime,\vartheta^\prime}^{\rm KD}\right)\right]^2}\frac{C_{ff^\prime,\varphi\varphi^\prime}^{\rm tot}}{\mathcal{E}_{f^\prime}^{\rm KD}\mathcal{E}_{f^{\prime\prime}}^{\rm KD}}\, .
\end{split}
\end{equation}
\end{small}
The minimum covariance we will have in estimating the kinetic dipole is then
\begin{small}
\begin{equation}
{\rm cov}_{\vartheta\vartheta^\prime}=\frac{2\delta_{\vartheta\vartheta^\prime}}{\sum_{f}{\rm Tr}\left(S_{f}^{-1}B_{f,\vartheta}^{\rm KD}S_{f}^{-1}B_{f,\vartheta}^{\rm KD}\right)}+\delta_{\vartheta\vartheta^\prime}\sum_{f,f^\prime,\varphi,\varphi^\prime}\frac{{\rm Tr}\left(S_{f}^{-1}B_{f,\varphi}^{\rm KD}S_{f}^{-1}B_{f,\vartheta^\prime}^{\rm KD}\right){\rm Tr}\left(S_{f^{\prime}}^{-1}B_{f^{\prime}\vartheta^\prime}^{\rm KD}S_{f^{\prime}}^{-1}B_{f^{\prime},\varphi^\prime}^{\rm KD}\right)}{\left[\sum_{f}{\rm Tr}\left(S_{f}^{-1}B_{f,\vartheta}^{\rm KD}S_{f}^{-1}B_{f,\vartheta}^{\rm KD}\right)\right]^2}\frac{C_{ff^\prime,\varphi\varphi^\prime}^{\rm tot}}{\mathcal{E}_{f}^{\rm KD}\mathcal{E}_{f^{\prime}}^{\rm KD}}\, .
\end{equation}
\end{small}


%% file: draft_jcap_resubmission.bbl
\begin{thebibliography}{}

\bibitem{Kogut:1993ag}
A.~Kogut, C.~Lineweaver, G.~F.~Smoot, C.~L.~Bennett, A.~Banday, N.~W.~Boggess, E.~S.~Cheng, G.~De Amici, D.~J.~Fixsen and G.~Hinshaw, \textit{et al.}
Astrophys. J. \textbf{419} (1993), 1
doi:10.1086/173453
[arXiv:astro-ph/9312056 [astro-ph]].

\bibitem{Lineweaver:1996xa}
C.~H.~Lineweaver, L.~Tenorio, G.~F.~Smoot, P.~Keegstra, A.~J.~Banday and P.~Lubin,
Astrophys. J. \textbf{470} (1996), 38-42
doi:10.1086/177846
[arXiv:astro-ph/9601151 [astro-ph]].

\bibitem{WMAP:2008ydk}
G.~Hinshaw \textit{et al.} [WMAP],
Astrophys. J. Suppl. \textbf{180} (2009), 225-245
doi:10.1088/0067-0049/180/2/225
[arXiv:0803.0732 [astro-ph]].

\bibitem{Planck:2013kqc}
N.~Aghanim \textit{et al.} [Planck],
Astron. Astrophys. \textbf{571} (2014), A27
doi:10.1051/0004-6361/201321556
[arXiv:1303.5087 [astro-ph.CO]].

\bibitem{Planck:2020qil}
Y.~Akrami \textit{et al.} [Planck],
Astron. Astrophys. \textbf{644} (2020), A100
doi:10.1051/0004-6361/202038053
[arXiv:2003.12646 [astro-ph.CO]].

\bibitem{Schwarz:2015cma}
D.~J.~Schwarz, C.~J.~Copi, D.~Huterer and G.~D.~Starkman,
Class. Quant. Grav. \textbf{33} (2016) no.18, 184001
doi:10.1088/0264-9381/33/18/184001
[arXiv:1510.07929 [astro-ph.CO]].

\bibitem{Planck:2019evm}
Y.~Akrami \textit{et al.} [Planck],
Astron. Astrophys. \textbf{641} (2020), A7
doi:10.1051/0004-6361/201935201
[arXiv:1906.02552 [astro-ph.CO]].

\bibitem{Gibelyou:2012ri}
C.~Gibelyou and D.~Huterer,
Mon. Not. Roy. Astron. Soc. \textbf{427} (2012), 1994-2021
doi:10.1111/j.1365-2966.2012.22032.x
[arXiv:1205.6476 [astro-ph.CO]].

\bibitem{Rubart:2013tx}
M.~Rubart and D.~J.~Schwarz,
Astron. Astrophys. \textbf{555} (2013), A117
doi:10.1051/0004-6361/201321215
[arXiv:1301.5559 [astro-ph.CO]].

\bibitem{Tiwari:2015tba}
P.~Tiwari and A.~Nusser,
JCAP \textbf{03} (2016), 062
doi:10.1088/1475-7516/2016/03/062
[arXiv:1509.02532 [astro-ph.CO]].

\bibitem{Chen:2015wga}
S.~Chen and D.~J.~Schwarz,
Astron. Astrophys. \textbf{591} (2016), A135
doi:10.1051/0004-6361/201526956
[arXiv:1507.02160 [astro-ph.CO]].

\bibitem{Bengaly:2017slg}
C.~A.~P.~Bengaly, R.~Maartens and M.~G.~Santos,
JCAP \textbf{04} (2018), 031
doi:10.1088/1475-7516/2018/04/031
[arXiv:1710.08804 [astro-ph.CO]].

\bibitem{Secrest:2020has}
N.~J.~Secrest, S.~von Hausegger, M.~Rameez, R.~Mohayaee, S.~Sarkar and J.~Colin,
Astrophys. J. Lett. \textbf{908} (2021) no.2, L51
doi:10.3847/2041-8213/abdd40
[arXiv:2009.14826 [astro-ph.CO]].

\bibitem{Crawford:2008nh}
F.~Crawford,
Astrophys. J. \textbf{692} (2009), 887-893
doi:10.1088/0004-637X/692/1/887
[arXiv:0810.4520 [astro-ph]].

\bibitem{Maartens:2017qoa}
R.~Maartens, C.~Clarkson and S.~Chen,
JCAP \textbf{01} (2018), 013
doi:10.1088/1475-7516/2018/01/013
[arXiv:1709.04165 [astro-ph.CO]].

\bibitem{Nadolny:2021hti}
T.~Nadolny, R.~Durrer, M.~Kunz and H.~Padmanabhan,
JCAP \textbf{11} (2021), 009
doi:10.1088/1475-7516/2021/11/009
[arXiv:2106.05284 [astro-ph.CO]].

\bibitem{Galloni:2022rgg}
G.~Galloni, N.~Bartolo, S.~Matarrese, M.~Migliaccio, A.~Ricciardone and N.~Vittorio,
[arXiv:2202.12858 [astro-ph.CO]].

\bibitem{Ferrari:1998jf}
V.~Ferrari, S.~Matarrese and R.~Schneider,
Mon. Not. Roy. Astron. Soc. \textbf{303} (1999), 258
doi:10.1046/j.1365-8711.1999.02207.x
[arXiv:astro-ph/9806357 [astro-ph]].

\bibitem{Ferrari:1998ut}
V.~Ferrari, S.~Matarrese and R.~Schneider,
Mon. Not. Roy. Astron. Soc. \textbf{303} (1999), 247
doi:10.1046/j.1365-8711.1999.02194.x
[arXiv:astro-ph/9804259 [astro-ph]].

\bibitem{Phinney:2001di}
E.~S.~Phinney,
[arXiv:astro-ph/0108028 [astro-ph]].

\bibitem{Regimbau:2011rp}
T.~Regimbau,
Res. Astron. Astrophys. \textbf{11} (2011), 369-390
doi:10.1088/1674-4527/11/4/001
[arXiv:1101.2762 [astro-ph.CO]].

\bibitem{LIGOScientific:2021psn}
R.~Abbott \textit{et al.} [LIGO Scientific, VIRGO and KAGRA],
[arXiv:2111.03634 [astro-ph.HE]].

\bibitem{Maggiore:2019uih}
M.~Maggiore, C.~Van Den Broeck, N.~Bartolo, E.~Belgacem, D.~Bertacca, M.~A.~Bizouard, M.~Branchesi, S.~Clesse, S.~Foffa and J.~Garc\'\i{}a-Bellido, \textit{et al.}
JCAP \textbf{03} (2020), 050
doi:10.1088/1475-7516/2020/03/050
[arXiv:1912.02622 [astro-ph.CO]].

\bibitem{Baker:2019ync}
J.~Baker, T.~Baker, C.~Carbone, G.~Congedo, C.~Contaldi, I.~Dvorkin, J.~Gair, Z.~Haiman, D.~F.~Mota and A.~Renzini, \textit{et al.}
Exper. Astron. \textbf{51} (2021) no.3, 1441-1470
doi:10.1007/s10686-021-09712-0
[arXiv:1908.11410 [astro-ph.HE]].

\bibitem{Seoane:2021kkk}
P.~A.~Seoane, M.~A.~Sedda, S.~Babak, C.~P.~L.~Berry, E.~Berti, G.~Bertone, D.~Blas, T.~Bogdanovi\'c, M.~Bonetti and K.~Breivik, \textit{et al.}
Gen. Rel. Grav. \textbf{54} (2022) no.1, 3
doi:10.1007/s10714-021-02889-x
[arXiv:2107.09665 [astro-ph.IM]].

\bibitem{Evans:2021gyd}
M.~Evans, R.~X.~Adhikari, C.~Afle, S.~W.~Ballmer, S.~Biscoveanu, S.~Borhanian, D.~A.~Brown, Y.~Chen, R.~Eisenstein and A.~Gruson, \textit{et al.}
[arXiv:2109.09882 [astro-ph.IM]].


\bibitem{KAGRA:2021kbb}
R.~Abbott \textit{et al.} [KAGRA, Virgo and LIGO Scientific],
Phys. Rev. D \textbf{104} (2021) no.2, 022004
doi:10.1103/PhysRevD.104.022004
[arXiv:2101.12130 [gr-qc]].

\bibitem{Cusin:2017fwz}
G.~Cusin, C.~Pitrou and J.~P.~Uzan,
Phys. Rev. D \textbf{96} (2017) no.10, 103019
doi:10.1103/PhysRevD.96.103019
[arXiv:1704.06184 [astro-ph.CO]].

\bibitem{Cusin:2017mjm}
G.~Cusin, C.~Pitrou and J.~P.~Uzan,
Phys. Rev. D \textbf{97} (2018) no.12, 123527
doi:10.1103/PhysRevD.97.123527
[arXiv:1711.11345 [astro-ph.CO]].

\bibitem{Cusin:2018rsq}
G.~Cusin, I.~Dvorkin, C.~Pitrou and J.~P.~Uzan,
Phys. Rev. Lett. \textbf{120} (2018), 231101
doi:10.1103/PhysRevLett.120.231101
[arXiv:1803.03236 [astro-ph.CO]].

\bibitem{Jenkins:2018kxc}
A.~C.~Jenkins, R.~O'Shaughnessy, M.~Sakellariadou and D.~Wysocki,
Phys. Rev. Lett. \textbf{122} (2019) no.11, 111101
doi:10.1103/PhysRevLett.122.111101
[arXiv:1810.13435 [astro-ph.CO]].

\bibitem{Jenkins:2018uac}
A.~C.~Jenkins, M.~Sakellariadou, T.~Regimbau and E.~Slezak,
Phys. Rev. D \textbf{98} (2018) no.6, 063501
doi:10.1103/PhysRevD.98.063501
[arXiv:1806.01718 [astro-ph.CO]].

\bibitem{Pitrou:2019rjz}
C.~Pitrou, G.~Cusin and J.~P.~Uzan,
Phys. Rev. D \textbf{101} (2020) no.8, 081301
doi:10.1103/PhysRevD.101.081301
[arXiv:1910.04645 [astro-ph.CO]].

\bibitem{Bertacca:2019fnt}
D.~Bertacca, A.~Ricciardone, N.~Bellomo, A.~C.~Jenkins, S.~Matarrese, A.~Raccanelli, T.~Regimbau and M.~Sakellariadou,
Phys. Rev. D \textbf{101} (2020) no.10, 103513
doi:10.1103/PhysRevD.101.103513
[arXiv:1909.11627 [astro-ph.CO]].

\bibitem{Jenkins:2019nks}
A.~C.~Jenkins, J.~D.~Romano and M.~Sakellariadou,
Phys. Rev. D \textbf{100} (2019) no.8, 083501
doi:10.1103/PhysRevD.100.083501
[arXiv:1907.06642 [astro-ph.CO]].

\bibitem{Jenkins:2019uzp}
A.~C.~Jenkins and M.~Sakellariadou,
Phys. Rev. D \textbf{100} (2019) no.6, 063508
doi:10.1103/PhysRevD.100.063508
[arXiv:1902.07719 [astro-ph.CO]].

\bibitem{Allen:1996gp}
B.~Allen and A.~C.~Ottewill,
Phys. Rev. D \textbf{56} (1997), 545-563
doi:10.1103/PhysRevD.56.545
[arXiv:gr-qc/9607068 [gr-qc]].

\bibitem{Jenkins:2018nty}
A.~C.~Jenkins and M.~Sakellariadou,
Phys. Rev. D \textbf{98} (2018) no.6, 063509
doi:10.1103/PhysRevD.98.063509
[arXiv:1802.06046 [astro-ph.CO]].

\bibitem{Cusin:2022cbb}
G.~Cusin and G.~Tasinato,
[arXiv:2201.10464 [astro-ph.CO]].

\bibitem{Kaiser1987}
N.~Kaiser,
Mon. Not. Roy. Astron. Soc. \textbf{227} (1987) no.1, 1-21
doi:10.1093/mnras/227.1.1

\bibitem{Maggiore:2007ulw}
M.~Maggiore,
``Gravitational Waves. Vol. 1: Theory and Experiments'',
Oxford Master Series in Physics (Oxford University Press 2007).

\bibitem{Flauger:2020qyi}
R.~Flauger, N.~Karnesis, G.~Nardini, M.~Pieroni, A.~Ricciardone and J.~Torrado,
JCAP \textbf{01} (2021), 059
doi:10.1088/1475-7516/2021/01/059
[arXiv:2009.11845 [astro-ph.CO]].

\bibitem{Orlando:2020oko}
G.~Orlando, M.~Pieroni and A.~Ricciardone,
JCAP \textbf{03} (2021), 069
doi:10.1088/1475-7516/2021/03/069
[arXiv:2011.07059 [astro-ph.CO]].

\bibitem{Amalberti:2021kzh}
L.~Amalberti, N.~Bartolo and A.~Ricciardone,
Phys. Rev. D \textbf{105} (2022) no.6, 064033
doi:10.1103/PhysRevD.105.064033
[arXiv:2105.13197 [astro-ph.CO]].

\bibitem{Alonso:2020rar}
D.~Alonso, C.~R.~Contaldi, G.~Cusin, P.~G.~Ferreira and A.~I.~Renzini,
Phys. Rev. D \textbf{101} (2020) no.12, 124048
doi:10.1103/PhysRevD.101.124048
[arXiv:2005.03001 [astro-ph.CO]].

\bibitem{Mentasti:2020yyd}
G.~Mentasti and M.~Peloso,
JCAP \textbf{03} (2021), 080
doi:10.1088/1475-7516/2021/03/080
[arXiv:2010.00486 [astro-ph.CO]].

\bibitem{LISACosmologyWorkingGroup:2022kbp}
N.~Bartolo \textit{et al.} [LISA Cosmology Working Group],
[arXiv:2201.08782 [astro-ph.CO]].

\bibitem{Bellomo:2021mer}
N.~Bellomo, D.~Bertacca, A.~C.~Jenkins, S.~Matarrese, A.~Raccanelli, T.~Regimbau, A.~Ricciardone and M.~Sakellariadou,
[arXiv:2110.15059 [gr-qc]].

\bibitem{Canas-Herrera:2019npr}
G.~Ca\~nas-Herrera, O.~Contigiani and V.~Vardanyan,
Phys. Rev. D \textbf{102} (2020) no.4, 043513
doi:10.1103/PhysRevD.102.043513
[arXiv:1910.08353 [astro-ph.CO]].


\bibitem{Mukherjee:2019oma}
S.~Mukherjee and J.~Silk,
Mon. Not. Roy. Astron. Soc. \textbf{491} (2020) no.4, 4690-4701
doi:10.1093/mnras/stz3226
[arXiv:1912.07657 [gr-qc]].

\bibitem{Alonso:2020mva}
D.~Alonso, G.~Cusin, P.~G.~Ferreira and C.~Pitrou,
Phys. Rev. D \textbf{102} (2020) no.2, 023002
doi:10.1103/PhysRevD.102.023002
[arXiv:2002.02888 [astro-ph.CO]].

\bibitem{Yang:2020usq}
K.~Z.~Yang, V.~Mandic, C.~Scarlata and S.~Banagiri,
Mon. Not. Roy. Astron. Soc. \textbf{500} (2020) no.2, 1666-1672
doi:10.1093/mnras/staa3159
[arXiv:2007.10456 [astro-ph.CO]].

\bibitem{Ricciardone:2021kel}
A.~Ricciardone, L.~V.~Dall'Armi, N.~Bartolo, D.~Bertacca, M.~Liguori and S.~Matarrese,
[arXiv:2106.02591 [astro-ph.CO]].

\bibitem{Capurri:2021prz}
G.~Capurri, A.~Lapi and C.~Baccigalupi,
[arXiv:2111.04757 [astro-ph.CO]].

\bibitem{Ajith:2007kx}
P.~Ajith, S.~Babak, Y.~Chen, M.~Hewitson, B.~Krishnan, A.~M.~Sintes, J.~T.~Whelan, B.~Bruegmann, P.~Diener and N.~Dorband, \textit{et al.}
Phys. Rev. D \textbf{77} (2008), 104017
[erratum: Phys. Rev. D \textbf{79} (2009), 129901]
doi:10.1103/PhysRevD.77.104017
[arXiv:0710.2335 [gr-qc]].

\bibitem{Ajith:2009bn}
P.~Ajith, M.~Hannam, S.~Husa, Y.~Chen, B.~Bruegmann, N.~Dorband, D.~Muller, F.~Ohme, D.~Pollney and C.~Reisswig, \textit{et al.}
Phys. Rev. Lett. \textbf{106} (2011), 241101
doi:10.1103/PhysRevLett.106.241101
[arXiv:0909.2867 [gr-qc]].

\bibitem{Ajith:2012mn}
P.~Ajith, N.~Fotopoulos, S.~Privitera, A.~Neunzert and A.~J.~Weinstein,
Phys. Rev. D \textbf{89} (2014) no.8, 084041
doi:10.1103/PhysRevD.89.084041
[arXiv:1210.6666 [gr-qc]].

\bibitem{Planck:2018yye}
Y.~Akrami \textit{et al.} [Planck],
Astron. Astrophys. \textbf{641} (2020), A4
doi:10.1051/0004-6361/201833881
[arXiv:1807.06208 [astro-ph.CO]].

\bibitem{Tegmark:1999ke}
M.~Tegmark, D.~J.~Eisenstein, W.~Hu and A.~de Oliveira-Costa,
Astrophys. J. \textbf{530} (2000), 133-165
doi:10.1086/308348
[arXiv:astro-ph/9905257 [astro-ph]].

\bibitem{Tegmark:2003ve}
M.~Tegmark, A.~de Oliveira-Costa and A.~Hamilton,
Phys. Rev. D \textbf{68} (2003), 123523
doi:10.1103/PhysRevD.68.123523
[arXiv:astro-ph/0302496 [astro-ph]].

\bibitem{ETsens}
\url{http://www.et-gw.eu/index.php/etsensitivities}

\bibitem{ce}
\url{https://cosmicexplorer.org/sensitivity.html}

\bibitem{Finn:1995ah}
L.~S.~Finn,
Phys. Rev. D \textbf{53} (1996), 2878-2894
doi:10.1103/PhysRevD.53.2878
[arXiv:gr-qc/9601048 [gr-qc]].

\bibitem{Chen:2018rzo}
Z.~C.~Chen, F.~Huang and Q.~G.~Huang,
Astrophys. J. \textbf{871} (2019) no.1, 97
doi:10.3847/1538-4357/aaf581
[arXiv:1809.10360 [gr-qc]].

\bibitem{Perigois:2021ovr}
C.~P\'erigois, F.~Santoliquido, Y.~Bouffanais, U.~N.~Di Carlo, N.~Giacobbo, S.~Rastello, M.~Mapelli and T.~Regimbau,
[arXiv:2112.01119 [astro-ph.CO]].

\bibitem{Sasaki:2018dmp}
M.~Sasaki, T.~Suyama, T.~Tanaka and S.~Yokoyama,
Class. Quant. Grav. \textbf{35} (2018) no.6, 063001
doi:10.1088/1361-6382/aaa7b4
[arXiv:1801.05235 [astro-ph.CO]].

\bibitem{Bird:2016dcv}
S.~Bird, I.~Cholis, J.~B.~Mu\~noz, Y.~Ali-Ha\"\i{}moud, M.~Kamionkowski, E.~D.~Kovetz, A.~Raccanelli and A.~G.~Riess,
Phys. Rev. Lett. \textbf{116} (2016) no.20, 201301
doi:10.1103/PhysRevLett.116.201301
[arXiv:1603.00464 [astro-ph.CO]].

\bibitem{Zhu:2011bd}
X.~J.~Zhu, E.~Howell, T.~Regimbau, D.~Blair and Z.~H.~Zhu,
Astrophys. J. \textbf{739} (2011), 86
doi:10.1088/0004-637X/739/2/86
[arXiv:1104.3565 [gr-qc]].

\bibitem{UniverseMachine}
P.~Behroozi, R.~H.~Wechsler, A.~P.~Hearin, and C.~Conroy, MNRAS \textbf{488} no.3, (05,2019) 3143-3194, [arXiv:1806.07893].

\bibitem{Mapelli:2017hqk}
M.~Mapelli, N.~Giacobbo, E.~Ripamonti and M.~Spera,
Mon. Not. Roy. Astron. Soc. \textbf{472} (2017) no.2, 2422-2435
doi:10.1093/mnras/stx2123
[arXiv:1708.05722 [astro-ph.GA]].

\bibitem{Tinker:2008ff}
J.~L.~Tinker, A.~V.~Kravtsov, A.~Klypin, K.~Abazajian, M.~S.~Warren, G.~Yepes, S.~Gottlober and D.~E.~Holz,
Astrophys. J. \textbf{688} (2008), 709-728
doi:10.1086/591439
[arXiv:0803.2706 [astro-ph]].

\bibitem{Ludlow:2016ifl}
A.~D.~Ludlow, S.~Bose, R.~E.~Angulo, L.~Wang, W.~A.~Hellwing, J.~F.~Navarro, S.~Cole and C.~S.~Frenk,
Mon. Not. Roy. Astron. Soc. \textbf{460} (2016) no.2, 1214-1232
doi:10.1093/mnras/stw1046
[arXiv:1601.02624 [astro-ph.CO]].

\bibitem{Lahav:1991wc}
O.~Lahav, P.~B.~Lilje, J.~R.~Primack and M.~J.~Rees,
Mon. Not. Roy. Astron. Soc. \textbf{251} (1991), 128-136
PRINT-91-0084 (UC,SANTA-CRUZ).

\bibitem{Schmidt:2012ne}
F.~Schmidt and D.~Jeong,
Phys. Rev. D \textbf{86} (2012), 083527
doi:10.1103/PhysRevD.86.083527
[arXiv:1204.3625 [astro-ph.CO]].

\bibitem{Challinor:2011bk}
A.~Challinor and A.~Lewis,
Phys. Rev. D \textbf{84} (2011), 043516
doi:10.1103/PhysRevD.84.043516
[arXiv:1105.5292 [astro-ph.CO]].

\bibitem{Jeong_2012}
D.~Jeong, F.~Schmidt and C.~M.~Hirata
Phys. Rev. D \textbf{85} (2012) no.2, 023504
doi.org/10.1103/PhysRevD.85.023504
[arXiv:1107.5427 [astro-ph.CO]].


\bibitem{DiDio:2013bqa}
E.~Di Dio, F.~Montanari, J.~Lesgourgues and R.~Durrer,
JCAP \textbf{11} (2013), 044
doi:10.1088/1475-7516/2013/11/044
[arXiv:1307.1459 [astro-ph.CO]].

\bibitem{Hamilton:1997zq}
A.~J.~S.~Hamilton,
doi:10.1007/978-94-011-4960-0\_17
[arXiv:astro-ph/9708102 [astro-ph]].

\bibitem{Siewert:2020krp}
T.~M.~Siewert, M.~Schmidt-Rubart and D.~J.~Schwarz,
Astron. Astrophys. \textbf{653} (2021), A9
doi:10.1051/0004-6361/202039840
[arXiv:2010.08366 [astro-ph.CO]].

\bibitem{Yahil1991}
A.~Yahil, M.~A.~Strauss, M. Davis and J.~P.~Hucra,
Astrophys. J. \textbf{381} (1991), 348
doi:10.1086/170657

\bibitem{Fisher:1994xm}
K.~Fisher, O.~Lahav, Y.~Hoffman, D.~Lynden-Bell and S.~Zaroubi,
[arXiv:astro-ph/9406009 [astro-ph]].

\bibitem{Elkhashab:2021lsk}
M.~Y.~Elkhashab, C.~Porciani and D.~Bertacca,
Mon. Not. Roy. Astron. Soc. \textbf{509} (2021) no.2, 1626-1645
doi:10.1093/mnras/stab3010
[arXiv:2108.13424 [astro-ph.CO]].

\bibitem{Bertacca:2019wyg}
D.~Bertacca,
Int. J. Mod. Phys. D \textbf{29} (2020) no.12, 2050085
doi:10.1142/S0218271820500856
[arXiv:1912.06887 [gr-qc]].

\bibitem{Bahr-Kalus:2021jvu}
B.~Bahr-Kalus, D.~Bertacca, L.~Verde and A.~Heavens,
JCAP \textbf{11} (2021), 027
doi:10.1088/1475-7516/2021/11/027
[arXiv:2107.00351 [astro-ph.CO]].


\bibitem{Peebles}
J.~E.~Peebles,
``The large-scale structure of the universe''

\bibitem{Tegmark:1997yq}
M.~Tegmark, A.~J.~S.~Hamilton, M.~A.~Strauss, M.~S.~Vogeley and A.~S.~Szalay,
Astrophys. J. \textbf{499} (1998), 555-576
doi:10.1086/305663
[arXiv:astro-ph/9708020 [astro-ph]].

\bibitem{Smith:2008ut}
R.~E.~Smith,
Mon. Not. Roy. Astron. Soc. \textbf{400} (2009), 851
doi:10.1111/j.1365-2966.2009.15490.x
[arXiv:0810.1960 [astro-ph]].

\bibitem{Pieroni:2022bbh}
M.~Pieroni, A.~Ricciardone and E.~Barausse,
[arXiv:2203.12586 [astro-ph.CO]].

\bibitem{Contaldi:2020rht}
C.~R.~Contaldi, M.~Pieroni, A.~I.~Renzini, G.~Cusin, N.~Karnesis, M.~Peloso, A.~Ricciardone and G.~Tasinato,
Phys. Rev. D \textbf{102} (2020) no.4, 043502
doi:10.1103/PhysRevD.102.043502
[arXiv:2006.03313 [astro-ph.CO]].

\bibitem{Hiratadipole}
C.~M.~Hirata,
JCAP \textbf{9} (2009), 011
doi:10.1088/1475-7516/2009/09/011
[arXiv:0907.0703 [astro-ph.CO]].

\bibitem{Maartens:2021dqy}
R.~Maartens, J.~Fonseca, S.~Camera, S.~Jolicoeur, J.~A.~Viljoen and C.~Clarkson,
JCAP \textbf{12} (2021) no.12, 009
doi:10.1088/1475-7516/2021/12/009
[arXiv:2107.13401 [astro-ph.CO]].

\end{thebibliography}
